\documentclass[a4paper,11pt]{article}
\usepackage{jcappub}             
\usepackage{amsmath}
\usepackage{amssymb}
\usepackage{graphicx}
\usepackage[utf8]{inputenc}
\usepackage{cancel}
\usepackage[colorlinks=true]{hyperref}
\usepackage{color}
\usepackage{array,multirow}
\usepackage{caption}
\usepackage{subcaption}
\usepackage{float}
\usepackage{comment}
\usepackage{tikz}
\usepackage{booktabs}
\usetikzlibrary{positioning}
\usetikzlibrary{decorations.pathmorphing}
\usepackage[dvipsnames]{xcolor}
\usepackage{colortbl}
\usepackage{pifont} 
\usepackage{bbold}
\definecolor{C0}{HTML}{1f77b4} 
\definecolor{C1}{HTML}{ff7f0e} 
\definecolor{C2}{HTML}{2ca02c} 
\definecolor{C3}{HTML}{d62728} 
\definecolor{C4}{HTML}{9467bd} 
\definecolor{C5}{HTML}{8c564b} 
\definecolor{C6}{HTML}{e377c2} 
\definecolor{C7}{HTML}{7f7f7f} 
\definecolor{C8}{HTML}{bcbd22} 
\definecolor{C9}{HTML}{17becf} 

\newcommand{\mchi}{\mchi}
\newcommand{\ms}{\ms}

\title{Interacting scalar--fermion two-component dark matter in a gauged
$\boldsymbol{U(1)_{B-L}}$ model}

\author[a,b]{Jacinto P. Neto,}
\author[a,c,d]{Farinaldo S. Queiroz,}
\author[e]{Carlos E. Yaguna}

\affiliation[a]{International Institute of Physics, Universidade Federal do Rio Grande do Norte, Campus Universit\'ario, Lagoa Nova, Natal-RN 59078-970, Brasil}
\affiliation[b]{University of Vienna, Faculty of Physics, Boltzmanngasse 5, A-1090 Vienna, Austria}
\affiliation[c]{Millennium Institute for Subatomic Physics at High-Energy Frontier (SAPHIR), Fernandez Concha 700, Santiago, Chile}
\affiliation[d]{Departamento de F\'isica, Facultad de Ciencias, Universidad de La Serena, Avenida Cisternas
1200, La Serena, Chile}
\affiliation[e]{Escuela de F\'isica, Universidad Pedagógica y Tecnol\'ogica de Colombia, Avenida Central del Norte, 39-115, Tunja, Colombia}

\emailAdd{jacinto.neto.100@ufrn.edu.br}
\emailAdd{farinaldo.queiroz@ufrn.br}
\emailAdd{carlos.yaguna@uptc.edu.co}

\abstract{
We propose an interacting two-component dark matter model based on a gauged $U(1)_{B-L}$ extension of the Standard Model, which also accommodates neutrino masses through the type-I seesaw mechanism. The dark sector consists of a Dirac fermion $\chi$ and a complex scalar $S$, which can both be stable without imposing an additional discrete symmetry and are directly coupled by a gauge-invariant Yukawa interaction. The resulting conversion, semi-conversion, and semi-annihilation processes lead to a genuinely coupled freeze-out. After imposing collider and dark matter constraints, we identify viable solutions away from the $Z'$ resonance for both mass hierarchies, $m_S > m_\chi$ and $m_S < m_\chi$. Despite the strong direct-detection constraints, part of the surviving parameter space remains within the reach of next-generation experiments.
}

\begin{document}
\maketitle

\section{Introduction}

The existence of dark matter (DM) is firmly established by astrophysical and cosmological observations, including galaxy rotation curves, gravitational lensing, large-scale structure formation, and measurements of the cosmic microwave background. Nevertheless, its particle nature remains unknown. Among the most extensively studied candidates are weakly interacting massive particles (WIMPs), whose abundance may be generated through thermal freeze-out in the early Universe. The apparent connection between a weak-scale annihilation cross section and the observed DM abundance---the so-called WIMP miracle---has traditionally motivated models in which a single particle accounts for the entire DM density. The absence of convincing non-gravitational signals, together with increasingly stringent limits from collider, direct-detection, and indirect-detection searches, has significantly restricted the simplest realizations of this paradigm~\cite{Abercrombie:2015wmb, Abdallah:2015ter, Arcadi:2017kky, Arcadi:2024ukq}.

An equally plausible possibility is that the dark sector contains more than one stable particle. In multi-component DM scenarios, the observed density is shared among two or more species, each contributing only a fraction of the total abundance~\cite{Tucker-Smith:2001myb, Zurek:2008qg, Esch:2014jpa, Yaguna:2019cvp, Belanger:2020hyh, Belanger:2022esk, Costa:2022oaa}. More importantly, the thermal histories of the different components need not be independent~\cite{Belanger:2011ww}. Interactions among the dark particles can induce conversion, semi-annihilation, and semi-conversion processes, thereby coupling their Boltzmann equations and qualitatively modifying the freeze-out dynamics~\cite{DiazSaez:2021pmg, Yaguna:2021rds, Escalona:2026trm}. These processes can weaken the usual connection between the relic abundance and the couplings probed by direct-detection experiments, opening viable regions that would not be available in the corresponding single-component models.

A central model-building question in such scenarios is the origin of the stability of the different DM particles. In many constructions, stability is imposed through additional discrete symmetries, such as $Z_N$ or products thereof~\cite{Yaguna:2019cvp, Yaguna:2021rds, Qi:2025jpm, Frank:2025jjt}. Although perfectly consistent at the phenomenological level, these symmetries are often introduced specifically for this purpose. It is therefore worthwhile to investigate frameworks in which the stability of multiple dark particles follows instead from an underlying gauge symmetry. Such a framework becomes especially attractive when the same gauge extension also addresses another open problem of the Standard Model (SM), such as the origin of neutrino masses.

The gauged $U(1)_{B-L}$ extension provides a simple realization of this idea~\cite{Basso:2010jm, Rodejohann:2015lca, Klasen:2016qux, Bernal:2018aon, Okada:2018ktp, Das:2022oyx}. In its minimal anomaly-free implementation, it contains three right-handed neutrinos and a complex scalar singlet $\Phi_{BL}$. The vacuum expectation value of $\Phi_{BL}$ spontaneously breaks the $U(1)_{B-L}$ symmetry, generates the mass of the new gauge boson $Z'$, and gives Majorana masses to the right-handed neutrinos. Together with their Dirac Yukawa interactions, this leads to light neutrino masses through the type-I seesaw mechanism~\cite{Mohapatra:1979ia, VanDong:2020bkg}. The scalar excitation associated with $\Phi_{BL}$ also mixes with the SM Higgs boson, giving rise to a second physical scalar, denoted by $h_2$.

In this work, we extend this framework by introducing only two additional fields: a Dirac fermion $\chi$ and a complex scalar $S$, both singlets under the SM gauge group and charged under $U(1)_{B-L}$. Their charges are chosen so that gauge invariance allows a Yukawa interaction that directly couples the two DM particles. This interaction is a defining feature of the model, as it enables conversion, semi-annihilation, and semi-conversion processes that couple the thermal evolution of the two components. For suitable $B-L$ charge assignments, gauge invariance forbids additional renormalizable operators that would destabilize the dark sector, while the kinematic condition $m_S < 2m_\chi$ prevents the decay of the scalar into a dark fermion pair. Both particles can therefore be stable without imposing an additional discrete symmetry, and their relic abundances are determined by a coupled network of dark-sector processes rather than by two independent freeze-out histories.

We perform a numerical exploration of the multidimensional parameter space, solving the coupled Boltzmann equations and imposing theoretical consistency conditions together with collider and dark matter constraints. We analyze separately the two mass hierarchies, $m_S>m_\chi$ and $m_S<m_\chi$, and identify viable regions in both regimes. In particular, viable solutions occur over broad regions away from the $Z'$ resonance, showing that the observed relic abundance does not require resonant enhancement. The coupled dark-sector processes can instead play a central role in determining the relic composition. Direct detection provides a strong constraint on the model, while part of the surviving parameter space remains within the projected reach of next-generation experiments.

The remainder of this paper is organized as follows. In Sec.~\ref{sec:model}, we introduce the model and discuss its scalar and dark matter sectors. Section~\ref{sec:relic} describes the coupled freeze-out dynamics, while the relevant experimental constraints are presented in Sec.~\ref{sec:exp-constraints}. The numerical results are discussed in Sec.~\ref{sec:results}, followed by a broader discussion in Sec.~\ref{sec:discussion}. We summarize our conclusions in Sec.~\ref{sec:conclusions}.

\section{The $U(1)_{B-L}$ two-component dark matter model}
\label{sec:model}

\subsection{Field content and Lagrangian}
\label{sec:lagrangian}

We consider a gauge extension of the SM based on
\begin{equation}
    SU(3)_C \otimes SU(2)_L \otimes U(1)_Y
    \otimes U(1)_{B-L},
\end{equation}
where $B$ and $L$ denote baryon and lepton number, respectively. The cancellation of the gauge and mixed gravitational anomalies requires the introduction of three right-handed neutrinos, $N_{iR}$, with $i=1,2,3$. The $U(1)_{B-L}$ symmetry is spontaneously broken by a complex scalar singlet $\Phi_{BL}$ with $B-L$ charge equal to $2$. The mass of the new gauge boson is $m_{Z'} = 2 g_{BL} v_\Phi$, where $g_{BL}$ is the $U(1)_{B-L}$ gauge coupling and $\langle\Phi_{BL}\rangle = v_\Phi/\sqrt{2}$ defines the $\Phi_{BL}$'s vacuum expectation value.

The dark sector contains a Dirac fermion $\chi$ and a complex scalar $S$, both singlets under the SM gauge group. The left- and right-handed components of $\chi$ carry the same $B-L$ charge $n_\chi$, so that $\chi$ is vector-like and does not introduce additional gauge anomalies. To obtain a genuinely interacting two-component dark sector, rather than a simple superposition of two independent single-component DM models, we require a renormalizable Yukawa interaction directly coupling $\chi$ and $S$. Gauge invariance of this interaction fixes their charges to satisfy $Q_{B-L}(S) = -2 Q_{B-L}(\chi) = - 2n_\chi$. The complete field content and charge assignments are summarized in Table~\ref{tab:particlecontent}.

\begin{table}[t]
    \centering
    \renewcommand{\arraystretch}{1.15}
    \begin{tabular}{ccccc}
        \toprule
        Field
        & $SU(3)_C$
        & $SU(2)_L$
        & $U(1)_Y$
        & $U(1)_{B-L}$ \\
        \midrule
        $q_{iL}$     & $\mathbf{3}$ & $\mathbf{2}$ & $1/6$  & $1/3$ \\
        $u_{iR}$     & $\mathbf{3}$ & $\mathbf{1}$ & $2/3$  & $1/3$ \\
        $d_{iR}$     & $\mathbf{3}$ & $\mathbf{1}$ & $-1/3$ & $1/3$ \\
        $\ell_{iL}$  & $\mathbf{1}$ & $\mathbf{2}$ & $-1/2$ & $-1$ \\
        $e_{iR}$     & $\mathbf{1}$ & $\mathbf{1}$ & $-1$   & $-1$ \\
        $N_{iR}$     & $\mathbf{1}$ & $\mathbf{1}$ & $0$    & $-1$ \\
        $H$           & $\mathbf{1}$ & $\mathbf{2}$ & $1/2$  & $0$ \\
        $\Phi_{BL}$   & $\mathbf{1}$ & $\mathbf{1}$ & $0$    & $2$ \\
        $\chi$        & $\mathbf{1}$ & $\mathbf{1}$ & $0$    & $n_\chi$ \\
        $S$           & $\mathbf{1}$ & $\mathbf{1}$ & $0$    & $-2n_\chi$ \\
        \bottomrule
    \end{tabular}
    \caption{Field content and gauge-charge assignments of the
    two-component DM model.}
    \label{tab:particlecontent}
\end{table}

The $B-L$ charge $n_\chi$ plays a crucial role in determining the structure of the dark sector. While most values lead to the minimal interacting two-component model considered here, certain special assignments either prevent the existence of two stable DM particles or allow additional interactions that modify the model. The relevant cases are summarized in Table~\ref{tab:exceptional-charges}.
Any value of $n_\chi$ not listed in Table~\ref{tab:exceptional-charges} defines a consistent realization of the minimal model at the renormalizable level. For the numerical analysis, we adopt the benchmark charge assignment $n_\chi=3/4$, which is an $\mathcal{O}(1)$ value compatible with the minimal two-component dark matter scenario defined above.

\begin{table}[t!]
\centering
\begin{tabular}{c c p{8.2cm}}
\hline\hline
Classification & $n_\chi$ & Consequence \\
\hline
Nonviable & $0, \ \pm \dfrac{1}{3},\ \pm 1,\ \pm 2,\ \pm 3$ & The two-component dark matter scenario considered here cannot be realized. \\[1mm]
Viable but nonminimal & $\pm\dfrac{1}{2}$ & Additional renormalizable interactions are allowed, leading to an extended two-component dark matter model. \\[1mm]
Conditionally viable & $\pm 4$ & The renormalizable model remains viable, but higher-dimensional operators can induce the decay of the scalar dark matter particle. \\
\hline\hline
\end{tabular}
\caption{Special values of the dark $B-L$ charge $n_\chi$ and their
implications for the two-component dark matter scenario.}
\label{tab:exceptional-charges}
\end{table}

The most general renormalizable gauge Lagrangian also contains kinetic mixing between $U(1)_Y$ and $U(1)_{B-L}$. For simplicity, we neglect gauge kinetic mixing throughout the phenomenological analysis. The terms relevant for neutrino masses and DM phenomenology can then be written as
\begin{align}
\mathcal{L} \supset {}&
    \mathcal{L}_{\rm SM}^{B-L}
    -\frac{1}{4}F'_{\mu\nu}F^{\prime\mu\nu}
    +\sum_{i=1}^{3}
        i\overline{N}_{iR}\gamma^\mu D_\mu N_{iR} + i\overline{\chi}\gamma^\mu D_\mu\chi -m_\chi\overline{\chi}\chi   \nonumber\\
    & +(D_\mu\Phi_{BL})^\dagger(D^\mu\Phi_{BL})
    +(D_\mu S)^\dagger(D^\mu S) \nonumber\\
    &-\left[
    y_{ij}^{D}\overline{L}_i\widetilde{H}N_{jR}
    +\frac{1}{2}y_{ij}^{M}
       \overline{N_{iR}^{c}}\,N_{jR}\Phi_{BL}
    +\frac{1}{2}S\,
       \overline{\chi^{c}}
       \left(y_s+y_p\gamma_5\right)\chi
    +\mathrm{H.c.}
  \right]
  \nonumber\\
&-V(H,\Phi_{BL},S),
\label{eq:lagrangian}
\end{align}
where $\mathcal{L}_{\rm SM}^{B-L}$ denotes the SM Lagrangian with the covariant derivatives extended to include the $U(1)_{B-L}$ gauge interaction, and $\widetilde H=i\sigma_2H^*$. For a field carrying $B-L$ charge $n_j$, the corresponding contribution to the covariant derivative is $D_\mu \supset \partial_\mu
    +i g_{BL}n_j Z'_\mu$.
After electroweak and $U(1)_{B-L}$ symmetry breaking, the Yukawa couplings $y^D_{ij}$ and $y^M_{ij}$ generate the Dirac and Majorana neutrino mass matrices, respectively, leading to light neutrino masses through the type-I seesaw mechanism. These parameters can be chosen to reproduce the observed neutrino masses and mixings while keeping the right-handed neutrinos sufficiently heavy that their production and decay channels are kinematically inaccessible throughout the parameter space considered. The neutrino sector therefore does not affect the dark matter phenomenology studied below.

The last Yukawa interaction in Eq.~\eqref{eq:lagrangian} is allowed because the dark-sector charges satisfy $Q_{B-L}(S)= -2 Q_{B-L}(\chi)$. It directly couples the two dark particles and induces the conversion, semi-annihilation, and semi-conversion processes that distinguish this model from two independently freezing-out DM components.
Finally, the scalar potential is
\begin{align}
V(H,\Phi_{BL},S)={}&
    \mu_H^2 H^\dagger H
    +\mu_\Phi^2\Phi_{BL}^\dagger\Phi_{BL}
    +\mu_S^2S^\dagger S
    \nonumber\\
&+\lambda_H(H^\dagger H)^2
 +\lambda_\Phi
    (\Phi_{BL}^\dagger\Phi_{BL})^2
 +\lambda_S(S^\dagger S)^2
    \nonumber\\
&+\lambda_{H\Phi}
    (H^\dagger H)(\Phi_{BL}^\dagger\Phi_{BL})
 +\lambda_{SH}(H^\dagger H)(S^\dagger S)
    \nonumber\\
&+\lambda_{S\Phi}
    (S^\dagger S)(\Phi_{BL}^\dagger\Phi_{BL}).
\label{eq:scalar-potential}
\end{align}
The mass-squared parameters $\mu_H^2$, $\mu_\Phi^2$, and $\mu_S^2$ are not assumed to be positive. Only $H$ and $\Phi_{BL}$ acquire vacuum expectation values (VEVs), whereas the vacuum is required to satisfy $\langle S\rangle=0$. The conditions under which this configuration is a consistent minimum are discussed in Sec.~\ref{sec:vacuum}.

\subsection{Symmetry breaking and scalar mixing}
\label{sec:symmetry}

After electroweak and $U(1)_{B-L}$ symmetry breaking, the Higgs doublet and the $B-L$ scalar acquire VEVs. In the unitary gauge, they can be parametrized as
\begin{equation}
    H= \frac{1}{\sqrt{2}} \begin{pmatrix}
                                0 \\
                                v_H+h
                            \end{pmatrix},
    \qquad
    \Phi_{BL} = \frac{1}{\sqrt{2}} \left(v_\Phi + \phi\right),
\end{equation}
where $v_H=246~\mathrm{GeV}$. We require the dark scalar not to acquire a VEV, $\langle S\rangle = 0$, because its CP-even excitation would generically mix with those of $H$ and $\Phi_{BL}$. Through this mixing, the predominantly-$S$ state would acquire couplings to visible particles and would no longer be stable, thereby spoiling the two-component DM scenario. In Sec.~\ref{sec:vacuum}, we discuss the conditions under which the vacuum with $\langle S\rangle=0$ is realized.

Assuming $v_H,v_\Phi\neq0$, the minimal conditions of the scalar potential are
\begin{align}
    &\mu_H^2 + \lambda_Hv_H^2 + \frac{1}{2}\lambda_{H\Phi}v_\Phi^2 = 0, \qquad
    \mu_\Phi^2 + \lambda_\Phi v_\Phi^2 + \frac{1}{2}\lambda_{H\Phi}v_H^2  = 0.
\end{align}
Since $S$ does not mix with the other neutral scalars, it remains a physical DM state with mass
\begin{equation}
m_S^2=
\mu_S^2
+\frac{1}{2}\lambda_{SH}v_H^2
+\frac{1}{2}\lambda_{S\Phi}v_\Phi^2.
\end{equation}

The remaining CP-even scalar fields, $h$ and $\phi$, mix through the portal coupling $\lambda_{H\Phi}$. Their mass-squared matrix in the $(h,\phi)$ basis is
\begin{equation}
    \mathcal{M}^2 = 
        \begin{pmatrix} 
            2\lambda_Hv_H^2 & \lambda_{H\Phi}v_Hv_\Phi \\
            \lambda_{H\Phi}v_Hv_\Phi & 2\lambda_\Phi v_\Phi^2
        \end{pmatrix}.
\end{equation}
The mass eigenstates are obtained through the orthogonal transformation
\begin{equation}
    h_1=\cos\theta\,h-\sin\theta\,\phi,
    \qquad
    h_2=\sin\theta\,h+\cos\theta\,\phi.
\end{equation}
where $h_1$ is identified with the observed Higgs boson, with $m_{h_1} = 125$~GeV, while $h_2$ is the predominantly $B-L$ scalar. With this rotation convention, the mixing angle satisfies
\begin{equation}
    \tan 2\theta= \frac{\lambda_{H\Phi}v_Hv_\Phi} {\lambda_\Phi v_\Phi^2-\lambda_Hv_H^2}.
\end{equation}

For the numerical analysis, it is convenient to express the quartic couplings $\lambda_H$, $\lambda_\Phi$, and $\lambda_{H\Phi}$ in terms of the physical scalar masses, the mixing angle, and the vacuum expectation values. The corresponding relations are
\begin{align}
    \lambda_H & = \frac{ m_{h_1}^2+m_{h_2}^2 + \left(m_{h_1}^2-m_{h_2}^2\right)\cos 2\theta} {4v_H^2}, \\
    \lambda_\Phi & = \frac{m_{h_1}^2+m_{h_2}^2 - \left(m_{h_1}^2-m_{h_2}^2\right)\cos 2\theta} {4v_\Phi^2}, \\
    \lambda_{H\Phi} & = \frac{\left(m_{h_2}^2-m_{h_1}^2\right)\sin 2\theta}{2v_Hv_\Phi}.
\end{align}
The mass parameters $\mu_H^2$ and $\mu_\Phi^2$ are then fixed by the minimal conditions. Thus, once $m_{h_1}$ and $v_H$ are fixed to their measured values, the mixed $h$--$\phi$ sector can be parametrized by $\{m_{h_2},m_{Z'},g_{BL},\theta\}$, with $v_\Phi=m_{Z'}/(2g_{BL})$.

\subsection{Boundedness and local vacuum stability}
\label{sec:vacuum}

A consistent scalar sector requires the potential to be bounded from below and the stationary point
\begin{equation}
    \langle H\rangle=\frac{v_H}{\sqrt{2}}, \qquad
    \langle\Phi_{BL}\rangle=\frac{v_\Phi}{\sqrt{2}}, \qquad 
    \langle S \rangle=0
\end{equation}
to be locally stable. These two requirements constrain different properties of the potential: boundedness from below controls its large-field behavior, whereas local stability is determined by the scalar mass matrix evaluated at the stationary point.

At large field values, the quartic part of the scalar potential dominates. Applying the copositivity criteria to the three non-negative field bilinears $H^\dagger H$, $\Phi_{BL}^\dagger\Phi_{BL}$, and $S^\dagger S$, the necessary and sufficient bounded-from-below conditions are~\cite{Kannike:2012pe} $\lambda_H\geq0$, $\lambda_\Phi\geq0$,  $\lambda_S\geq0$, together with
\begin{align*}
    & \overline{\lambda}_{H\Phi} \equiv
    \lambda_{H\Phi} + 2\sqrt{\lambda_H\lambda_\Phi} \geq 0,
    \quad
    \overline{\lambda}_{SH} \equiv \lambda_{SH}+ 2\sqrt{\lambda_S\lambda_H} \geq 0,
    \quad \overline{\lambda}_{S\Phi} \equiv \lambda_{S\Phi} +2\sqrt{\lambda_S\lambda_\Phi} \geq 0,
\end{align*}
and
\begin{align}
    \sqrt{\lambda_H\lambda_\Phi\lambda_S} 
    + \frac{1}{2}\Bigl( \lambda_{H\Phi}\sqrt{\lambda_S} + \lambda_{SH}\sqrt{\lambda_\Phi} + \lambda_{S\Phi}\sqrt{\lambda_H} + \sqrt{ \overline{\lambda}_{H\Phi} \overline{\lambda}_{SH} \overline{\lambda}_{S\Phi} }\Bigr) \geq 0.
\label{eq:bfb-conditions}
\end{align}
All these inequalities must be imposed simultaneously; the last condition is additional to the three pairwise conditions.
Local stability in the $h$--$\phi$ sector follows from the positivity of the physical scalar masses, $m_{h_1}^2>0$, $m_{h_2}^2>0$. 
Since $S$ does not mix with the other scalars at this stationary point, local stability along both real components of the complex field $S$ requires $m_S^2 > 0$.

In the numerical analysis, we impose the bounded-from-below conditions together with positive physical scalar masses. 

\subsection{Dark matter stability}
\label{sec:dm-stability}

Assuming the vacuum configuration $\langle S\rangle=0$, the stability of the dark particles can be inferred directly from the renormalizable Lagrangian in Eq.~\eqref{eq:lagrangian}. We first consider the dark fermion $\chi$. Its gauge interactions are bilinear in $\chi$, and the dark Yukawa interaction also contains two fermionic fields. Moreover, the Lagrangian contains no operator involving a single $\chi$ and only lighter particles. Consequently, any process mediated by the dark Yukawa interaction with one $\chi$ in the initial state would require another particle of mass $m_\chi$ in the final state and is therefore kinematically forbidden. The fermion $\chi$ is thus always stable within the renormalizable model.

The situation is different for the complex scalar $S$. Its gauge and portal interactions contain $S$ only through the bilinear $S^\dagger S$ and hence cannot mediate the decay of a single $S$ particle. However, the Yukawa interaction
\begin{equation}
    \frac{1}{2}S\,\overline{\chi^c}    \left(y_s+y_p\gamma_5\right)\chi+\mathrm{H.c.}
\end{equation}
allows $S$ and $S^\dagger$ to decay into pairs of dark antifermions and fermions, respectively, whenever these channels are kinematically open. Therefore, two qualitatively different regimes are possible:
\begin{equation}
\begin{aligned}
    m_S>2m_\chi  & : \qquad S\ \text{is unstable and $\chi$ is the only DM particle}, \\
    m_S<2m_\chi & : \qquad S\ \text{and $\chi$ are both stable}.
\end{aligned}
\end{equation}
Throughout this work, we focus exclusively on the two-component DM regime and therefore impose $m_S<2m_\chi$. No additional discrete symmetry is then required to stabilize the two dark matter particles. Hence, our work differs from previous studies in the literature, where more symmetries are often invoked.

\subsection{Interaction structure and model parameters}
\label{sec:interactions}

The phenomenology of the model is governed by three types of interactions. The $U(1)_{B-L}$ gauge coupling $g_{BL}$ connects both DM particles to the $Z'$ and mediates annihilation into SM states as well as conversion between the two dark components. The scalar portal couplings $\lambda_{SH}$ and $\lambda_{S\Phi}$ connect $S$ to the physical scalars $h_1$ and $h_2$, providing an additional link between the scalar DM particle and the visible sector. The fermion $\chi$ has no direct coupling to $h_1$ or $h_2$.

Finally, the dark Yukawa couplings $y_s$ and $y_p$ directly connect $\chi$ and $S$ and generate conversion, semi-conversion, and semi-annihilation processes. These interactions couple the thermal evolution of the two dark matter components and constitute a defining feature of the model.

After fixing $n_\chi=\frac{3}{4}$, $m_{h_1}=125~\mathrm{GeV}$, $v_H=246~\mathrm{GeV}$, the independent parameters relevant to the DM phenomenology
are
\begin{equation}
\left\{
    m_\chi,\,
    m_S,\,
    m_{Z'},\,
    m_{h_2},\,
    g_{BL},\,
    y_s,\,
    y_p,\,
    \lambda_{SH},\,
    \lambda_{S\Phi},\,
    \sin\theta
\right\}.
\label{eq:model-parameters}
\end{equation}
The $B-L$ symmetry-breaking scale is fixed by $v_\Phi=m_{Z'}/(2g_{BL})$, while $\lambda_H$, $\lambda_\Phi$, $\lambda_{H\Phi}$, $\mu_H^2$, and $\mu_\Phi^2$ are determined by the physical scalar masses, mixing angle, and VEVs through the relations in Sec.~\ref{sec:symmetry}. The parameter $\mu_S^2$ is similarly fixed by the physical mass $m_S$. The scalar self-coupling $\lambda_S$ does not enter directly into the relic-density or detection observables considered below, but it enters the bounded-from-below conditions of Sec.~\ref{sec:vacuum}; throughout the numerical analysis we fix $\lambda_S=1$.

For the phenomenological analysis, the model was implemented in \textsc{FeynRules}~\cite{Alloul:2013bka} and cross-checked against an independent implementation in \textsc{LanHEP}~\cite{Semenov:2014rea}.

\section{Relic abundance and dark-sector dynamics}
\label{sec:relic}

We assume that both DM particles were initially in thermal and chemical equilibrium with the primordial plasma. As the temperature decreases below their masses, their abundances depart from equilibrium and eventually freeze out through a network of annihilation, conversion, semi-conversion, and semi-annihilation processes. Because the scalar and fermionic DM particles interact directly, their thermal histories are coupled and must be described by a system of Boltzmann equations.

We assume vanishing particle-antiparticle asymmetries and define the total comoving abundances of the two dark sectors as
\begin{equation}
    Y_\chi
    \equiv
    \frac{\mathbb{n}_\chi + \mathbb{n}_{\bar{\chi}}}{\mathbb{s}},
    \qquad
    Y_S
    \equiv
    \frac{\mathbb{n}_S + \mathbb{n}_{S^\dagger}}{\mathbb{s}},
\end{equation}
where $\mathbb{s}$ is the entropy density. The corresponding equilibrium abundances are denoted by $Y_\chi^{\rm eq}$ and $Y_S^{\rm eq}$. We use the dimensionless inverse-temperature variable
\begin{equation}
    z\equiv\frac{\mu}{T},
    \qquad
    \mu\equiv\frac{m_\chi m_S}{m_\chi+m_S},
\end{equation}
where $T$ is the temperature of the thermal bath and $\mu$ is the reduced mass of the two DM particles.

Following the sector notation adopted in \textsc{micrOMEGAs}, we label $\chi$ and $\bar{\chi}$ as sector 1, $S$ and $S^\dagger$ as sector 2, and all remaining particles as sector 0~\cite{Belanger:2014vza,Alguero:2023zol}. In this notation, the coupled Boltzmann equations are
\begin{align}
    \frac{dY_\chi}{dz}={}&
    -\lambda_{1100}\!\left[Y_\chi^2-(Y_\chi^{\rm eq})^2\right]
    -\lambda_{1120}\!\left[
    Y_\chi^2-Y_S\frac{(Y_\chi^{\rm eq})^2}{Y_S^{\rm eq}}
    \right] 
    -\lambda_{1122}\!\left[Y_\chi^2-Y_S^2 \left(\frac{Y_\chi^{\rm eq}}{Y_S^{\rm eq}}\right)^2 \right], \label{eq:BEQchi} \\[1mm]
    \frac{dY_S}{dz}={}&
    -\lambda_{2200}\!\left[Y_S^2-(Y_S^{\rm eq})^2\right]
    -\lambda_{2211}\!\left[
    Y_S^2-Y_\chi^2
    \left(\frac{Y_S^{\rm eq}}{Y_\chi^{\rm eq}}\right)^2 \right] \nonumber\\
    &-\frac{1}{2}\lambda_{1210}Y_\chi
    \left(Y_S-Y_S^{\rm eq}\right)
    +\frac{1}{2}\lambda_{1120}\!\left[
    Y_\chi^2-Y_S\frac{(Y_\chi^{\rm eq})^2}{Y_S^{\rm eq}} \right].
\label{eq:BEQS}
\end{align}

The temperature-dependent coefficients appearing in Eqs.~\eqref{eq:BEQchi} and \eqref{eq:BEQS} are defined as
\begin{equation}
    \lambda_{ijkl}(z)
    \equiv
    \frac{
        \mathbb{s}(z)\,
        \langle\sigma v\rangle_{ijkl}(z)
    }{
        z\,\overline{H}(z)
    },
\label{eq:lambda-ijkl}
\end{equation}
where $\langle\sigma v\rangle_{ijkl}$ is the thermally averaged cross section for a reaction involving particles from sectors $i,j\rightarrow k,l$~\cite{Gondolo:1990dk}. The quantity
\begin{equation}
    \overline{H}(T)
    \equiv
    \frac{H(T)}
    {
        1+\dfrac{1}{3}
        \dfrac{d\ln g_{*s}(T)}{d\ln T}
    }
\end{equation}
accounts for the temperature dependence of the effective number of entropy degrees of freedom, $g_{*s}$. In the limit in which $g_{*s}$ is constant, one has $\overline{H}=H$. The reaction classes contributing to the coupled thermal evolution are summarized in Table~\ref{tab:dmprocesses}. Here $X_0$ denotes any kinematically accessible particle belonging to sector 0. Charge-conjugate reactions are included implicitly.

\begin{table}[t]
\centering
\renewcommand{\arraystretch}{1.15}
\begin{tabular}{lll}
\toprule
\textbf{Process} &
\textbf{Label} &
\textbf{Classification}
\\
\midrule
$\chi\bar{\chi}\rightarrow X_0X_0$ & $1100$ & Annihilation \\
$\chi\chi\rightarrow S^\dagger X_0$ and c.c. & $1120$ & Semi-conversion \\
$\chi\bar{\chi}\rightarrow SS^\dagger$ & $1122$ & Conversion \\
$SS^\dagger\rightarrow X_0X_0$ & $2200$ & Annihilation \\
$SS^\dagger\rightarrow\chi\bar{\chi}$ & $2211$ & Conversion \\
$\chi S\rightarrow\bar{\chi}X_0$ and c.c. & $1210$ & Semi-annihilation \\
\bottomrule
\end{tabular}
\caption{Reaction classes contributing to the thermal evolution of the two DM components. Sector 1 contains $\chi$ and $\bar{\chi}$, sector 2 contains $S$ and $S^\dagger$, and sector 0 contains all remaining particles.}
\label{tab:dmprocesses}
\end{table}

The annihilation processes $1100$ and $2200$ deplete the corresponding DM component into particles in the thermal bath. The conversion processes $1122$ and $2211$ redistribute the abundance between the two dark sectors without changing the total number of dark particles. By contrast, the semi-conversion process $1120$ transforms two fermionic DM particles into one scalar DM particle and one thermal plasma state. It therefore removes two particles from the fermionic sector while producing only one particle in the scalar sector, accounting for the relative factor of $1/2$ in Eq.~\eqref{eq:BEQS}. Finally, the $1210$ process depletes the scalar sector while leaving the total fermionic abundance unchanged and therefore contributes only to the evolution equation for $Y_S$.

We solve the coupled system in Eqs.~\eqref{eq:BEQchi} and \eqref{eq:BEQS} numerically using \textsc{micrOMEGAs}~7.1.1 \cite{Belanger:2014vza, Alguero:2023zol, alguero_2024_10844949, Belanger:2026asz}. The numerical calculation automatically includes all kinematically accessible channels belonging to the reaction classes listed in Table~\ref{tab:dmprocesses}, including channels containing a $Z'$ boson in the final state whenever they are open~\cite{Feng:2017drg}.

The present-day DM abundance is the sum of the scalar and fermionic contributions,
\begin{equation}
    \Omega_{\rm DM}h^2
    =
    \Omega_\chi h^2+\Omega_S h^2.
\end{equation}
In terms of the asymptotic abundances obtained by solving the coupled Boltzmann equations,
\begin{equation}
    \Omega_{\rm DM}h^2
    \simeq
    2.74\times10^8
    \left(
        m_\chi Y_\chi^0
        +
        m_S Y_S^0
    \right)
    \mathrm{GeV}^{-1},
\label{eq:total-relic}
\end{equation}
where $Y_\chi^0$ and $Y_S^0$ denote the present-day values of the corresponding comoving abundances. The observed cold DM abundance is~\cite{Planck:2018vyg}

   $ \Omega_{\rm DM}h^2
    =
    0.120\pm0.001 $.
We retain parameter points for which the sum of the two contributions is compatible with this value. It is also convenient to define the fractional contribution of each component to the total DM density,
\begin{equation}
    \xi_\chi
    \equiv
    \frac{\Omega_\chi}{\Omega_{\rm DM}},
    \qquad
    \xi_S
    \equiv
    \frac{\Omega_S}{\Omega_{\rm DM}},
    \qquad
    \xi_\chi+\xi_S=1.
\label{eq:dm-fractions}
\end{equation}
These fractions will be used below to rescale the predicted direct- and indirect-detection signals of the two DM components.

In the next section, we confront the model with the relevant collider, direct-detection, and indirect-detection constraints.

\section{Experimental constraints}
\label{sec:exp-constraints}

\subsection{Collider constraints}
\label{sec:collider}

Collider searches constrain the mediators connecting the dark and visible sectors, namely the scalar states $h_1$ and $h_2$ and the $U(1)_{B-L}$ gauge boson $Z'$. Although the scalar and gauge sectors are related through the spontaneous breaking of $U(1)_{B-L}$, it is convenient to discuss their collider constraints separately.

\subsubsection{Higgs and scalar-sector constraints}
\label{sec:scalar-collider}

The scalar singlet $\Phi_{BL}$ mixes with the neutral CP-even component of the SM Higgs doublet, giving rise to the physical states $h_1$ and $h_2$. We identify $h_1$ with the observed SM-like Higgs boson and fix $m_{h_1}=125~\mathrm{GeV}$, whereas $h_2$ denotes the additional, predominantly $B-L$ scalar.
The couplings of $h_1$ to SM particles are universally rescaled by $\cos\theta$ relative to their SM values~\cite{Bhattiprolu:2022ycf}. We restrict the analysis to the region in which decays of $h_1$ into dark-sector particles are kinematically forbidden.  Its production cross sections and visible partial decay widths are therefore rescaled by $\cos^2\theta$, while its total width is modified by the same factor. Consequently, the signal-strength modifier for any visible SM final state is
\begin{equation}
    \mu_{h_1}
    \equiv
    \frac{\sigma(pp\to h_1)}
         {\sigma_{\mathrm{SM}}(pp\to h)}
    \frac{\operatorname{BR}(h_1\to X_{\mathrm{SM}})}
         {\operatorname{BR}_{\mathrm{SM}}(h\to X_{\mathrm{SM}})}
    =
    \cos^2\theta.
\label{eq:higgs-signal-strength}
\end{equation}
The ATLAS and CMS Higgs signal-strength measurements adopted in this analysis are~\cite{ATLAS:2022vkf, CMS:2022dwd} $\mu_{\mathrm{ATLAS}}=1.05\pm0.06$, $\mu_{\mathrm{CMS}}=1.00\pm0.06$, which yield the approximate combined value $\mu_{\mathrm{ATLAS+CMS}}
    \simeq
    1.03\pm0.06$. Requiring consistency at the 95\% confidence level gives the bound
\begin{equation}
    |\sin\theta|\lesssim0.27.
\label{eq:higgs-mixing-bound}
\end{equation}

Direct searches for the additional scalar $h_2$ provide complementary constraints that depend on its mass, mixing angle, and branching fractions. Since $h_2$ couples to SM particles through its Higgs-doublet component, its production cross sections through SM mechanisms are suppressed by $\sin^2\theta$: $\sigma(pp\to h_2) = \sin^2\theta \, \sigma_{\mathrm{SM}}(m_{h_2})$,
where $\sigma_{\mathrm{SM}}(m_{h_2})$ is the production cross section of an SM Higgs boson evaluated at the mass $m_{h_2}$. Gluon fusion provides the dominant production mechanism over most of the mass range, while vector-boson fusion and associated production offer additional sensitivity.

The most relevant searches involve the decay channels
    $pp\to h_2\to WW$,  $pp\to h_2\to ZZ$,
    and $pp\to h_2\to h_1h_1$.
We impose the 95\% confidence-level limits from the ATLAS and CMS heavy-scalar searches~\cite{CMS:2015hra, CMS:2018amk, ATLAS:2018sbw, ATLAS:2020fry, ATLAS:2020tlo, CMS:2021yci, ATLAS:2022xzm}, using the recast described in Ref.~\cite{Ferber:2023iso}. For every parameter point, the predicted production rate multiplied by the corresponding branching fraction is compared with the experimental upper limit:
\begin{equation}
    \sigma(pp\to h_2)\, \operatorname{BR}(h_2\to X)
    \leq
    \left[
    \sigma(pp\to h_2)\,
    \operatorname{BR}(h_2\to X)
    \right]_{\mathrm{95\%\,CL}},
\end{equation}
with $X=WW,\, ZZ,\, h_1h_1$. The branching fractions of $h_2$ are evaluated within the full model. In addition to its decays into SM particles, the channels $h_2\to h_1h_1$ and $h_2\to SS^\dagger$ may modify its total width whenever they are kinematically open. Possible decays into other non-SM states must likewise be included when allowed.

\subsubsection{Searches for the \texorpdfstring{$Z'$}{Z'} gauge boson}
\label{sec:zprime-collider}

The $Z'$ boson couples to SM quarks and leptons according to their $B-L$ charges and can therefore be resonantly produced at hadron colliders through the Drell--Yan process, $pp\to Z'\to f\bar f$. Searches in dilepton final states provide particularly strong constraints because of their clean experimental signature, $pp\to Z'\to \ell^+\ell^-, \qquad \ell=e,\mu$, with dedicated searches performed by both ATLAS and CMS~\cite{ATLAS:2019erb,CMS:2021ctt}. Dijet and top-quark final states provide complementary, though typically weaker, sensitivity~\cite{ATLAS:2019fgd,ATLAS:2019npw}.

The predicted dilepton rate, $\sigma(pp\to Z')\,{\rm BR}(Z'\to\ell^+\ell^-)$, depends on $m_{Z'}$, $g_{BL}$, and the total $Z'$ width. In particular, the branching fraction into charged leptons can be reduced when decays into dark matter particles or other non-SM states are kinematically accessible. Collider constraints on $m_{Z'}$ must therefore be applied as a function of $g_{BL}$ and of the complete particle spectrum, rather than as a universal lower bound on the $Z'$ mass~\cite{Arcadi:2023lwc, Queiroz:2024ipo}. For every parameter point, we compare the predicted dilepton signal with the 95\% confidence-level upper limits reported by ATLAS and CMS~\cite{ATLAS:2019erb, CMS:2021ctt},
\begin{equation}
    \sigma(pp\to Z')\,
    \operatorname{BR}(Z'\to\ell^+\ell^-)
    \leq
    \left[
    \sigma(pp\to Z')\,
    \operatorname{BR}(Z'\to\ell^+\ell^-)
    \right]_{\mathrm{95\%\,CL}}.
\label{eq:zprime-collider-bound}
\end{equation}
Motivated by the strong sensitivity of current dilepton searches to $B-L$ gauge bosons in the multi-TeV mass range~\cite{Arcadi:2023lwc, Queiroz:2024ipo}, we restrict the numerical scan to $m_{Z'} \geq 4~\mathrm{TeV}$. This lower limit defines the mass range explored in the present analysis; it should not be interpreted as a coupling-independent experimental exclusion, since the collider sensitivity decreases rapidly for small values of $g_{BL}$ and depends on the available $Z'$ decay modes.

\subsection{Direct detection bounds}
\label{sec:direct-detection}

\begin{figure}[t]
\centering
    \begin{subfigure}[b]{0.45\textwidth}
        \centering
        \includegraphics[width=\textwidth]{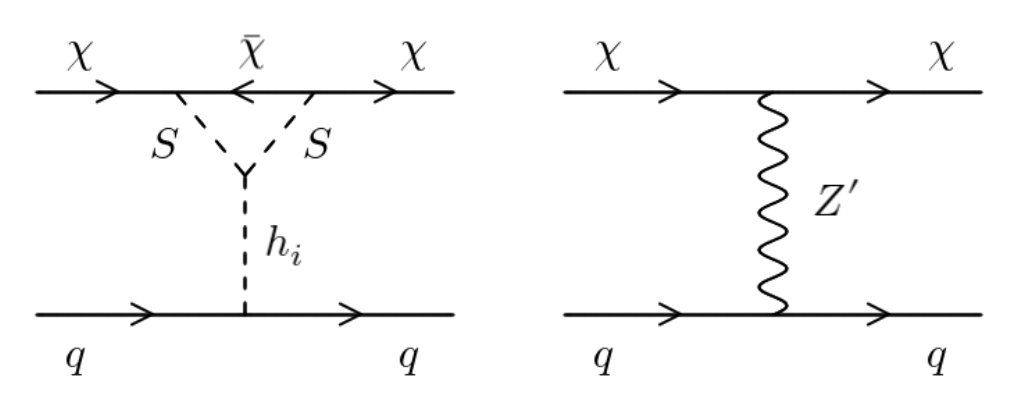}
        \caption{Fermion, $\chi$.}
        \label{fig:ddchi}
    \end{subfigure}
    \hfill
    \begin{subfigure}[b]{0.45\textwidth}
        \centering
        \includegraphics[width=\textwidth]{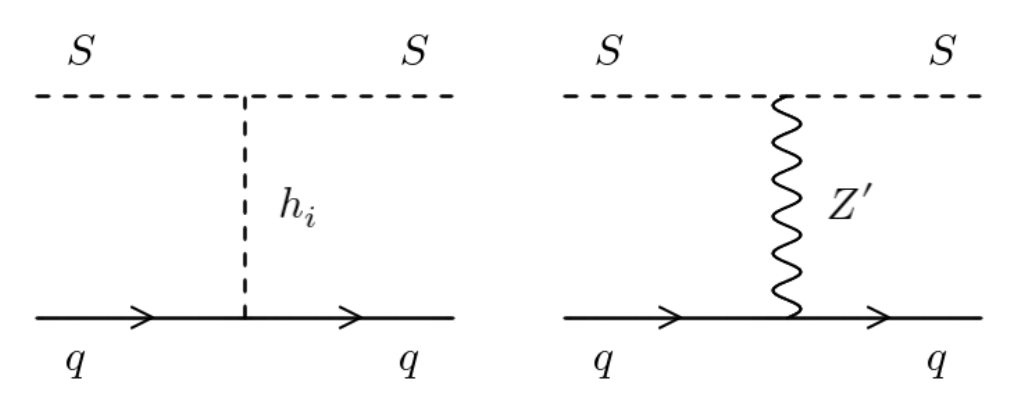}
        \caption{Scalar, $S$.}
        \label{fig:dds}
    \end{subfigure}
    \caption{Relevant contributions to the SI dark matter--nucleon scattering amplitudes. For the fermionic component $\chi$, the leading contributions arise from tree-level $Z'$ exchange and one-loop scalar exchange. For the scalar component $S$, scattering proceeds at tree level through $Z'$ and $h_{1,2}$ exchange.}
    \label{fig:ddfeyndiag}
\end{figure}

Direct-detection experiments search for nuclear recoils induced by dark matter scattering off nucleons~\cite{Bhattacharya:2025eym}. In the present model, both dark matter components give rise to spin-independent (SI) interactions. For the fermion $\chi$, the relevant contributions arise from tree-level $Z'$ exchange and from one-loop scalar exchange induced by the dark Yukawa interaction. For the scalar $S$, both $Z'$ and $h_{1,2}$ exchange contribute already at tree level~\cite{Chang:2017gla}. These processes are illustrated in Fig.~\ref{fig:ddfeyndiag}.

We define the trilinear scalar interactions through
\begin{equation}
    \mathcal{L}\supset
    -\sum_{i=1,2} g_{SSh_i}\,h_iS^\dagger S,
\end{equation}
with
\begin{align}
    g_{SSh_1} & = \lambda_{SH}v_H\cos\theta -\lambda_{S\Phi}v_\Phi\sin\theta, \\
    g_{SSh_2} & = \lambda_{SH}v_H\sin\theta +\lambda_{S\Phi}v_\Phi\cos\theta.
\label{eq:SSh-couplings}
\end{align}
The couplings of the physical scalars to SM particles are controlled by $\kappa_1=\cos\theta$ and  $\kappa_2=\sin\theta$.

For the fermionic component, the effective coefficient associated with tree-level $Z'$ exchange is
\begin{equation}
    C_\chi^{Z'} = \frac{g_{BL}^2 n_\chi n_N}{m_{Z'}^2},
\label{eq:Cchi-Zp}
\end{equation}
where $n_N=1$ is the effective $B-L$ charge of either a proton or a neutron~\cite{Arcadi:2023lwc}. The scalar contribution is generated at one loop and can be written as
\begin{equation}
    C_\chi^{h} = \frac{f_Nm_N}{v_H} \sum_{i=1,2}\frac{\kappa_i g_{SSh_i}}{m_{h_i}^2}\frac{|y_s|^2 f(r)+|y_p|^2 g(r)}{16\pi^2m_\chi},
\label{eq:Cchi-h}
\end{equation}
where $m_N$ is the nucleon mass, $f_N\simeq0.3$ is the scalar nucleon form factor~\cite{Hoferichter:2017olk}, and $r=\frac{m_S^2}{m_\chi^2}$. The loop functions are~\cite{Haisch:2013uaa,Yaguna:2021rds,Amiri:2025ras}
\begin{align}
    f(r)
    &=
    \frac{r^2-5r+4}{\sqrt{(4-r)r}}
    \arctan\!\left(\sqrt{\frac{4-r}{r}}\right)
    +\frac{1}{2}\left[2-(r-3)\ln r\right],
    \\
    g(r)
    &=
    (r-3)\sqrt{\frac{r}{4-r}}
    \arctan\!\left(\sqrt{\frac{4-r}{r}}\right)
    +\frac{1}{2}\left[2-(r-1)\ln r\right].
\label{eq:dd-loop-functions}
\end{align}
These expressions are real throughout the two-component dark matter regime, $m_S<2m_\chi$. The contribution proportional to $|y_p|^2$ is not velocity suppressed, whereas the mixed term proportional to $y_sy_p$ is suppressed in the non-relativistic limit.

For the scalar component, the corresponding tree-level coefficients are
\begin{align}
    C_S^{Z'}
    &=
    \frac{g_{BL}^2 n_S n_N}{m_{Z'}^2},
    &
    C_S^h
    &=
    \frac{m_N f_N}{2m_S v_H}
    \sum_{i=1,2}
    \frac{\kappa_i g_{SSh_i}}{m_{h_i}^2},
\label{eq:CS-coefficients}
\end{align}
with $n_S=-2n_\chi$.

Assuming vanishing particle--antiparticle asymmetries, the observable SI cross sections are obtained by averaging over $\chi,\bar\chi$ and $S, S^\dagger$. Since the vector contribution changes sign under charge conjugation whereas the scalar contribution does not, the interference between them cancels in the averaged rates. The resulting cross sections are
\begin{align}
    \overline{\sigma}_{\chi N}^{\rm SI}
    &=
    \frac{\mu_{\chi N}^2}{\pi}
    \left(
        |C_\chi^{Z'}|^2+|C_\chi^h|^2
    \right),
    \\
    \overline{\sigma}_{SN}^{\rm SI}
    &=
    \frac{\mu_{SN}^2}{\pi}
    \left(
        |C_S^{Z'}|^2+|C_S^h|^2
    \right),
\label{eq:dd-sigSI}
\end{align}
where
\begin{equation}
    \mu_{XN}
    =
    \frac{m_Xm_N}{m_X+m_N},
    \qquad X=\chi,S.
\end{equation}
The interference between $h_1$ and $h_2$ remains encoded in the coherent scalar amplitudes $C_\chi^h$ and $C_S^h$.

We compare the predicted SI cross sections with the XENONnT and LZ results and use the projected DARWIN/XLZD sensitivity to illustrate the future experimental reach~\cite{XENON:2023cxc, LZ:2024zvo, Baudis:2024jnk}. Since each species constitutes only a fraction of the local dark matter density, the quantities relevant for comparison with the single-species experimental limits are
\begin{equation}
    \sigma_{X,N}^{\rm th}
    =
    \xi_X\,\overline{\sigma}_{XN}^{\rm SI},
    \qquad
    \xi_X=\frac{\Omega_X}{\Omega_{\rm DM}},
    \qquad X = \chi,S.
\label{eq:dd-rescaling}
\end{equation}
The rescaling is linear in $\xi_X$ because the direct-detection event rate is proportional to the local density of the incident dark matter component.

\subsection{\label{sec:indirect-detection} Indirect detection limits}

Dark matter annihilation in regions of large density can produce gamma-ray signals observable by present and future telescopes. We use continuum gamma-ray limits from \textit{Fermi}-LAT~\cite{Fermi-LAT:2015att} and H.E.S.S.~\cite{HESS:2022ygk}, together with the projected sensitivity of CTAO~\cite{CTA:2020qlo,Silverwood:2014yza}, to estimate the indirect-detection reach of the model. Since our aim is not to perform a dedicated gamma-ray likelihood analysis, these bounds are applied as approximate constraints on the present-day annihilation and semi-conversion rates.

For conventional pair annihilation of either dark matter component,
we define
\begin{equation}
    \langle\sigma v\rangle^{\rm tot}_{f\bar f}
    \equiv
    \sum_f
    \langle\sigma v\rangle_{X\bar X\to f\bar f},
    \qquad X=\chi,S,
\end{equation}
where the sum runs over kinematically accessible charged SM fermions. Because each species constitutes only a fraction of the total dark matter density, the corresponding gamma-ray flux is suppressed by the square of its relic fraction. The effective rate is therefore~\cite{Yaguna:2021rds,DiazSaez:2021pmg,Queiroz:2026tah}
\begin{equation}
    \langle\sigma v\rangle^{\rm eff}_{X\bar X\to f\bar f} 
    = 
    \xi_X^2\, \langle\sigma v\rangle^{\rm tot}_{f\bar f}.
\end{equation}

The model also contains semi-conversion channels of the form $\chi\chi\to S^\dagger h_i$, $i=1,2$, together with their charge-conjugate processes. These reactions can provide an important contribution to the gamma-ray signal through the subsequent decays of $h_i$. Their comparison with standard annihilation limits requires accounting for the modified kinematics and decay spectra of the visible state in the final state~\cite{Elor:2015bho, Queiroz:2019acr, DiazSaez:2021pmg}. We therefore use the corresponding semi-annihilation recasting prescription as an approximate estimate of the constraint, with the overall normalization rescaled by the fermionic relic fraction,
\begin{equation}
    \langle\sigma v\rangle^{\rm eff}_{\chi\chi\to S^\dagger h_i}
    =
    \xi_\chi^2\,
    \langle\sigma v\rangle_{\chi\chi\to S^\dagger h_i}.
\end{equation}
The resulting indirect-detection limits should consequently be interpreted as indicative rather than as a dedicated spectral recast.

\section{Numerical Results}
\label{sec:results}

We explore the phenomenology of the model through a global scan of its multidimensional parameter space using the differential-evolution algorithm implemented in \texttt{SciPy}. The scan ranges are chosen to satisfy perturbativity requirements and the kinematic condition $m_S<2m_\chi$, which guarantees the stability of the scalar DM component. We consider
\begin{align}
    m_\chi &\in [100,3000]~{\rm GeV}, & \frac{m_S}{m_\chi} & \in [0.05,1.95], \nonumber\\
    m_{Z'} &\in [4000,10000]~{\rm GeV}, & m_{h_2} &\in [200,4000]~{\rm GeV}, \nonumber\\
    g_{BL} &\in [10^{-3},1], & |y_s|,|y_p| &\in [10^{-2},3], \nonumber\\
    |\lambda_{SH}|,|\lambda_{S\Phi}| & \in [10^{-3},3], & \sin\theta &\in [10^{-6},1]. \label{eq:scan-ranges}
\end{align}
For each trial point, we first impose the theoretical consistency conditions discussed in Sec.~\ref{sec:model} together with the collider constraints of Sec.~\ref{sec:collider}. We then pass points that satisfy these requirements to \textsc{micrOMEGAs} to solve the coupled Boltzmann equations and compute the individual relic abundances $\Omega_\chi h^2$ and $\Omega_S h^2$. The scan is guided by a $\chi^2$-like objective function that favors points satisfying Planck observation~\cite{Planck:2018vyg}
\begin{equation}
    \Omega_\chi h^2 + \Omega_S h^2 = 0.120 \pm 0.001.
\end{equation}
Several independent differential-evolution runs are combined in order to improve the coverage of the allowed multidimensional parameter space.

Once the relic abundances are known, the corresponding fractions $\xi_\chi$ and $\xi_S$ are used to rescale the direct- and indirect-detection rates. Current direct-detection constraints are then imposed on both DM components following the prescription described in Sec.~\ref{sec:direct-detection}.

Throughout this section, we refer to a parameter point as \emph{viable} if it satisfies the theoretical consistency conditions, the collider constraints, the observed total DM abundance, and the current direct-detection bounds. Indirect-detection constraints are not included in this definition; instead, they are applied subsequently to the viable sample.

The scatter plots shown below represent projections of this viable multidimensional parameter space. Since differential evolution is an optimization algorithm rather than a statistical sampling procedure, a statistical interpretation should not be assigned to the density of points in a particular region.

To quantify the effect of the different reaction classes on the relic abundance, we also recompute the freeze-out evolution after switching off one process at a time. For a DM component $X=\chi,S$, we define
\begin{equation}
    \zeta_X^{ijkl} \equiv \frac{\Omega_X^{\,(-ijkl)}}{\Omega_X^{\rm full}},
\label{eq:off-channel-ratio}
\end{equation}
where $\Omega_X^{\,(-ijkl)}$ is the relic abundance obtained after removing the reaction class $ijkl\in\{1100, 2200, 1120, 1210, 1122, 2211\}$, and $\Omega_X^{\rm full}$ denotes the abundance obtained with all processes included. Thus, $\zeta_X^{ijkl}=1$ indicates that removing the corresponding process does not affect the final abundance, whereas deviations from unity quantify its impact.

The numerical results are presented separately for the two possible mass hierarchies, $m_S > m_\chi$ and $m_S < m_\chi$.

\subsection{The \texorpdfstring{$\boldsymbol{m_S>m_\chi}$}{mS > mchi} regime}
\label{sec:results-heavyS}
\begin{figure}[t!]
    \centering
    \includegraphics[width=0.45\linewidth]{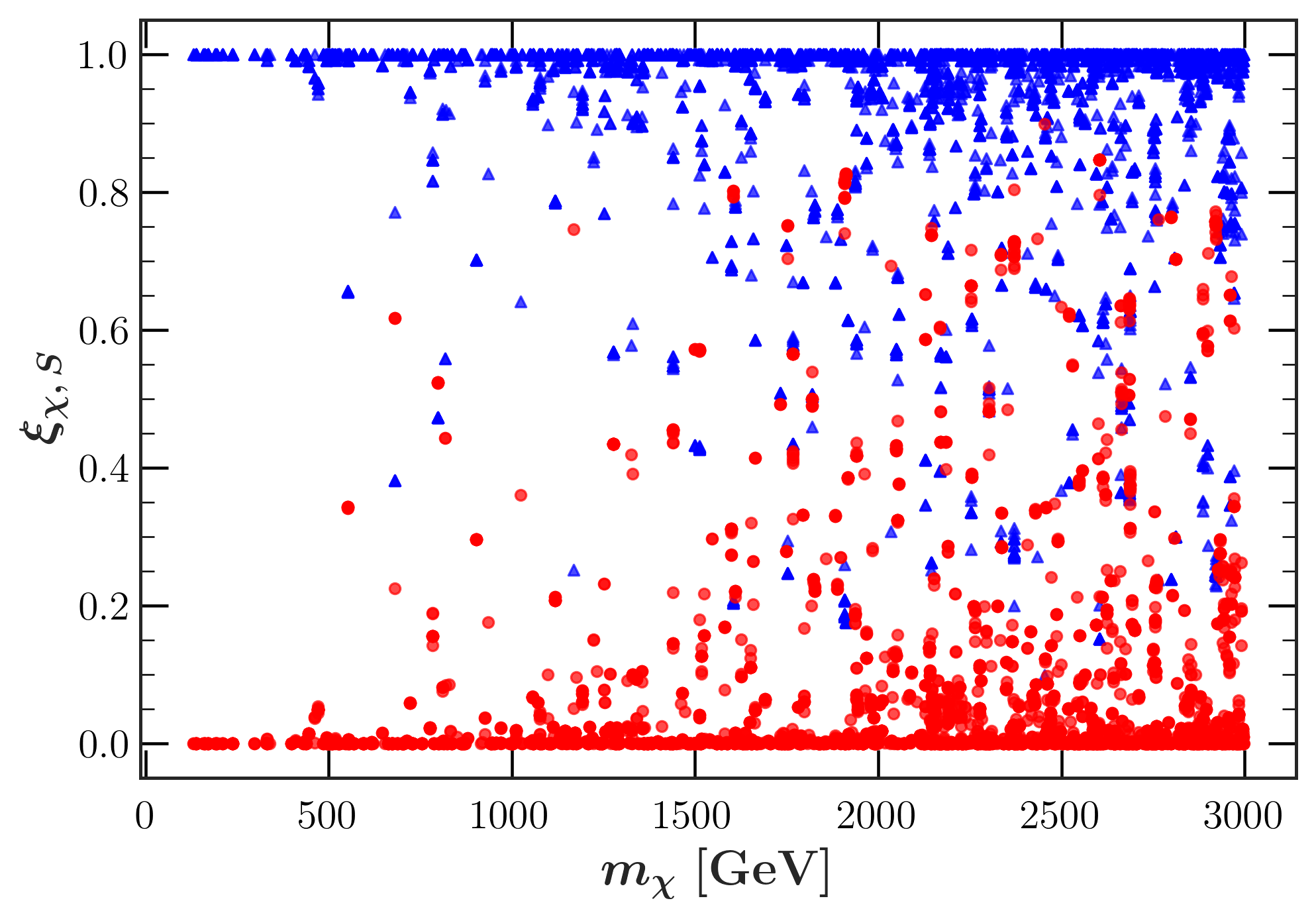}
    \includegraphics[width=0.45\linewidth]{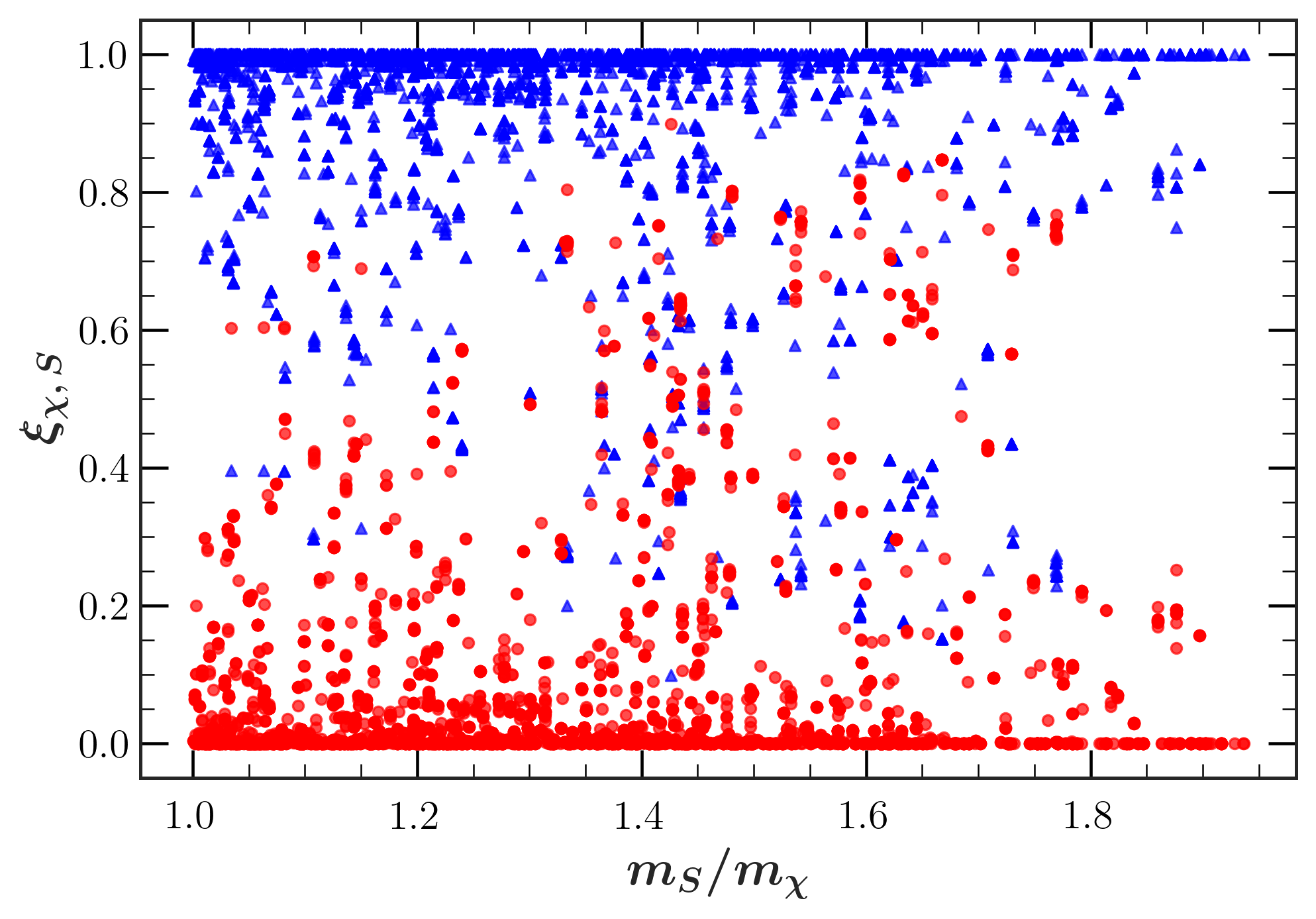}
    \includegraphics[width=0.45\linewidth]{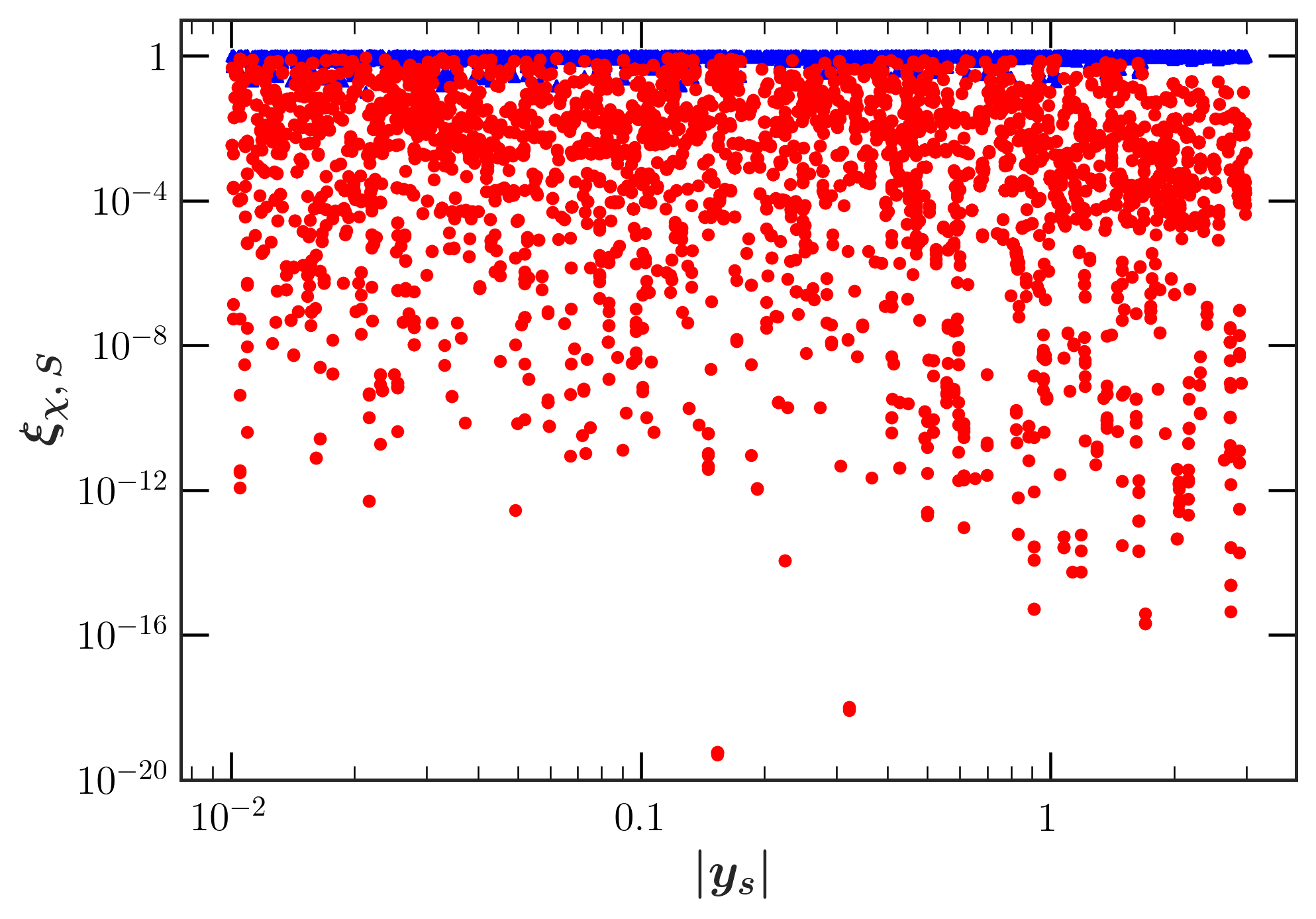}
    \includegraphics[width=0.45\linewidth]{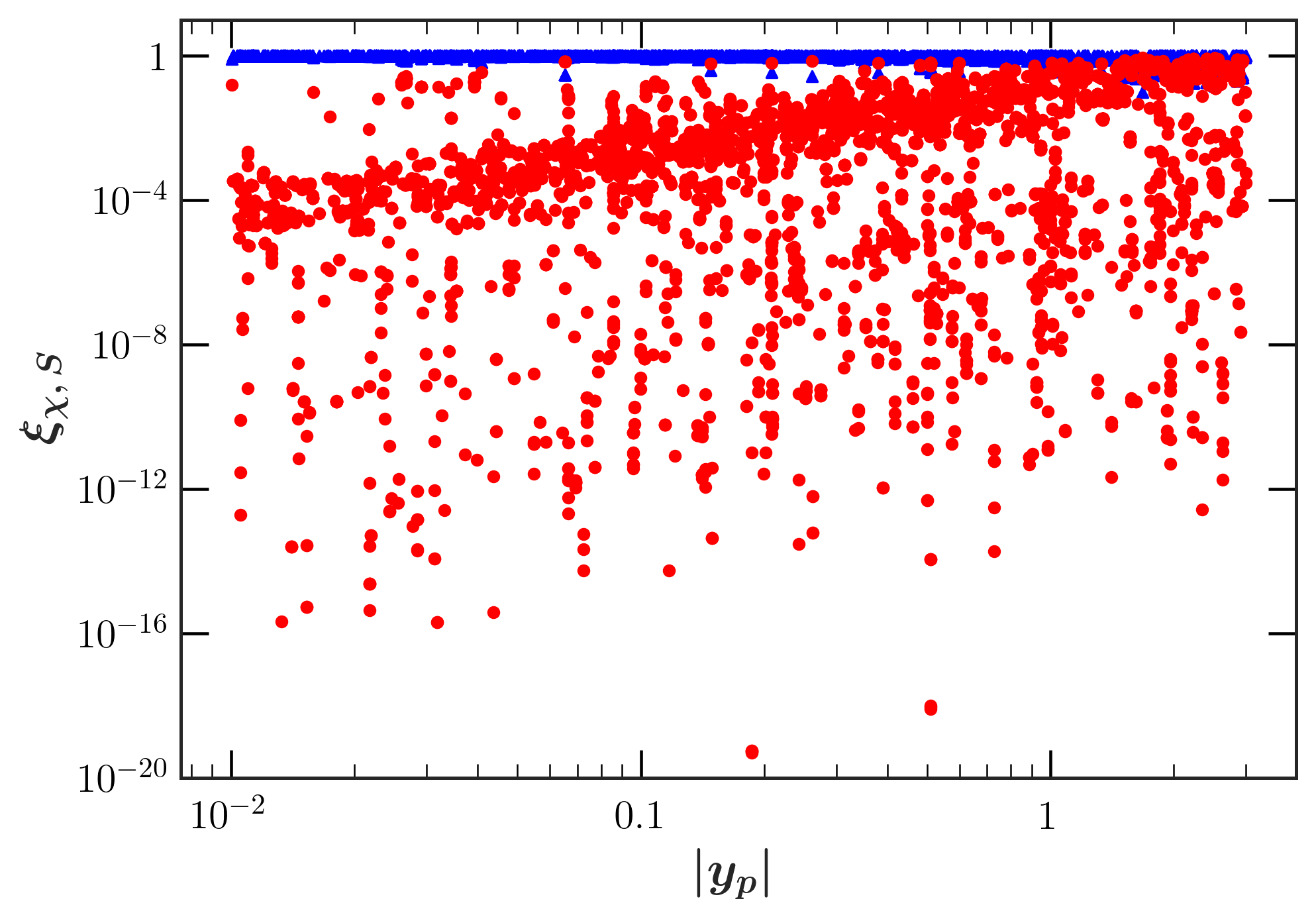}
    \begin{tabular}{c}
        \textbf{DM component:} \hspace{0.1cm} \textcolor{blue}{\ding{115}} $\boldsymbol{\xi_\chi}$ \hspace{0.2cm} \textcolor{red}{\ding{108}} $\boldsymbol{\xi_S}$
    \end{tabular}
    \caption{Relic-density fractions $\xi_\chi$ and $\xi_S$ as functions of selected model parameters for $m_S>m_\chi$. Blue triangles and red circles denote the fermionic and scalar components, respectively. The upper panels use linear scales and show the dependence on $m_\chi$ and $m_S/m_\chi$, while the lower panels use logarithmic scales and show the dependence on the dark Yukawa couplings. All displayed points satisfy the theoretical constraints, collider bounds, the observed total relic abundance, and current direct-detection limits. Indirect-detection constraints are not imposed.}
    \label{fig:reg1_omega}
\end{figure}
We first consider the hierarchy $m_S>m_\chi$, for which the fermion is the lighter DM particle. Figure~\ref{fig:reg1_omega} shows the relic-density fractions $\xi_\chi$ and $\xi_S$ for the viable points projected onto selected model parameters.

A clear feature of this regime is that the fermionic component dominates the relic density throughout most of the viable parameter space. Nevertheless, viable points with a sizable scalar contribution are also present, and the two components can contribute comparable fractions in restricted regions. Large mass splittings tend to favor a strongly fermion-dominated relic density, while the dependence on the dark Yukawa couplings is more pronounced for $|y_p|$ than for $|y_s|$. Moreover, the scalar fraction can become highly suppressed, spanning several orders of magnitude.

\begin{figure}[t!]
    \centering
    \includegraphics[width=0.48\linewidth]
    {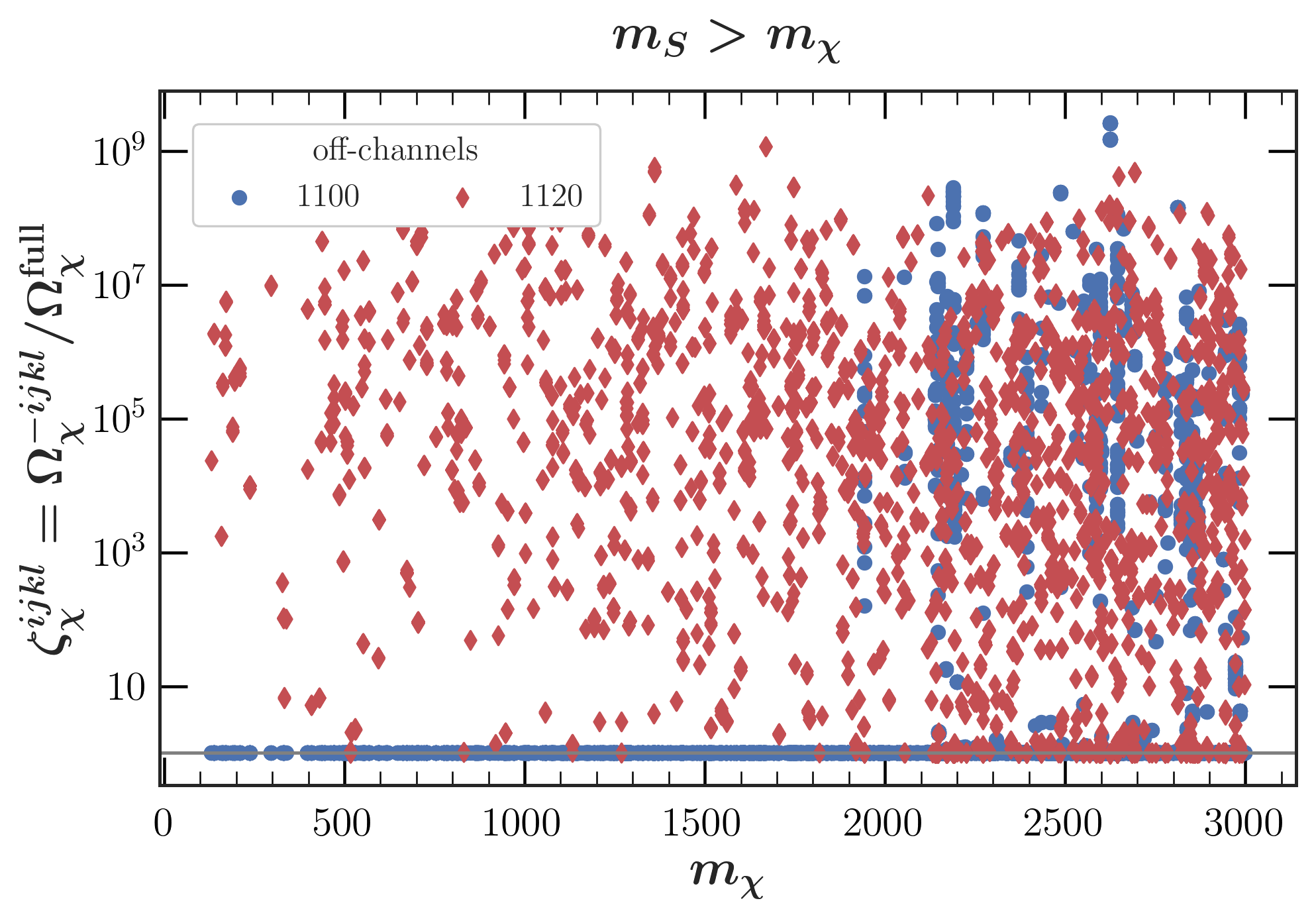}
    \includegraphics[width=0.48\linewidth]
    {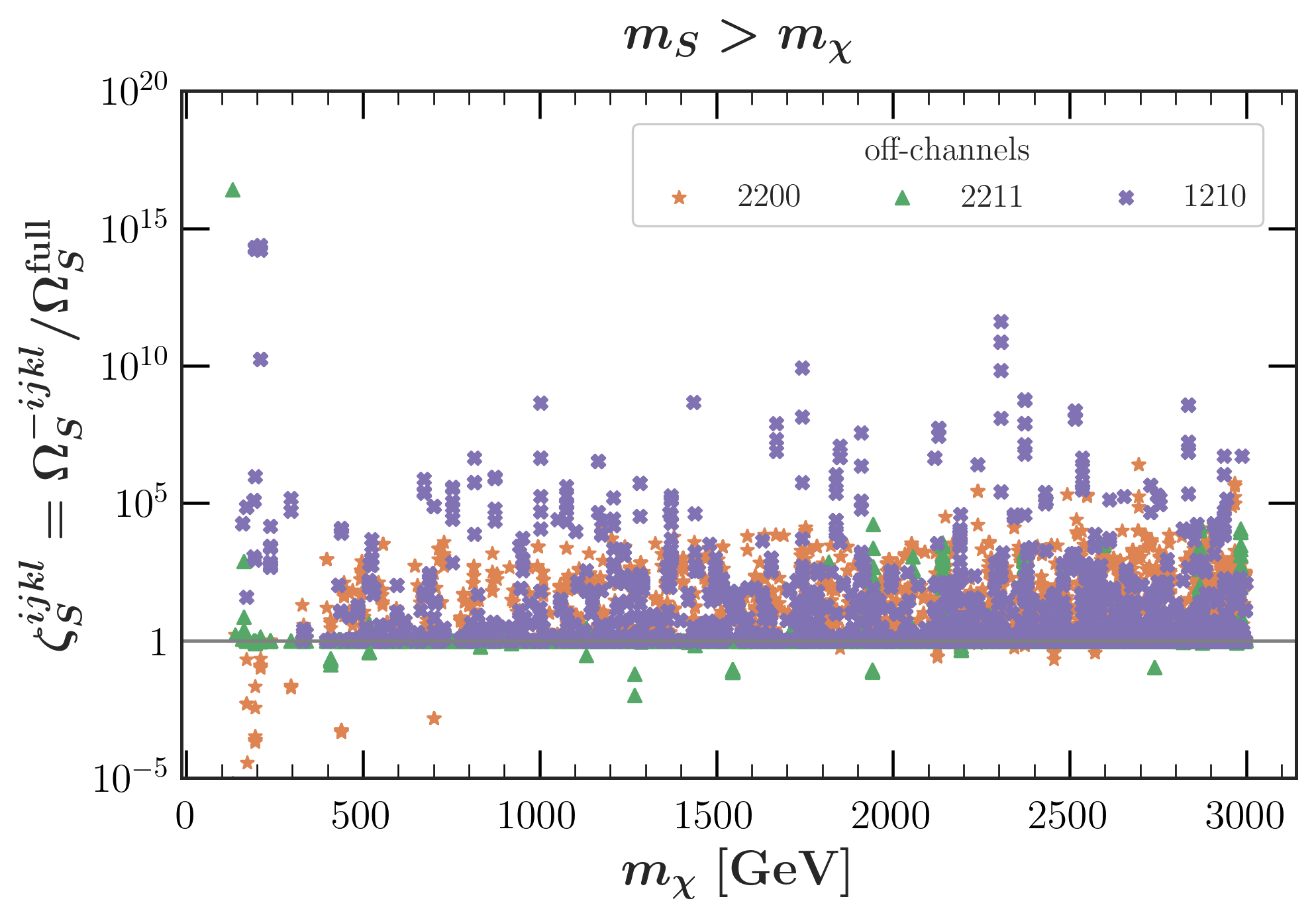}
    \caption{Impact of individual reaction classes on the relic abundances for $m_S>m_\chi$. The left and right panels show $\zeta_\chi^{ijkl}=\Omega_\chi^{(-ijkl)}/\Omega_\chi^{\rm full}$ and $\zeta_S^{ijkl}=\Omega_S^{(-ijkl)}/\Omega_S^{\rm full}$, respectively, as functions of the lighter DM mass $m_\chi$. The superscript $(-ijkl)$ denotes a calculation in which reaction class $ijkl$ has been removed. The reference value $\zeta_X^{ijkl}=1$ corresponds to no change relative to the full calculation. Values above unity indicate that the removed process normally depletes component $X$, whereas values below unity indicate that it contributes, directly or indirectly, to its production. The 1120 contribution is not displayed in the $S$ panel because removing this process leads to an extremely strong suppression of the scalar relic abundance.}
    \label{fig:reg1_offchannels}
\end{figure}

The processes responsible for this pattern can be identified more directly through the off-channel analysis shown in Figure~\ref{fig:reg1_offchannels}. For the fermionic component, the contrast between the 1100 annihilation and the 1120 semi-conversion processes is particularly clear. Removing the 1100 channel leaves $\Omega_\chi$ close to its full value for a large fraction of the viable points. By contrast, switching off 1120 can increase $\Omega_\chi$ by several orders of magnitude. Thus, over a substantial part of the viable parameter space, the fermionic abundance is controlled primarily by the semi-conversion reaction $\chi\chi\rightarrow S^\dagger X_0$ 
and its charge-conjugate process, rather than by conventional $\chi \bar\chi$ annihilation.

The same reaction has the opposite effect on the scalar component: the 1120 process produces one scalar DM particle and therefore acts as a source for $S$. Indeed, removing this channel leads to an extremely strong suppression of $\Omega_S$ and is not shown in the right panel of Fig.~\ref{fig:reg1_offchannels} because of the resulting dynamic range. On the other hand, removing the 2200 annihilation, 2211 conversion, or 1210 semi-annihilation processes generally increases $\Omega_S$, showing that these reactions contribute to the depletion of the scalar population. The 1120 semi-conversion therefore plays a dual role in this hierarchy, simultaneously reducing the fermionic abundance and feeding the scalar component.

\begin{figure}[t!]
    \centering
    \includegraphics[width=0.45\linewidth]{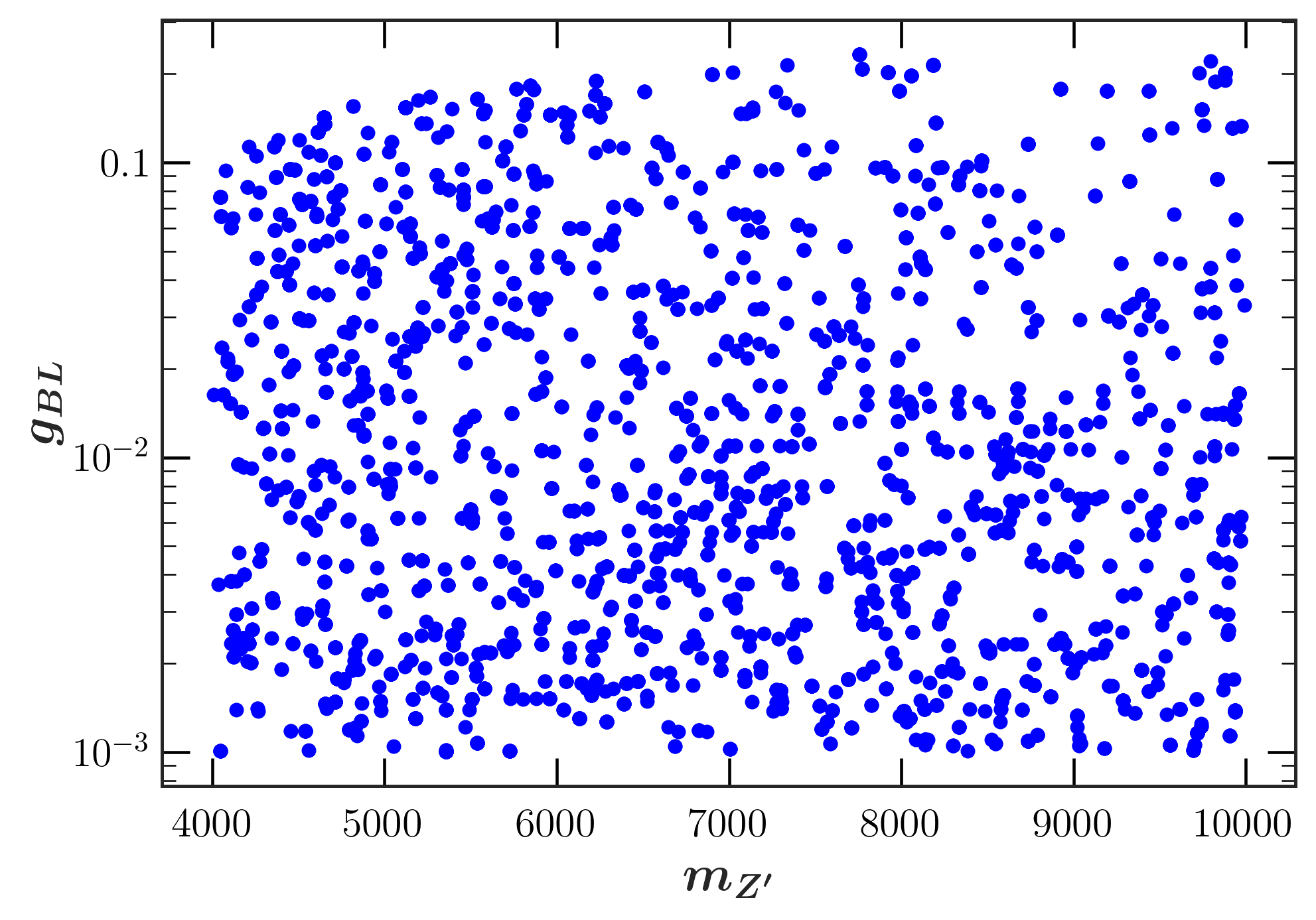}
    \includegraphics[width=0.45\linewidth]{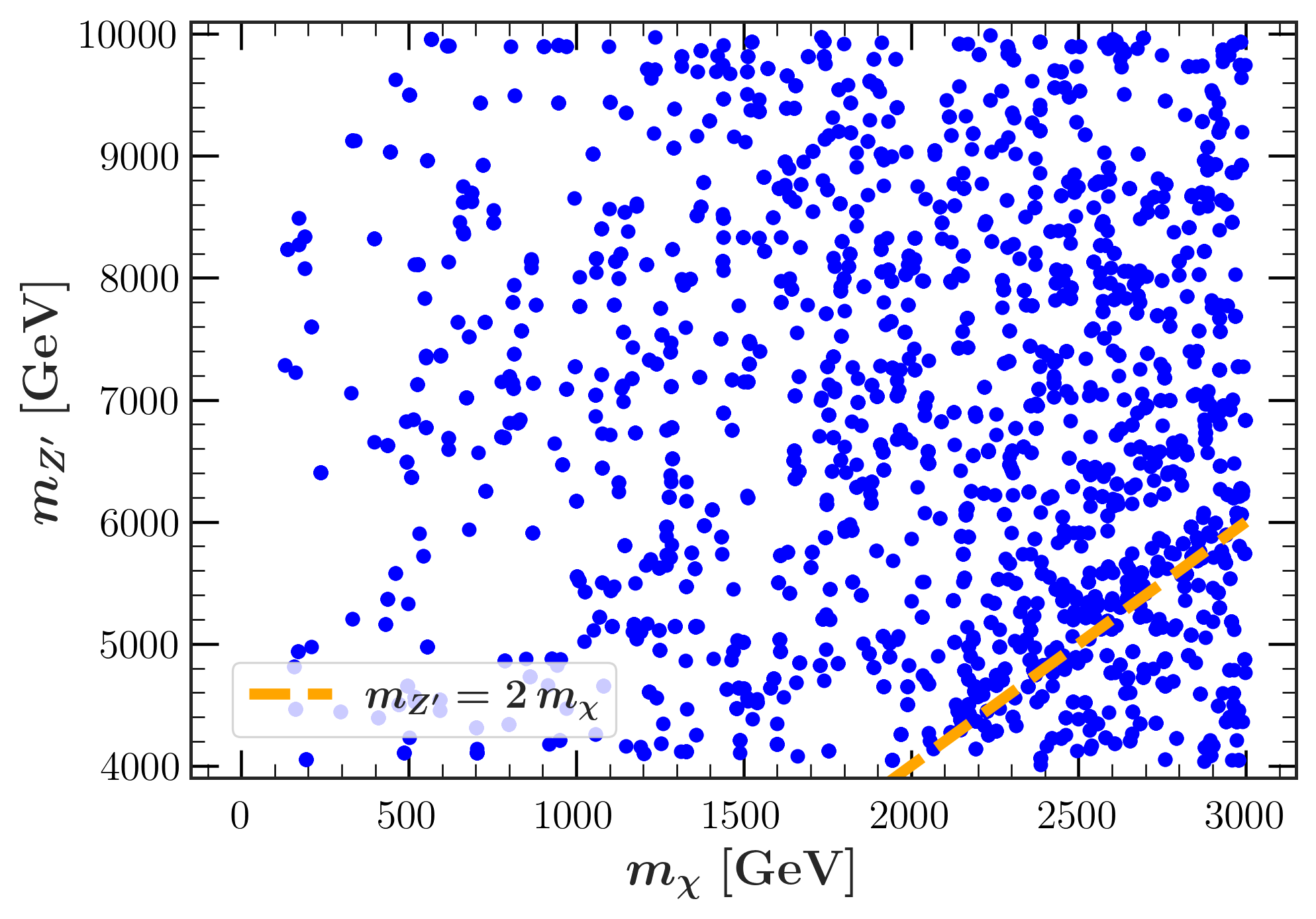}
    \includegraphics[width=0.45\linewidth]{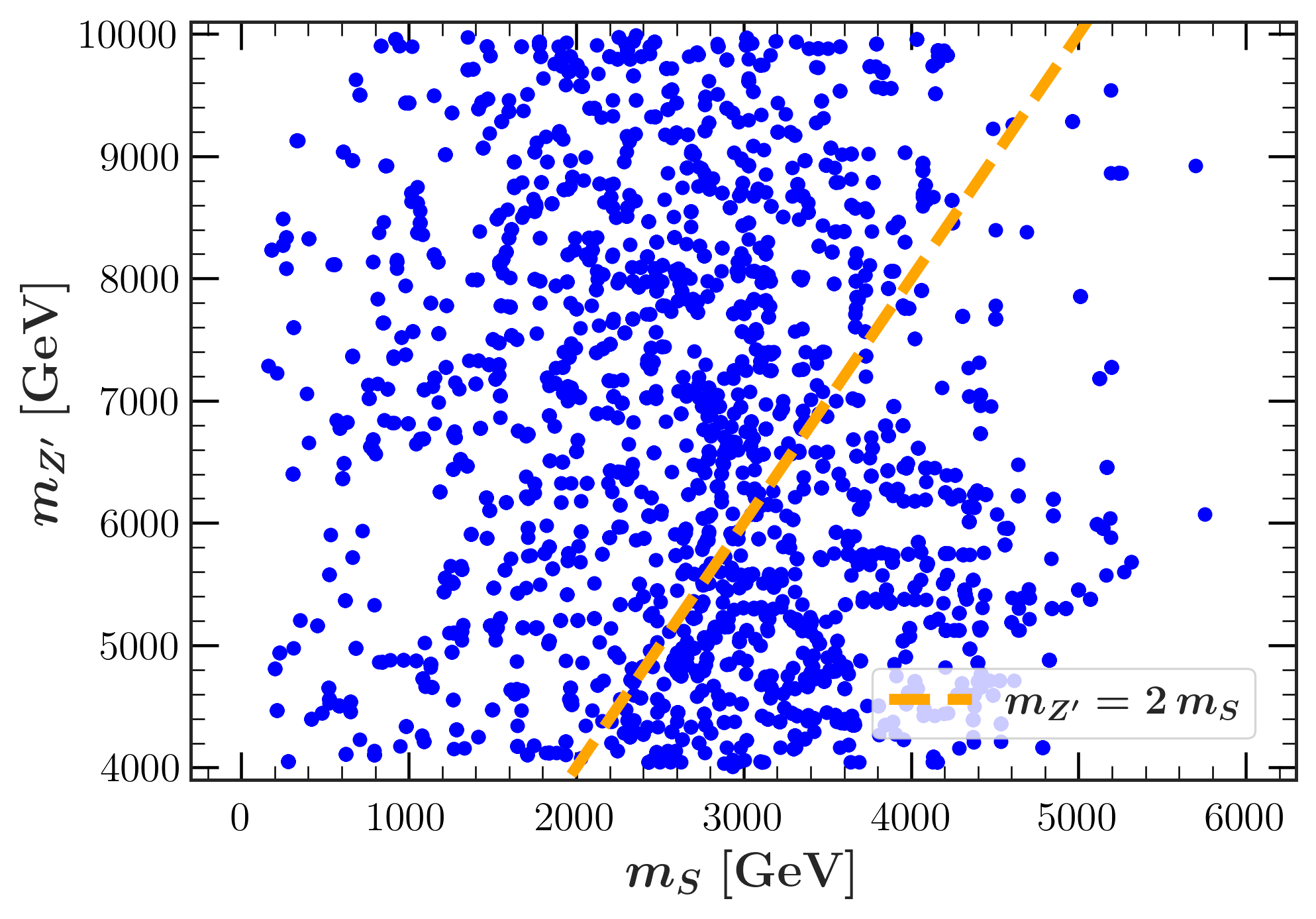}
    \includegraphics[width=0.45\linewidth]{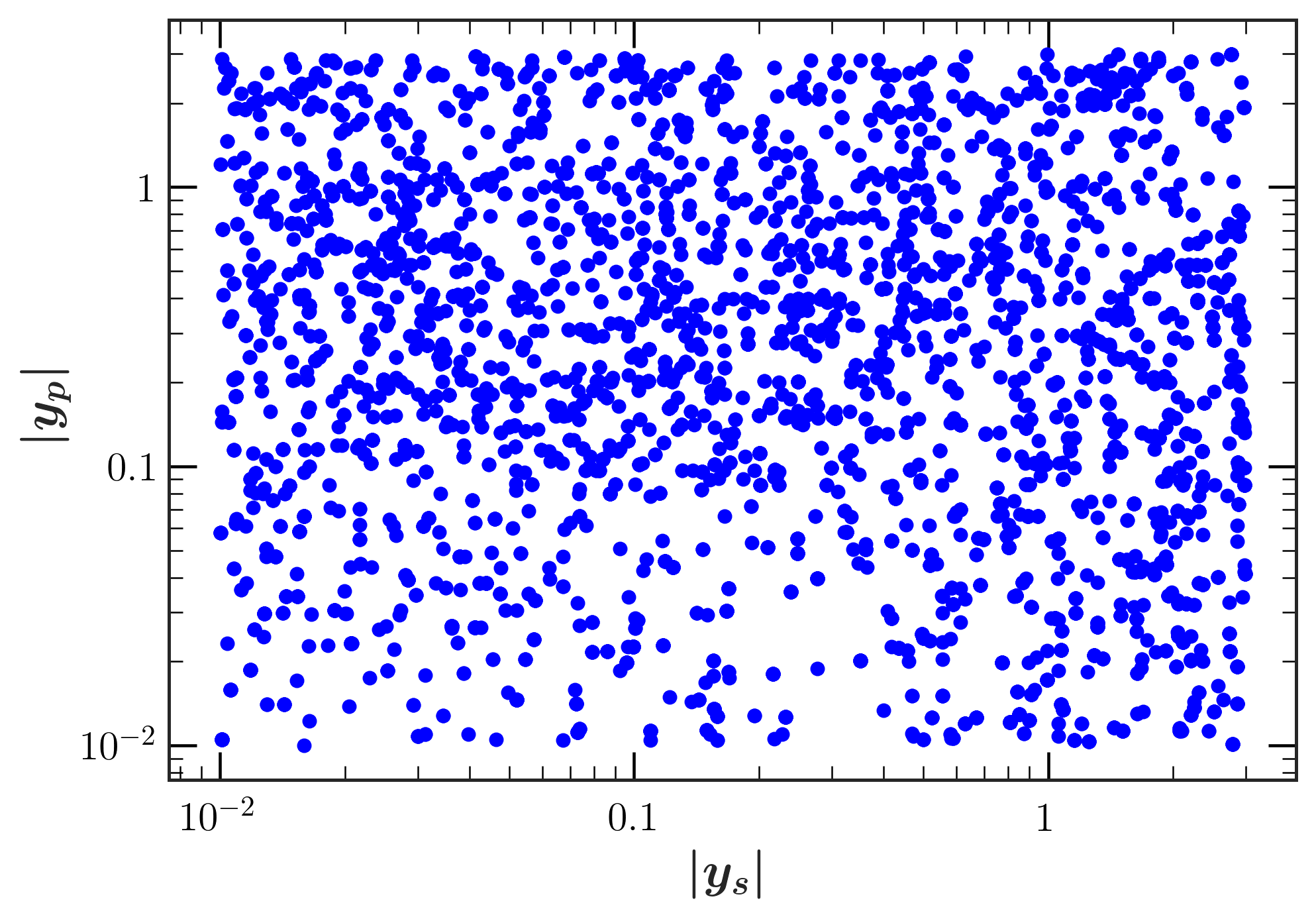}
    \caption{Selected projections of the viable parameter space for $m_S > m_\chi$. The upper-left panel shows the gauge sector in the $(m_{Z'},g_{BL})$ plane. The upper-right and lower-left panels show $m_{Z'}$ as a function of $m_\chi$ and $m_S$, respectively, with reference lines indicating the resonance conditions $m_{Z'} = 2m_\chi$ and $m_{Z'}=2m_S$. The lower-right panel shows the correlation between the dark Yukawa couplings $|y_s|$ and $|y_p|$. All displayed points satisfy the theoretical constraints, collider bounds, the observed total relic abundance, and current direct-detection limits.}
    \label{fig:reg1_par}
\end{figure}

Figure~\ref{fig:reg1_par} shows complementary projections of the viable parameter space. Viable solutions occur both near and well away from the $Z'$ resonance conditions, demonstrating that the observed relic abundance does not require resonant annihilation. The dark Yukawa couplings also span broad ranges, with no narrow correlation between $|y_s|$ and $|y_p|$ selected by the combined constraints.

\begin{figure*}[t!]
    \centering
    \includegraphics[width=0.45\linewidth]{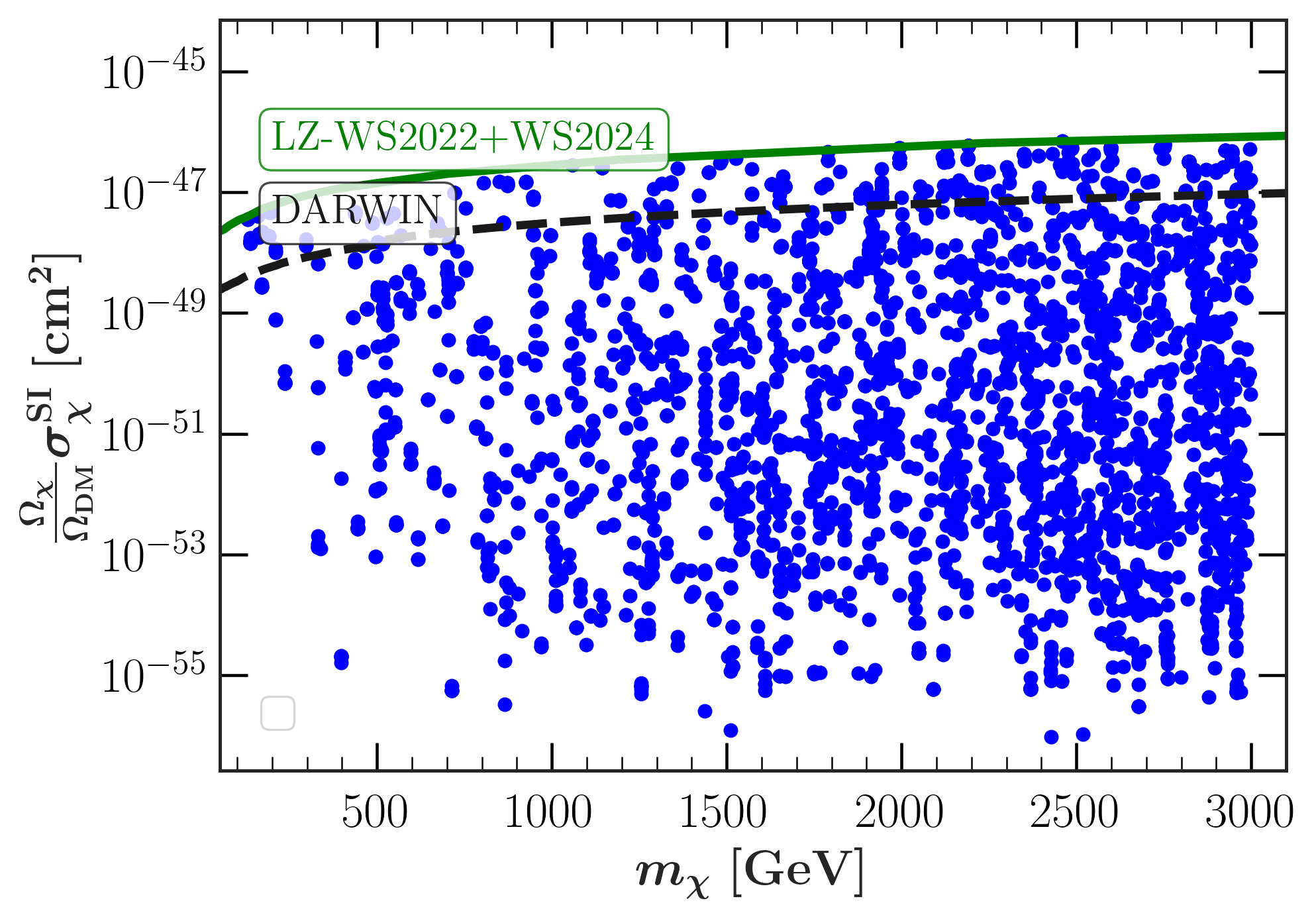}
    \includegraphics[width=0.45\linewidth]{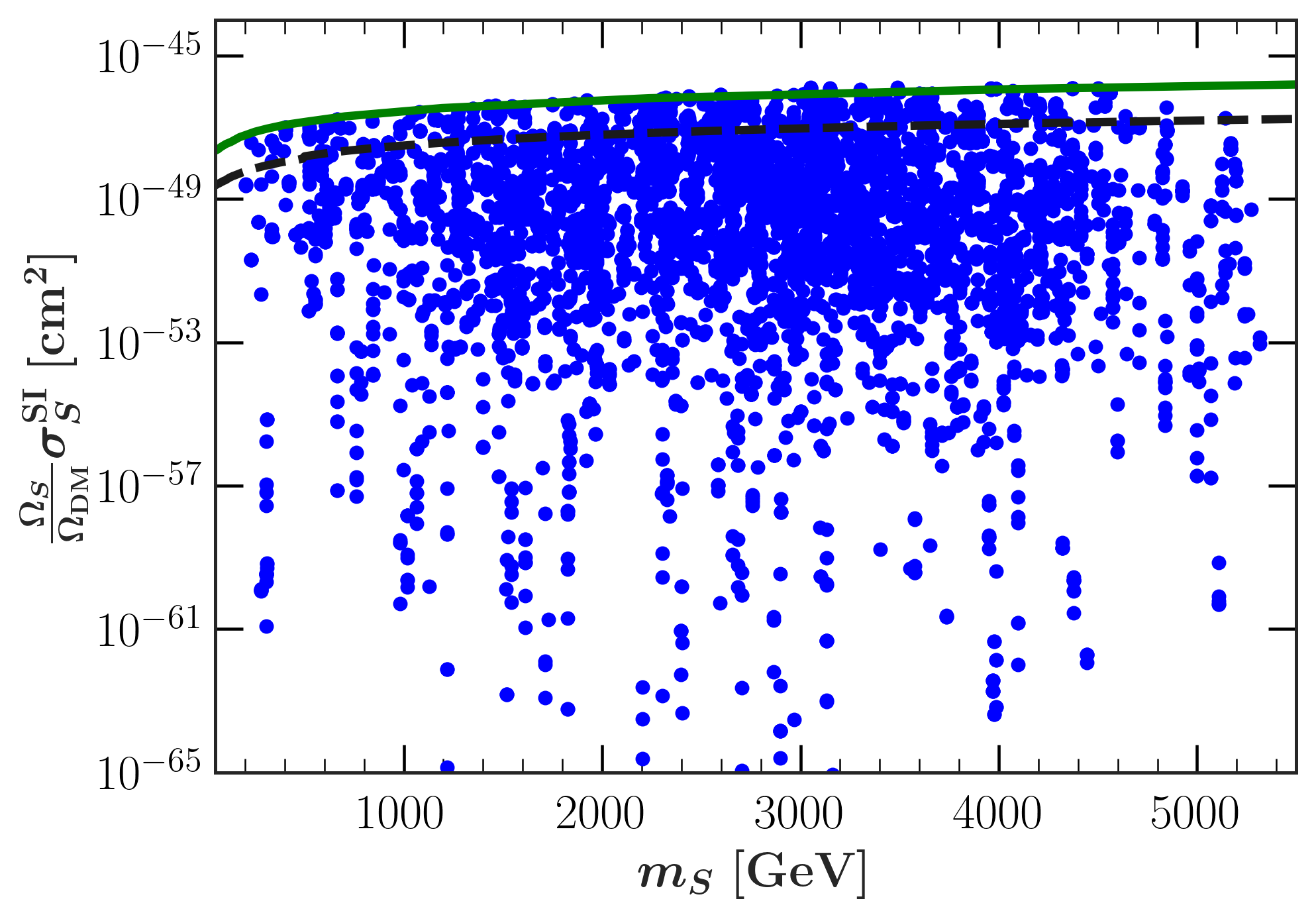}
    \includegraphics[width=0.5\linewidth]{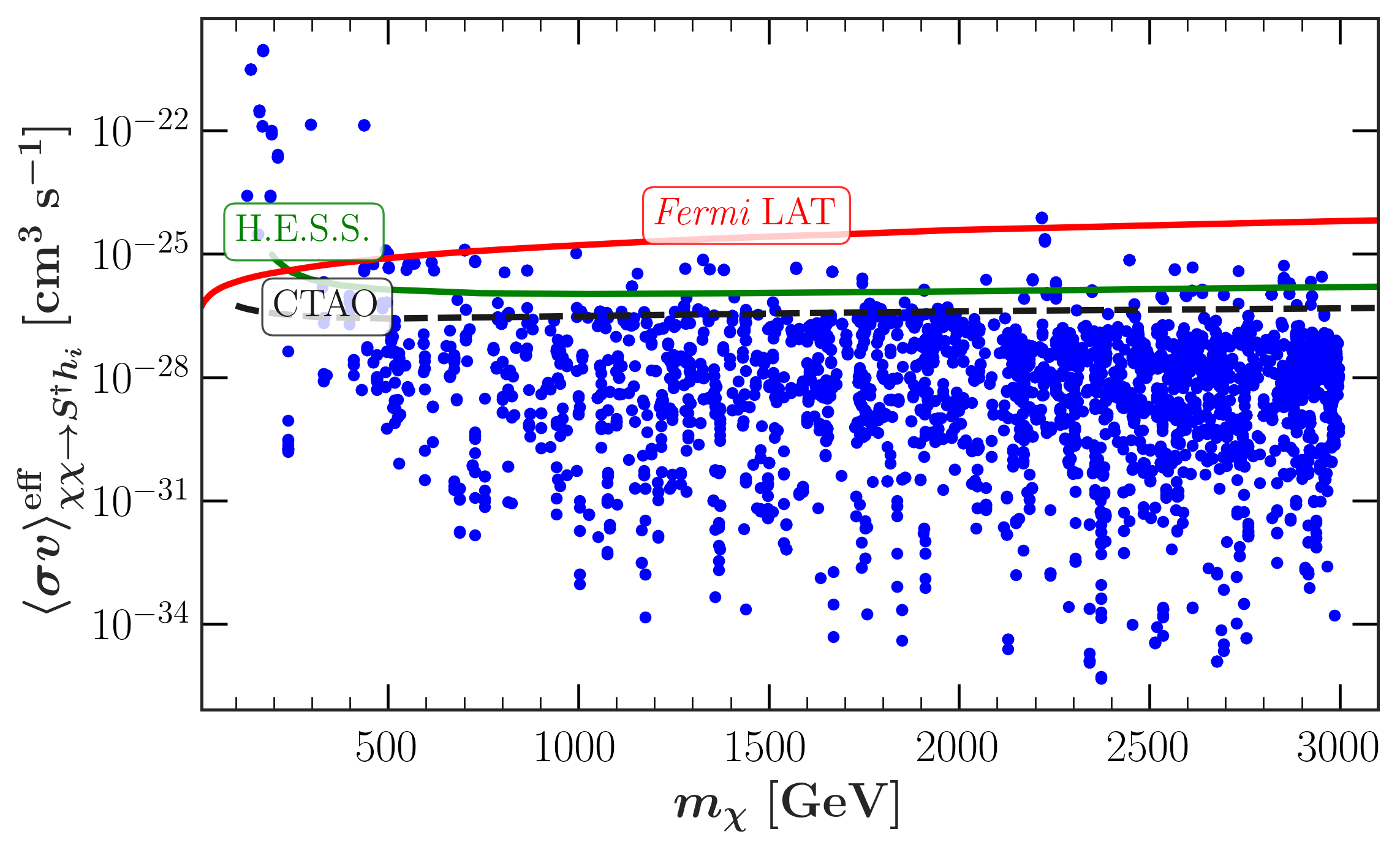}
    \caption{Direct- and indirect-detection predictions for the viable points with $m_S > m_\chi$. The upper panels show the rescaled SI scattering cross sections for $\chi$ (left) and $S$ (right), together with the current LZ limit and the projected DARWIN/XLZD sensitivity. The lower panel shows the effective gamma-ray annihilation rate together with the current \textit{Fermi}-LAT and H.E.S.S. limits and the projected CTAO sensitivity. Current direct-detection constraints are already imposed on the displayed sample, whereas indirect-detection constraints are shown but not imposed.}
    \label{fig:reg1_bounds}
\end{figure*}

Figure~\ref{fig:reg1_bounds} summarizes the detection prospects in this hierarchy. Current direct-detection bounds are already imposed on the displayed sample, but part of the surviving parameter space remains within the projected sensitivity of DARWIN/XLZD. Current gamma-ray limits provide only an additional restriction on part of this sample, while CTAO offers complementary future sensitivity.

\subsection{The \texorpdfstring{$\boldsymbol{m_S < m_\chi}$}{mS < mchi} regime}
\label{sec:results-lightS}

\begin{figure}[t!]
    \centering
    \includegraphics[width=0.45\linewidth]{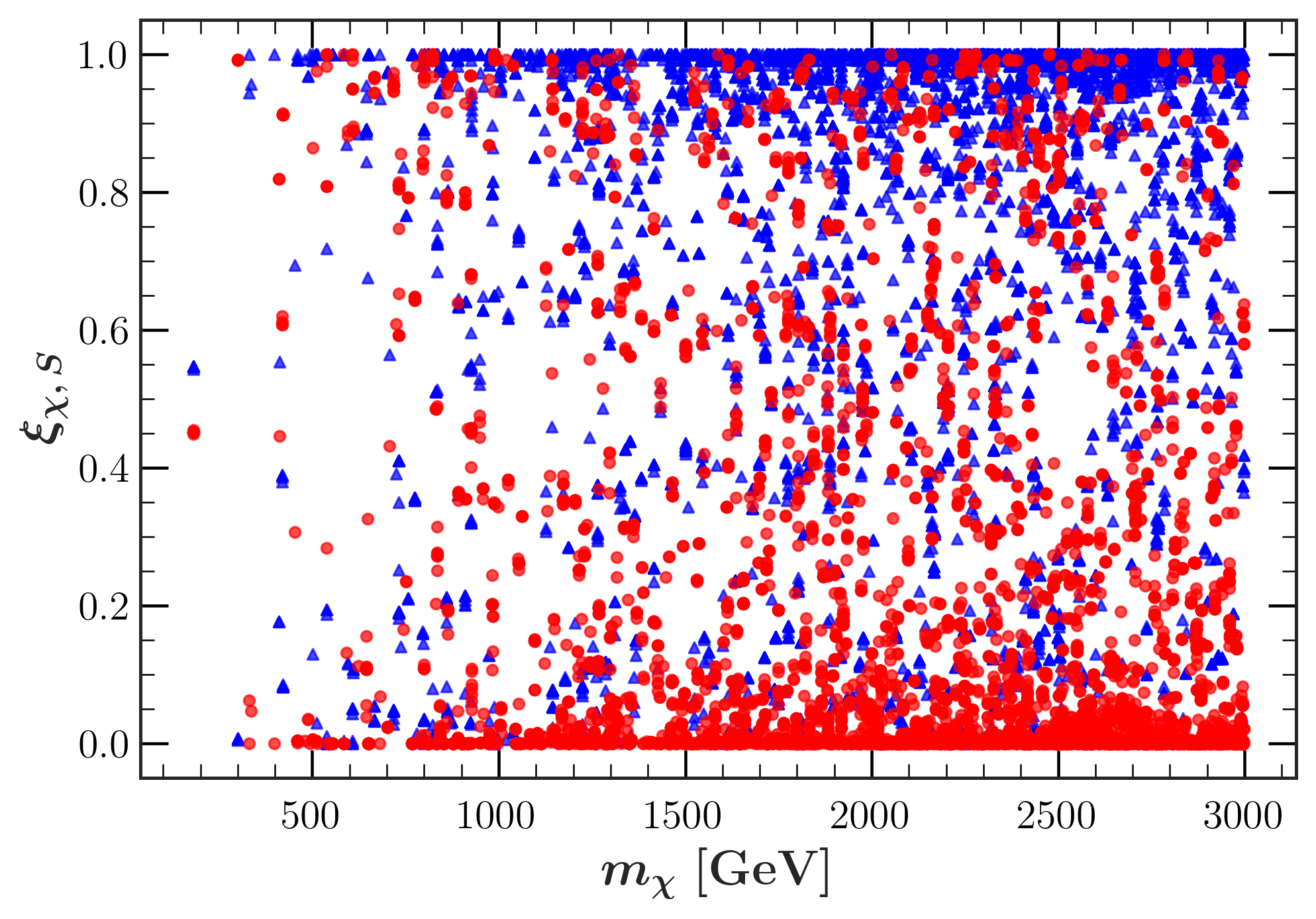}
    \includegraphics[width=0.45\linewidth]{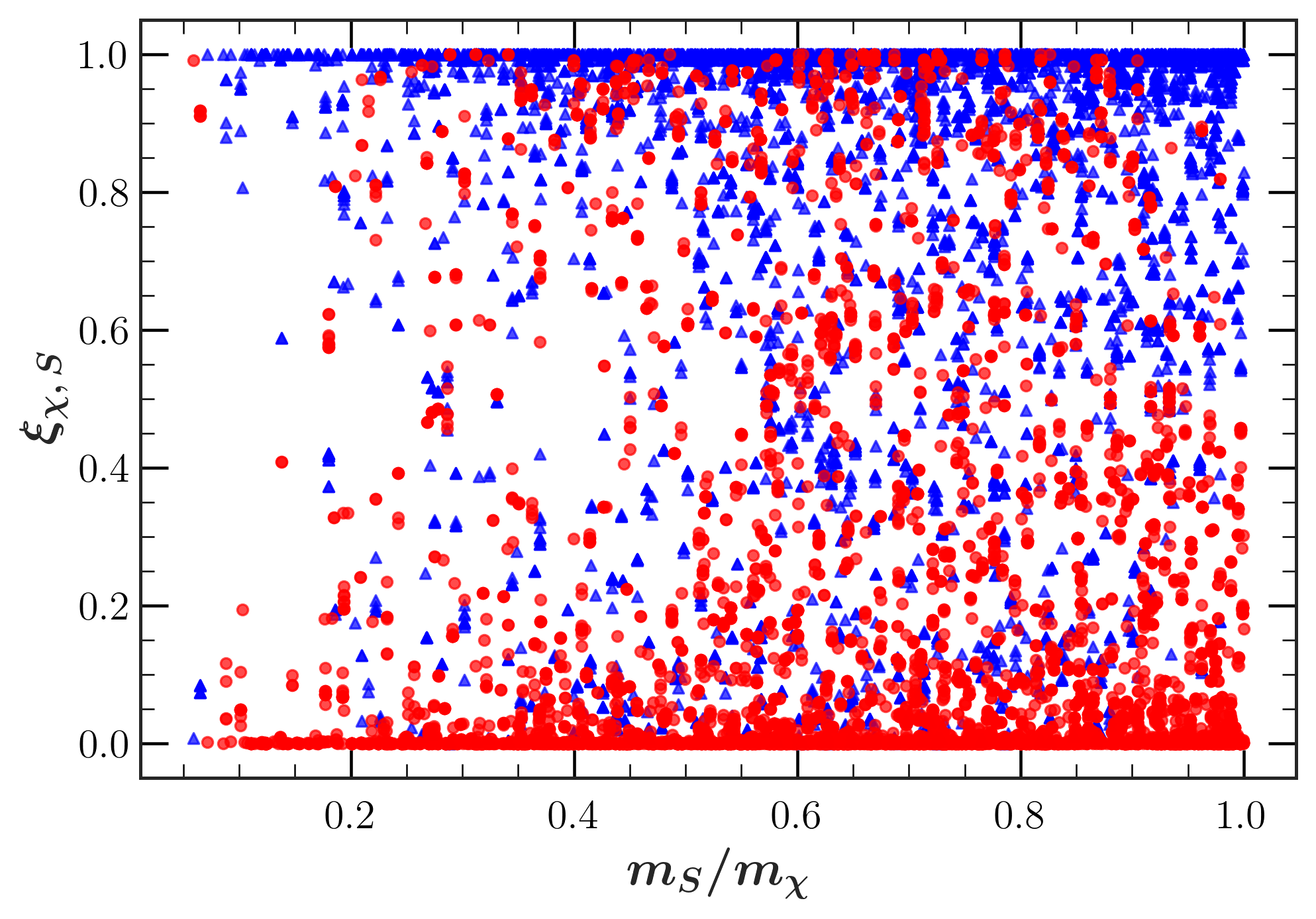}
    \includegraphics[width=0.45\linewidth]{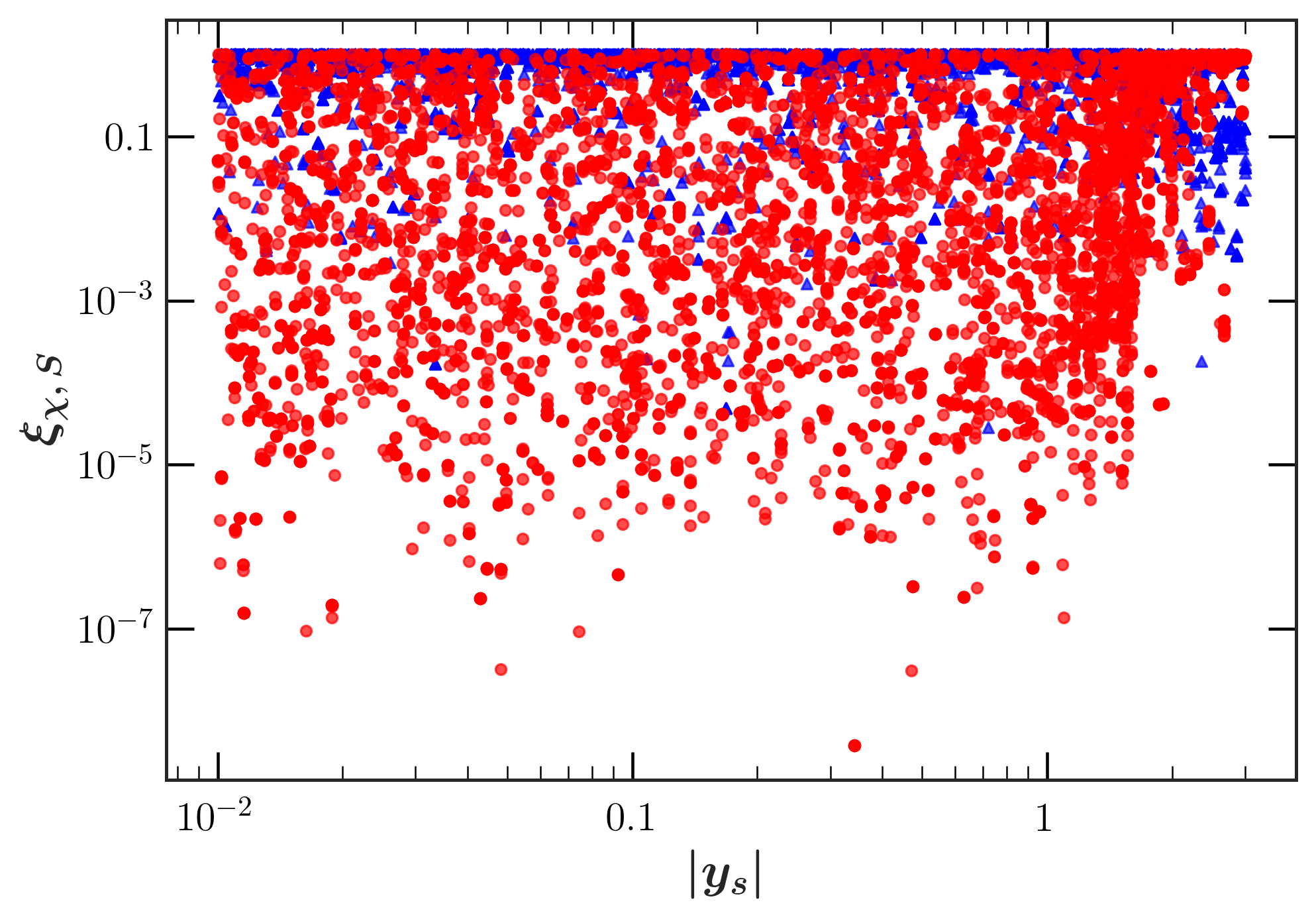}
    \includegraphics[width=0.45\linewidth]{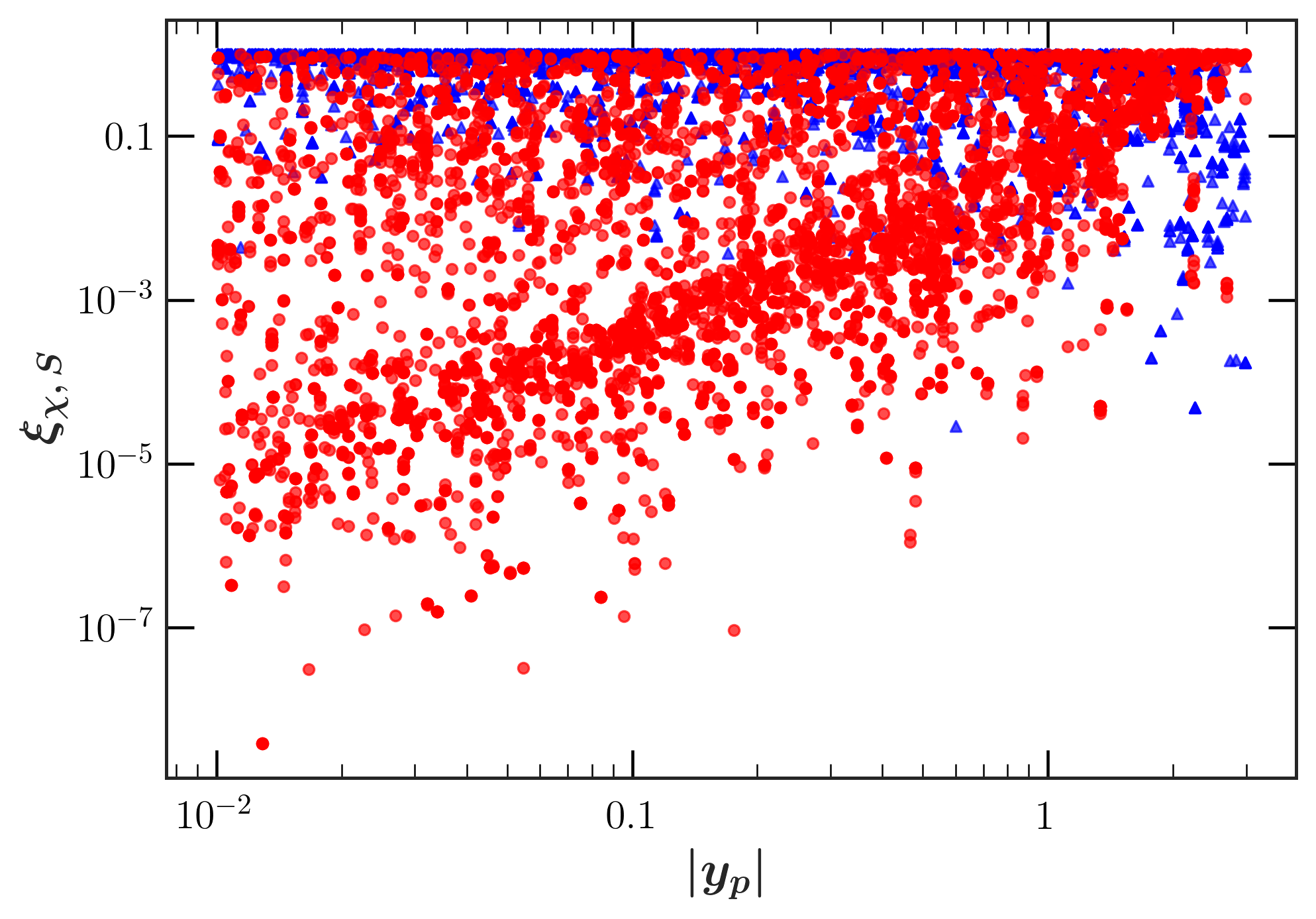}
    \begin{tabular}{c}
        \textbf{DM component:} \hspace{0.1cm} \textcolor{blue}{\ding{115}} $\boldsymbol{\xi_\chi}$ \hspace{0.2cm} \textcolor{red}{\ding{108}} $\boldsymbol{\xi_S}$
    \end{tabular}
    \caption{Relic-density fractions $\xi_\chi$ and $\xi_S$ as functions of selected model parameters for $m_S<m_\chi$. Blue triangles and red circles denote the fermionic and scalar components, respectively. The upper panels use linear scales and show the dependence on $m_\chi$ and $m_S/m_\chi$, while the lower panels use logarithmic scales and show the dependence on the dark Yukawa couplings. All displayed points satisfy the theoretical constraints, collider bounds, the observed total relic abundance, and current direct-detection limits. Indirect-detection constraints are not imposed.}
    \label{fig:reg2_omega}
\end{figure}

We now consider the opposite hierarchy, $m_S<m_\chi$, in which the scalar is the lighter DM particle. Figure~\ref{fig:reg2_omega} shows the relic-density fractions $\xi_\chi$ and $\xi_S$ for the viable points projected onto the same set of representative model parameters used in the previous subsection.

In contrast to the $m_S>m_\chi$ regime, viable solutions now span a much broader range of relic compositions. Scalar-dominated, fermion-dominated, and mixed configurations are all possible, even though $S$ is the lighter state. Thus, the mass ordering alone does not determine which particle dominates the present-day abundance. Either component can become highly subdominant, and the relic composition shows a more pronounced dependence on $|y_p|$ than on $|y_s|$.

\begin{figure}[t!]
    \centering
    \includegraphics[width=0.48\linewidth]
    {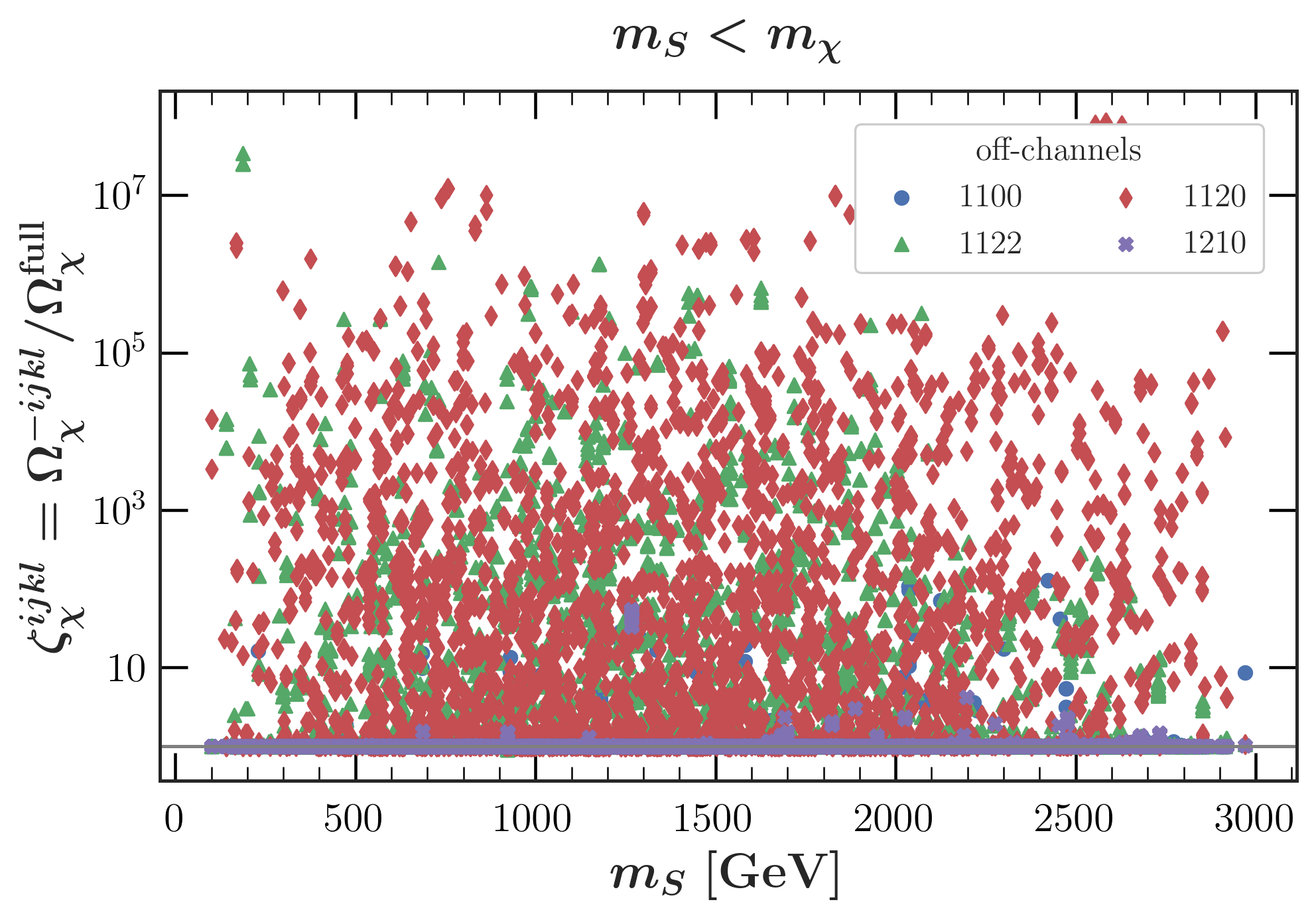}
    \includegraphics[width=0.48\linewidth]
    {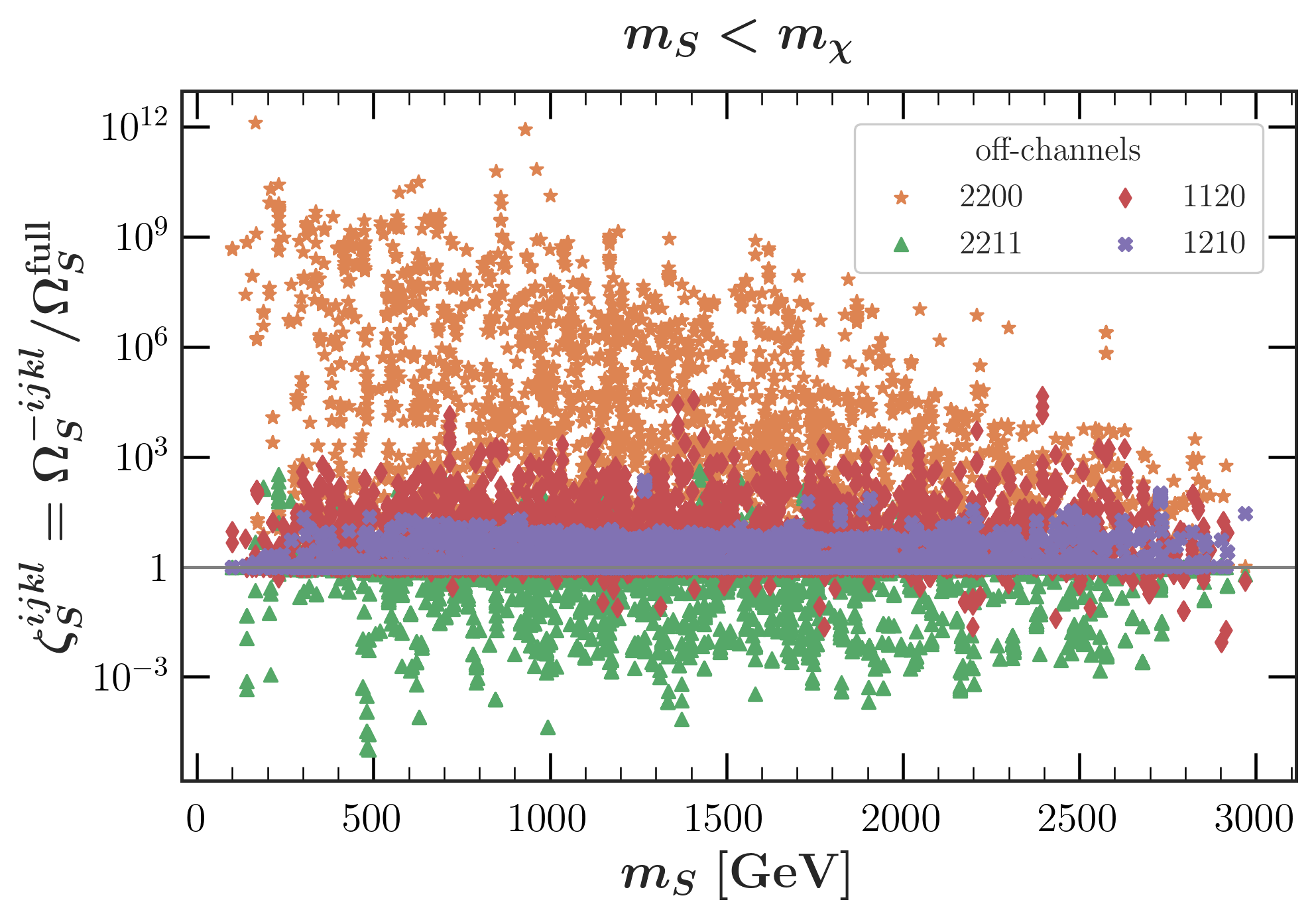}
    \caption{Impact of individual reaction classes on the relic abundances for $m_S<m_\chi$. The left and right panels show $\zeta_\chi^{ijkl}=\Omega_\chi^{(-ijkl)}/\Omega_\chi^{\rm full}$ and $\zeta_S^{ijkl}=\Omega_S^{(-ijkl)}/\Omega_S^{\rm full}$, respectively, as functions of the lighter DM mass $m_S$. The superscript $(-ijkl)$ denotes a calculation in which reaction class $ijkl$ has been removed. The reference value $\zeta_X^{ijkl}=1$ corresponds to no change relative to the full calculation. Values above unity indicate that the removed process normally contributes to the depletion of component $X$, whereas values below unity indicate that it contributes, directly or indirectly, to its production.}
    \label{fig:reg2_offchannels}
\end{figure}

The off-channel analysis in Fig.~\ref{fig:reg2_offchannels} identifies the processes responsible for these relic compositions. For the fermionic component, both the 1120 semi-conversion and 1122 conversion processes can substantially affect the final abundance. Removing either process can significantly increase $\Omega_\chi$, showing that reactions transferring fermionic dark matter into the scalar sector provide important depletion mechanisms for $\chi$. Conventional 1100 annihilation generally has a smaller impact.

The scalar abundance is instead particularly sensitive to the 2200 annihilation channel, whose removal can strongly increase $\Omega_S$. The 1120 semi-conversion plays the opposite role by feeding the scalar population, while 2211 conversion and 1210 semi-annihilation provide additional, point-dependent modifications. The final composition reflects the competition between scalar depletion and fermion-to-scalar transfer.

\begin{figure}[t!]
    \centering
    \includegraphics[width=0.45\linewidth]{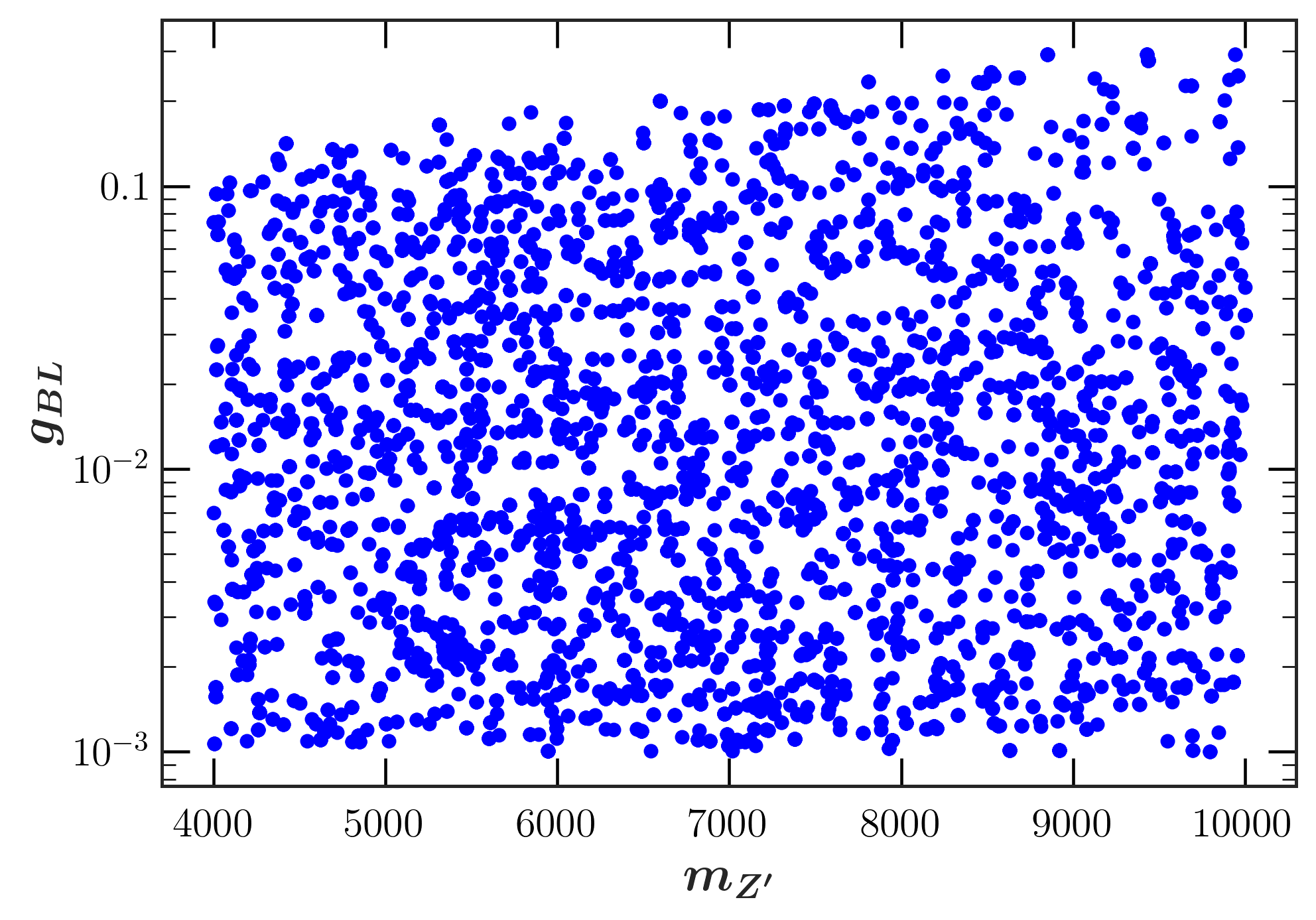}
    \includegraphics[width=0.45\linewidth]{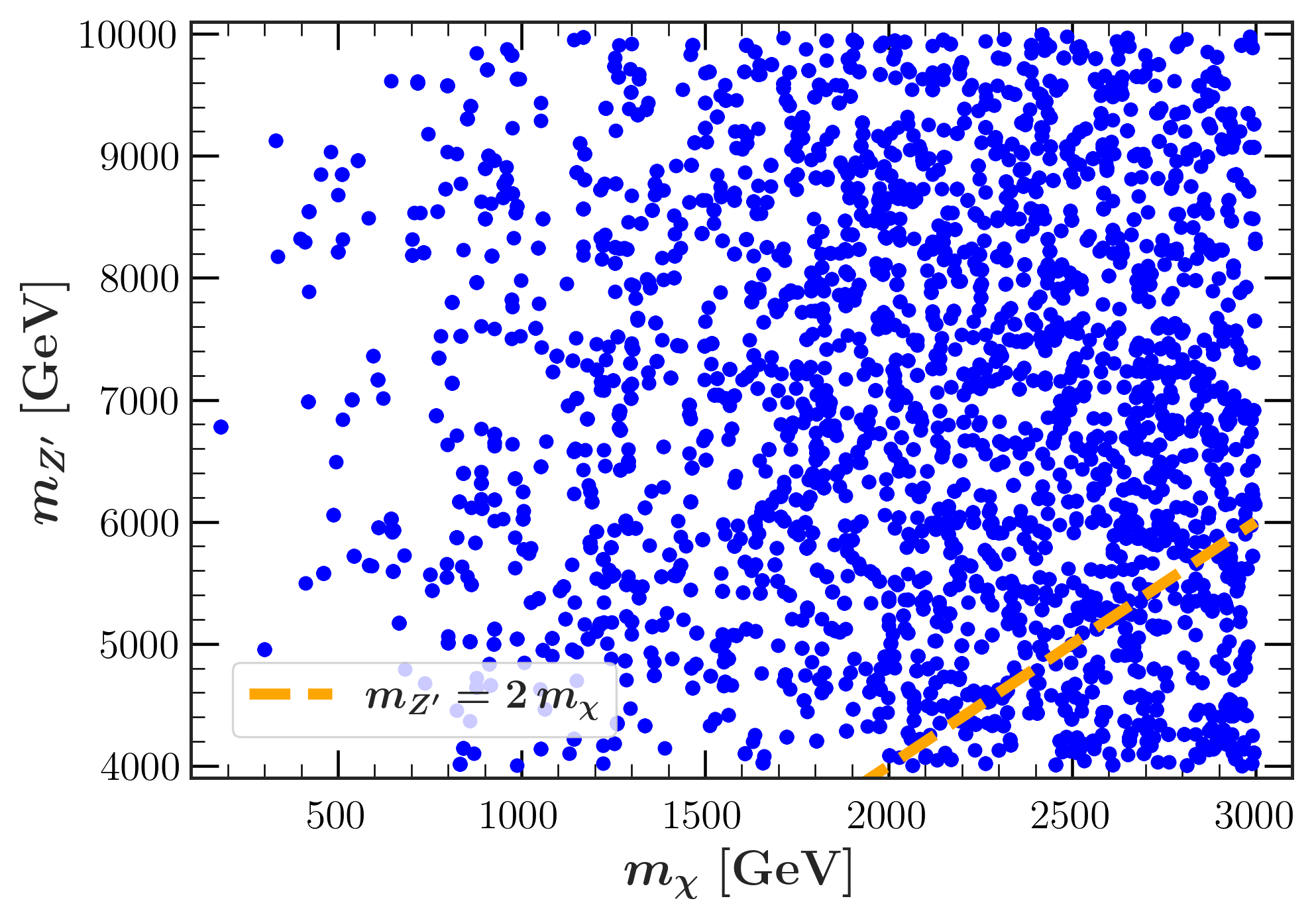}
    \includegraphics[width=0.45\linewidth]{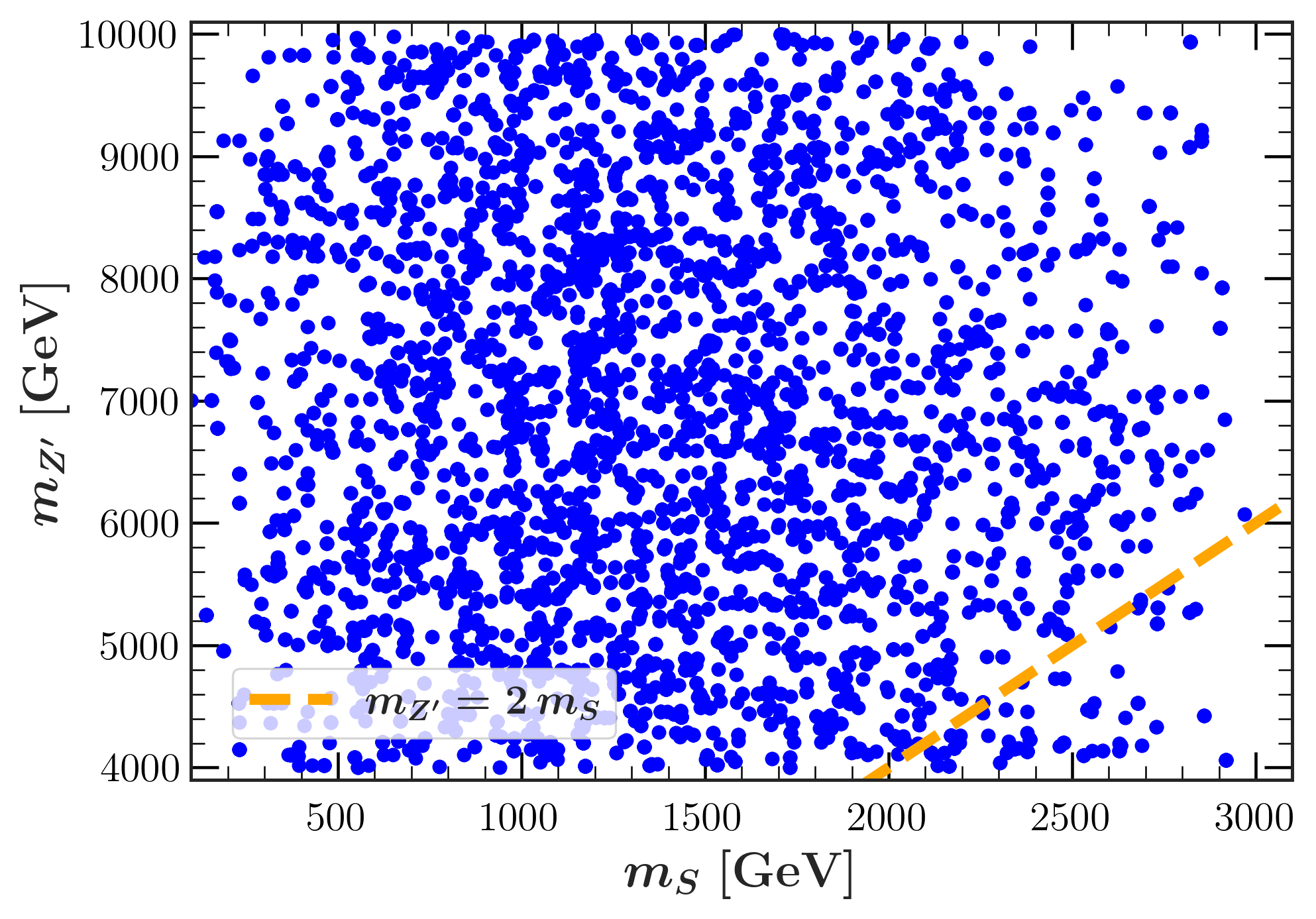}
    \includegraphics[width=0.45\linewidth]{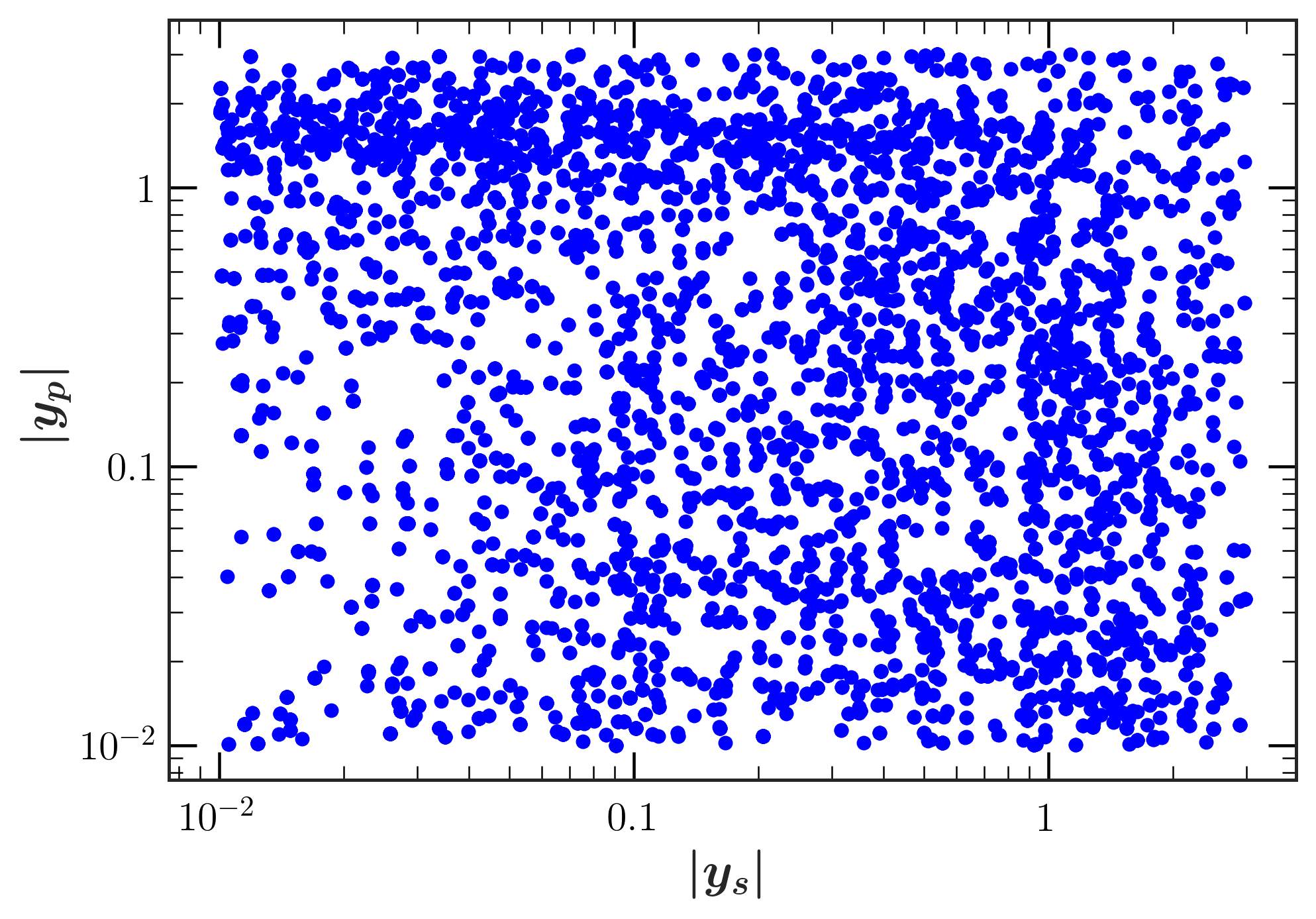}
    \caption{Selected projections of the viable parameter space for $m_S<m_\chi$. The upper-left panel shows the gauge sector in the $(m_{Z'},g_{BL})$ plane. The upper-right and lower-left panels show $m_{Z'}$ as a function of $m_\chi$ and $m_S$, respectively, with reference lines indicating the resonance conditions $m_{Z'}=2m_\chi$ and $m_{Z'}=2m_S$. The lower-right panel shows the correlation between the dark Yukawa couplings $|y_s|$ and $|y_p|$. All displayed points satisfy the theoretical constraints, collider bounds, the observed total relic abundance, and current direct-detection limits.}
    \label{fig:reg2_par}
\end{figure}

Figure~\ref{fig:reg2_par} shows complementary projections of the viable parameter space for the $m_S<m_\chi$ hierarchy. As in the opposite hierarchy, viable solutions occur both near and well away from the $Z'$ resonance conditions, confirming that resonant annihilation is not required. The dark Yukawa couplings also span broad ranges, with no preferred relation between $|y_s|$ and $|y_p|$ selected by the combined constraints.

\begin{figure}[t!]
    \centering
    \includegraphics[width=0.45\linewidth]{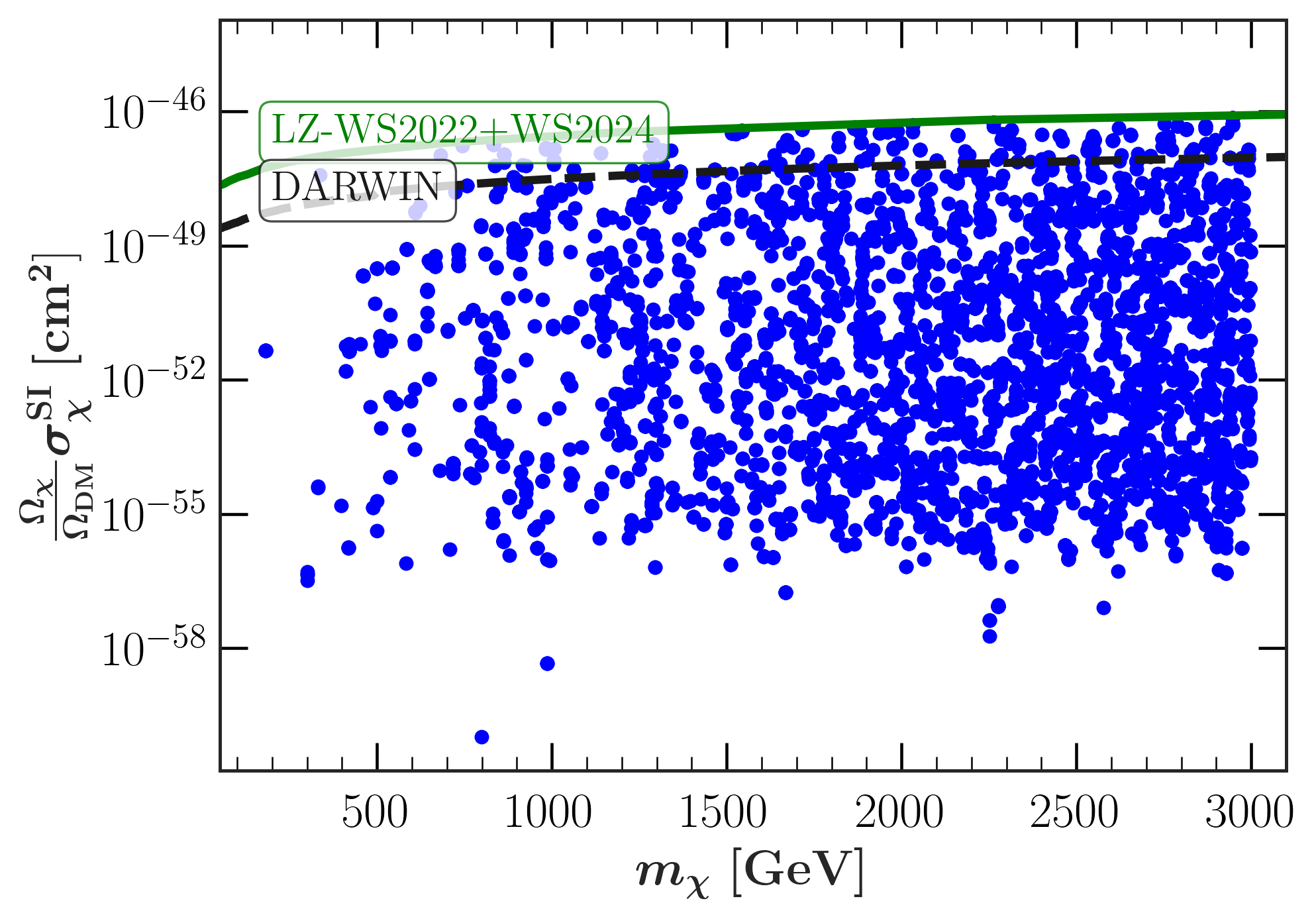}
    \includegraphics[width=0.45\linewidth]{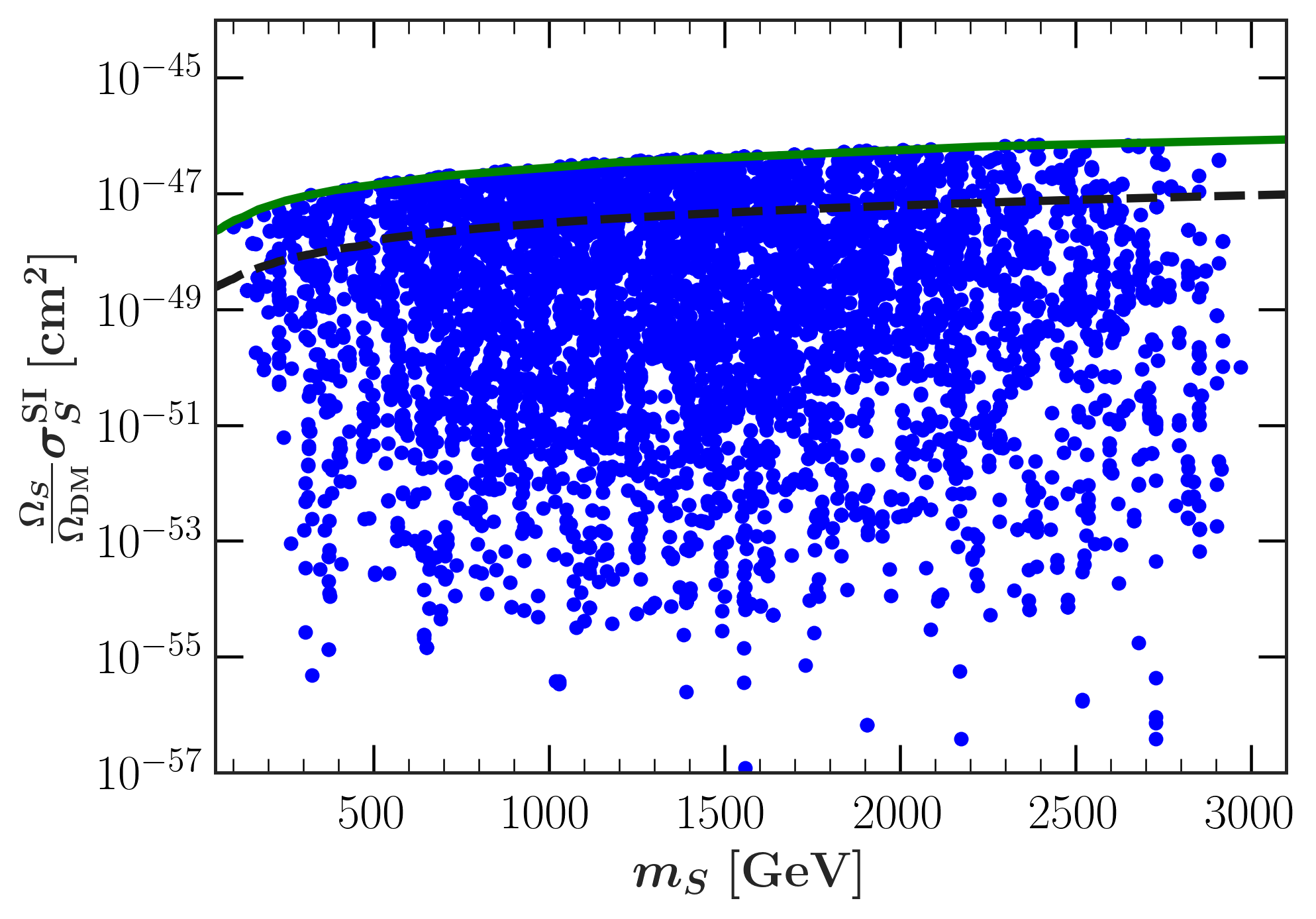}
    \includegraphics[width=0.5\linewidth]{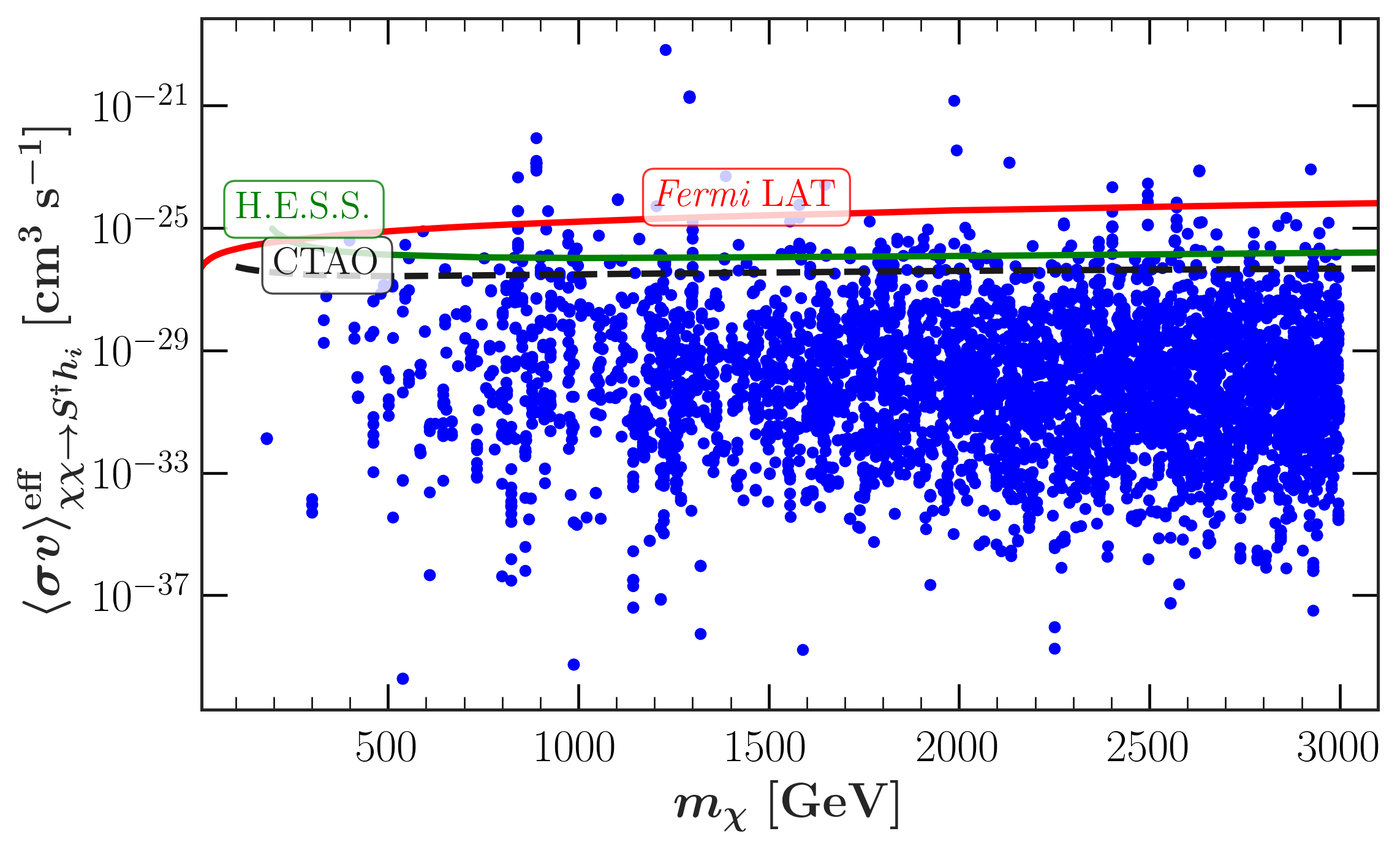}
    \caption{Direct- and indirect-detection predictions for the viable points with $m_S < m_\chi$. The upper panels show the rescaled SI scattering cross sections for $\chi$ (left) and $S$ (right), together with the current LZ limit and the projected DARWIN/XLZD sensitivity. The lower panel shows the effective gamma-ray annihilation rate together with the current \textit{Fermi}-LAT and H.E.S.S. limits and the projected CTAO sensitivity. Current direct-detection constraints are already imposed on the displayed sample, whereas indirect-detection constraints are shown but not imposed.}
    \label{fig:reg2_bounds}
\end{figure}

Figure~\ref{fig:reg2_bounds} summarizes the detection prospects in this hierarchy. Current direct-detection bounds are already imposed on the displayed sample, while part of the surviving parameter space remains within the projected sensitivity of DARWIN/XLZD. Current gamma-ray limits exclude only part of this sample, and CTAO provides complementary future sensitivity.

\section{Discussion}
\label{sec:discussion}

\begin{figure}[t!]
    \centering
    \includegraphics[width=0.65\linewidth]{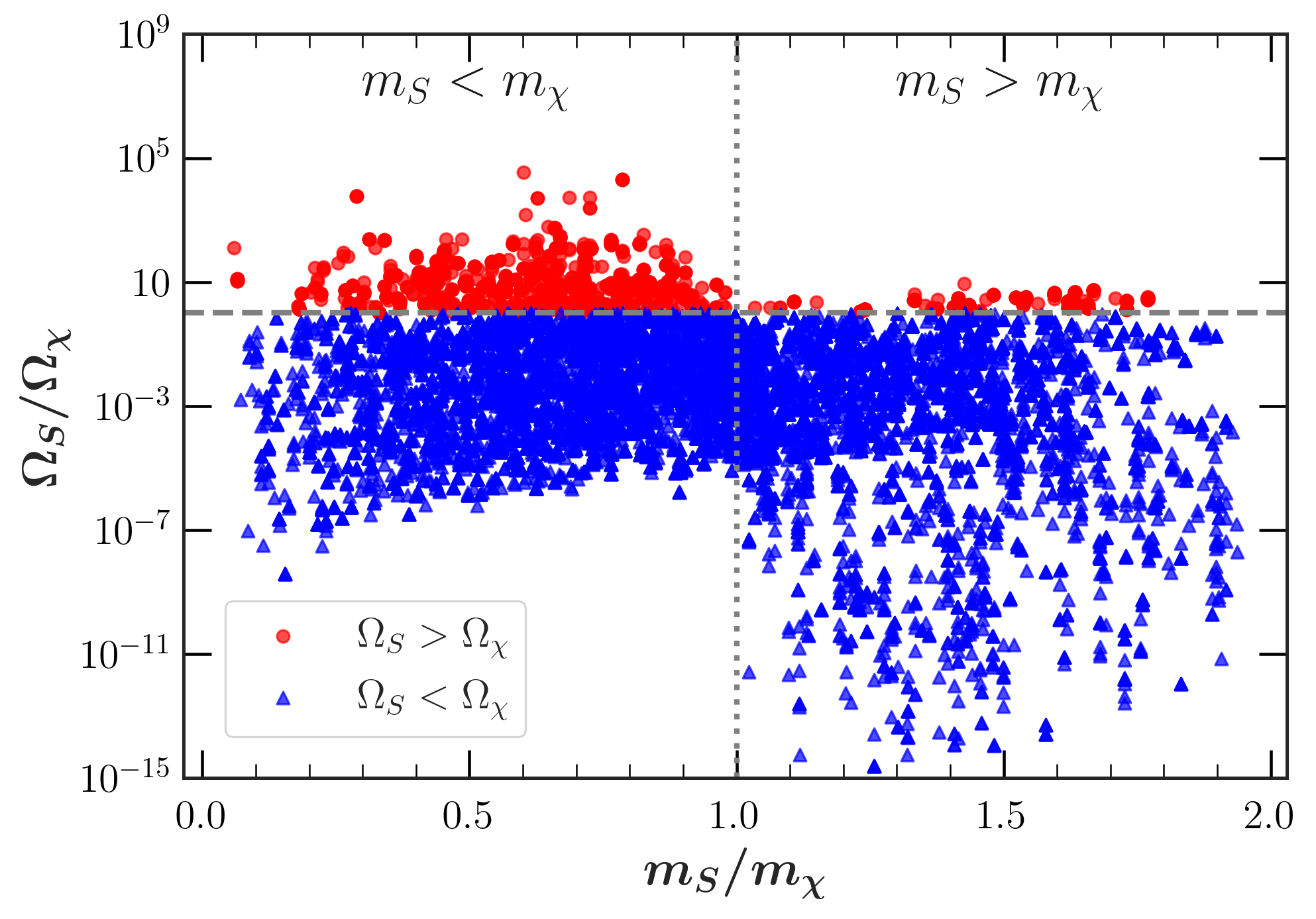}
    \caption{Ratio of the scalar and fermionic relic abundances, $\Omega_S/\Omega_\chi$, as a function of the dark matter mass ratio $m_S/m_\chi$, combining the viable points from both mass hierarchies. The line $m_S/m_\chi=1$ separates the $m_S<m_\chi$ and $m_S>m_\chi$ regimes, while $\Omega_S/\Omega_\chi=1$ corresponds to equal relic abundances. Points above (below) this value are scalar- (fermion-) dominated. All displayed points satisfy the theoretical constraints, collider bounds, the observed total relic abundance, and current direct-detection limits.} 
    \label{fig:omega-ratio}
\end{figure}

The numerical results reveal a common physical picture across the two mass hierarchies. The relic abundances of $\chi$ and $S$ cannot, in general, be understood as the outcome of two approximately independent freeze-out processes, since reactions connecting the two dark sectors can both deplete one component and feed the other. Consequently, the mass ordering alone does not determine the relic composition. Figure~\ref{fig:omega-ratio} illustrates this result directly by combining the viable points from both hierarchies: scalar- and fermion-dominated solutions occur on either side of $m_S=m_\chi$, and the lighter dark matter particle need not be the dominant component.

This behavior is particularly clear for $m_S > m_\chi$, where the 1120 semi-conversion process can strongly deplete the fermionic abundance while simultaneously sourcing the scalar population. For $m_S < m_\chi$, the final composition instead results from a broader competition among annihilation, conversion, semi-annihilation, and semi-conversion processes.

A second important feature is that the correct relic abundance does not require resonant enhancement. Viable solutions populate broad regions away from the $Z'$ resonance, where the freeze-out dynamics is controlled by the interplay among the gauge portal, the scalar portal, and the dark-Yukawa interaction. The gauge coupling $g_{BL}$ connects both dark matter particles to the $Z'$, the scalar portal couplings connect $S$ to the visible sector through $h_1$ and $h_2$, and the couplings $y_s$ and $y_p$ directly connect $\chi$ and $S$. The relic composition is therefore determined by the combined action of these interactions rather than by any single coupling or by the mass hierarchy alone.

The numerical analysis presented above was performed for the benchmark charge assignment $n_\chi=3/4$. It is therefore important to determine which of these features are specific to this choice and which persist for other $B-L$ charges. For generic values of $n_\chi$ outside the special cases listed in Table~\ref{tab:exceptional-charges}, the renormalizable interaction structure of the model remains unchanged: the relation $Q_{B-L}(S)=-2n_\chi$ continues to allow the dark Yukawa interaction, and the same classes of annihilation, conversion, semi-annihilation, and semi-conversion processes remain present. Changing $n_\chi$, however, modifies the strength of the gauge interactions of both dark matter particles and can therefore alter the balance between the $Z'$ portal and the Yukawa-driven dark-sector dynamics. To assess this dependence, we compare the benchmark with one smaller and one larger non-exceptional charge assignment,
\begin{equation}
    n_\chi=\frac{2}{5},\qquad
    \frac{3}{4},\qquad
    \frac{7}{5}.
    \label{eq:nchi-values}
\end{equation}
Since $n_S=-2n_\chi$, varying $n_\chi$ directly affects the interactions mediated by the $Z'$ gauge boson, whereas the Yukawa and scalar interactions do not depend explicitly on $n_\chi$.

The relative importance of the $Z'$ interaction is not the same for all reaction classes. The fermionic annihilation process 1100 is particularly simple. Since $\chi$ has no tree-level coupling to $h_1$ or $h_2$, its annihilation into sector-0 states proceeds predominantly through the $Z'$ portal. Moreover, throughout the scan $m_\chi<m_{Z'}$, so annihilation into a pair of on-shell $Z'$ bosons is kinematically forbidden. Away from the $Z'$ resonance and kinematic thresholds, the corresponding rate therefore scales as $\langle\sigma v\rangle_{1100}^{Z'} \propto n_\chi^2 g_{BL}^4$.

\begin{figure}[t]
    \centering

    \begin{subfigure}[t]{0.49\textwidth}
        \centering
        \includegraphics[width=\linewidth]{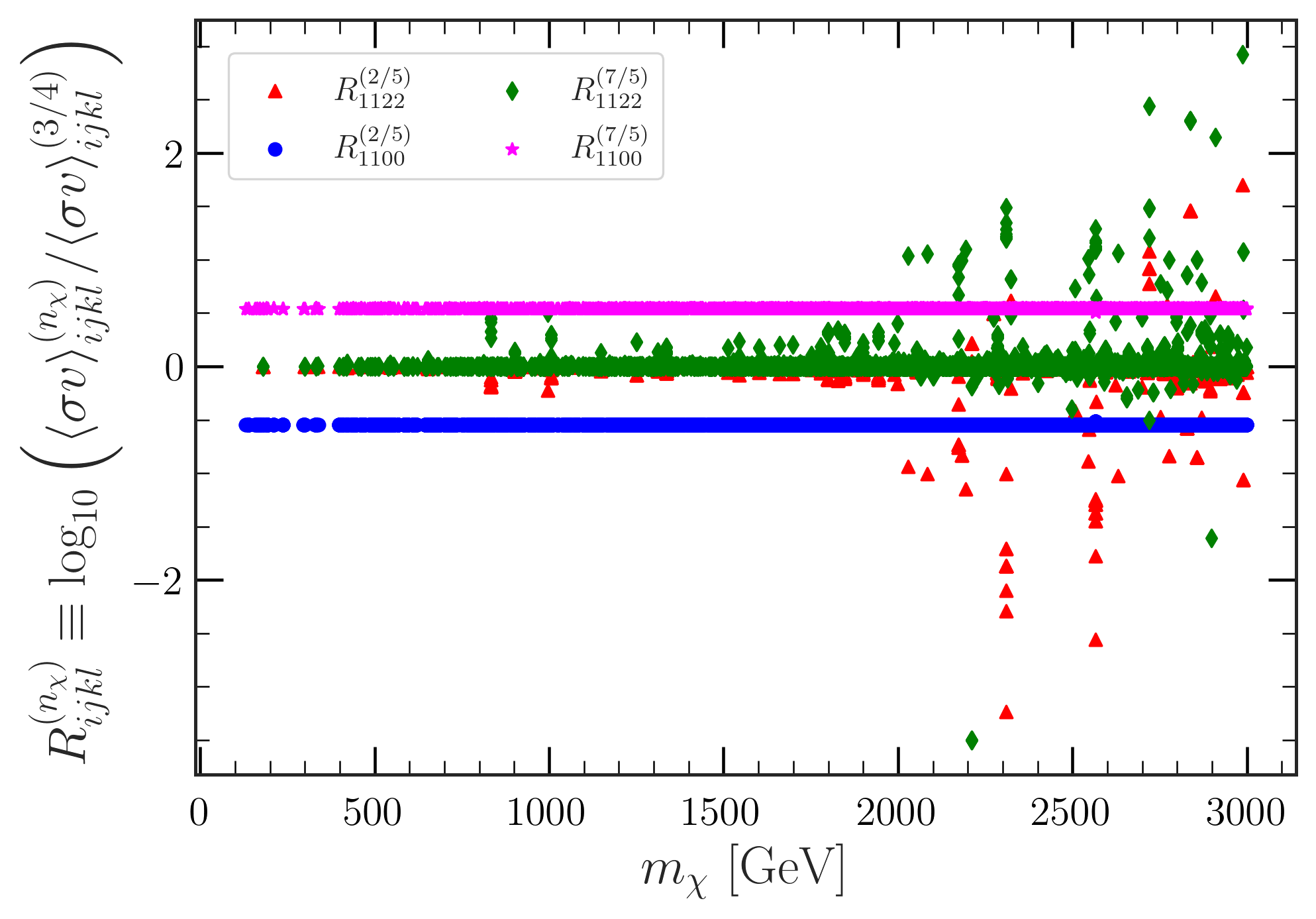}
        \caption{$1100$ and $1122$ processes.}
        \label{fig:nchi-rates-chi}
    \end{subfigure}
    \hfill
    \begin{subfigure}[t]{0.49\textwidth}
        \centering
        \includegraphics[width=\linewidth]{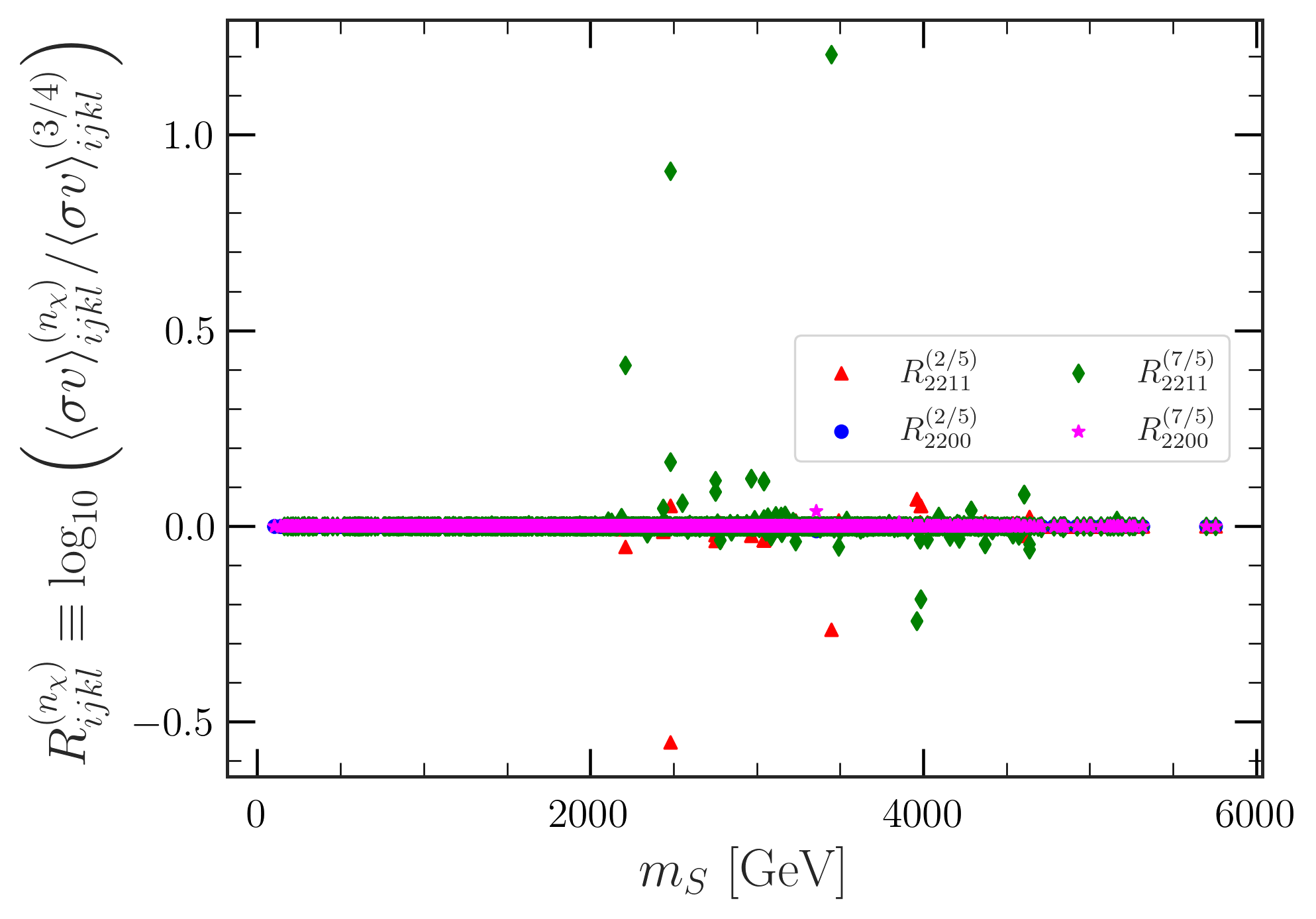}
        \caption{$2200$ and $2211$ processes.}
        \label{fig:nchi-rates-S}
    \end{subfigure}

    \caption{Dependence of representative annihilation and conversion rates on the dark $B-L$ charge. We show the logarithmic ratios $R_{ijkl}^{(n_\chi)}$ defined in Eq.~\eqref{eq:Rnchi}, using $n_\chi=3/4$ as the reference value and comparing it with $n_\chi=2/5$ and $7/5$. The left panel shows the 1100 annihilation and 1122 conversion processes as functions of $m_\chi$, while the right panel shows the 2200 annihilation and 2211 conversion processes as functions of $m_S$.}
    \label{fig:nchi-rates}
\end{figure}

To isolate this charge dependence numerically, we take each viable point from the benchmark scan with $n_\chi=3/4$ and recompute the reaction rates after changing only $n_\chi$, while keeping all masses and couplings fixed. We define
\begin{equation}
    R_{ijkl}^{(n_\chi)}
    =
    \log_{10}\left[
    \frac{\langle\sigma v\rangle_{ijkl}^{(n_\chi)}}
    {\langle\sigma v\rangle_{ijkl}^{(3/4)}}
    \right].
    \label{eq:Rnchi}
\end{equation}
The ratios $R_{ijkl}^{(n_\chi)}$ therefore isolate the explicit dependence of each reaction rate on the dark $B-L$ charge. For the two alternative charge assignments, this scaling predicts
\begin{align}
    R_{1100}^{(2/5)}
    =
    2\log_{10}\left(\frac{2/5}{3/4}\right)
    \simeq -0.55,
    \qquad
    R_{1100}^{(7/5)}
    =
    2\log_{10}\left(\frac{7/5}{3/4}\right)
    \simeq +0.54.
    \label{eq:R1100-expected}
\end{align}
Thus, if the 1100 process is indeed controlled by the $Z'$ portal, the numerical results should form approximately horizontal bands at these two values. The corresponding rate ratios are shown in Fig.~\ref{fig:nchi-rates}.

The situation is different for the scalar annihilation and conversion processes. The 2200 rate receives both gauge and scalar-portal contributions, while $\chi\bar\chi\leftrightarrow SS^\dagger$ receives, in addition to $s$-channel $Z'$ exchange, contributions involving two dark-Yukawa vertices. These latter contributions carry no explicit dependence on $n_\chi$. Since the $Z'$ is heavy and $g_{BL}$ is restricted by collider and direct-detection constraints, whereas the scalar-portal and dark-Yukawa couplings can remain sizable, there is no a priori reason for the 2200, 1122, or 2211 rates to be dominated by the gauge contribution.

The behavior of the 1100 process provides a particularly clear verification of the expected charge dependence. The points form two approximately horizontal bands at $R_{1100}^{(2/5)}\simeq-0.55$ and $R_{1100}^{(7/5)}\simeq+0.54$, in excellent agreement with Eq.~\eqref{eq:R1100-expected}. This confirms that the dependence of the fermionic annihilation rate on $n_\chi$ is controlled primarily by the $Z'$ portal.

The remaining processes behave quite differently. The 2200 scalar annihilation rate shows essentially no systematic dependence on $n_\chi$, while the 1122 and 2211 conversion rates remain clustered around $R_{ijkl}^{(n_\chi)}=0$, apart from isolated points with larger deviations. Thus, although their gauge-mediated contributions depend explicitly on the dark $B-L$ charge, their total rates are largely insensitive to $n_\chi$. This behavior reflects the competing scalar and dark-Yukawa contributions, which carry no explicit charge dependence and dilute the dependence associated with the $Z'$ portal.

\begin{figure}[t]
    \centering

    \begin{subfigure}[t]{0.49\textwidth}
        \centering
        \includegraphics[width=\linewidth]{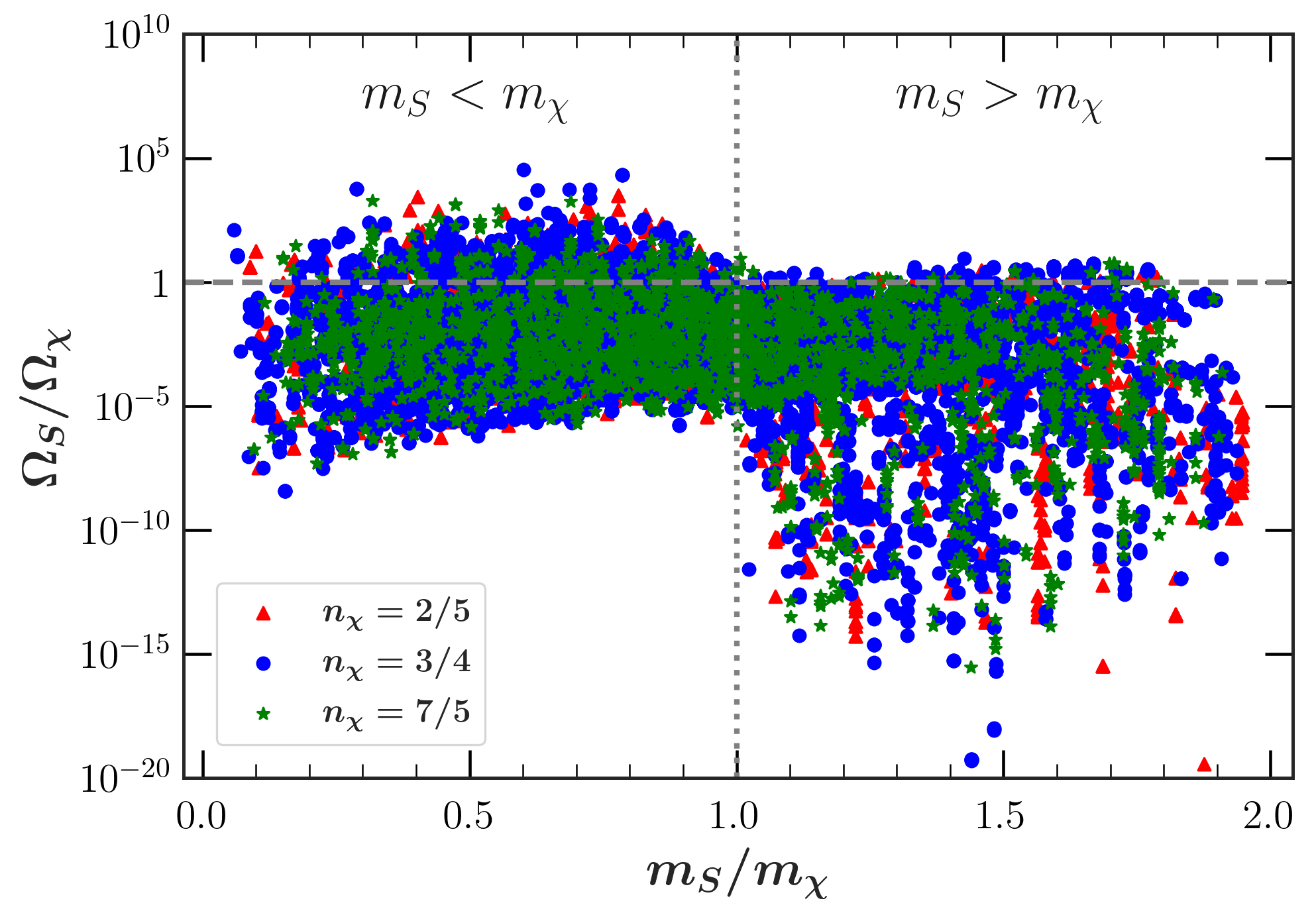}
        \caption{Relic composition.}
        \label{fig:nchi-scan-omega}
    \end{subfigure}
    \hfill
    \begin{subfigure}[t]{0.49\textwidth}
        \centering
        \includegraphics[width=\linewidth]{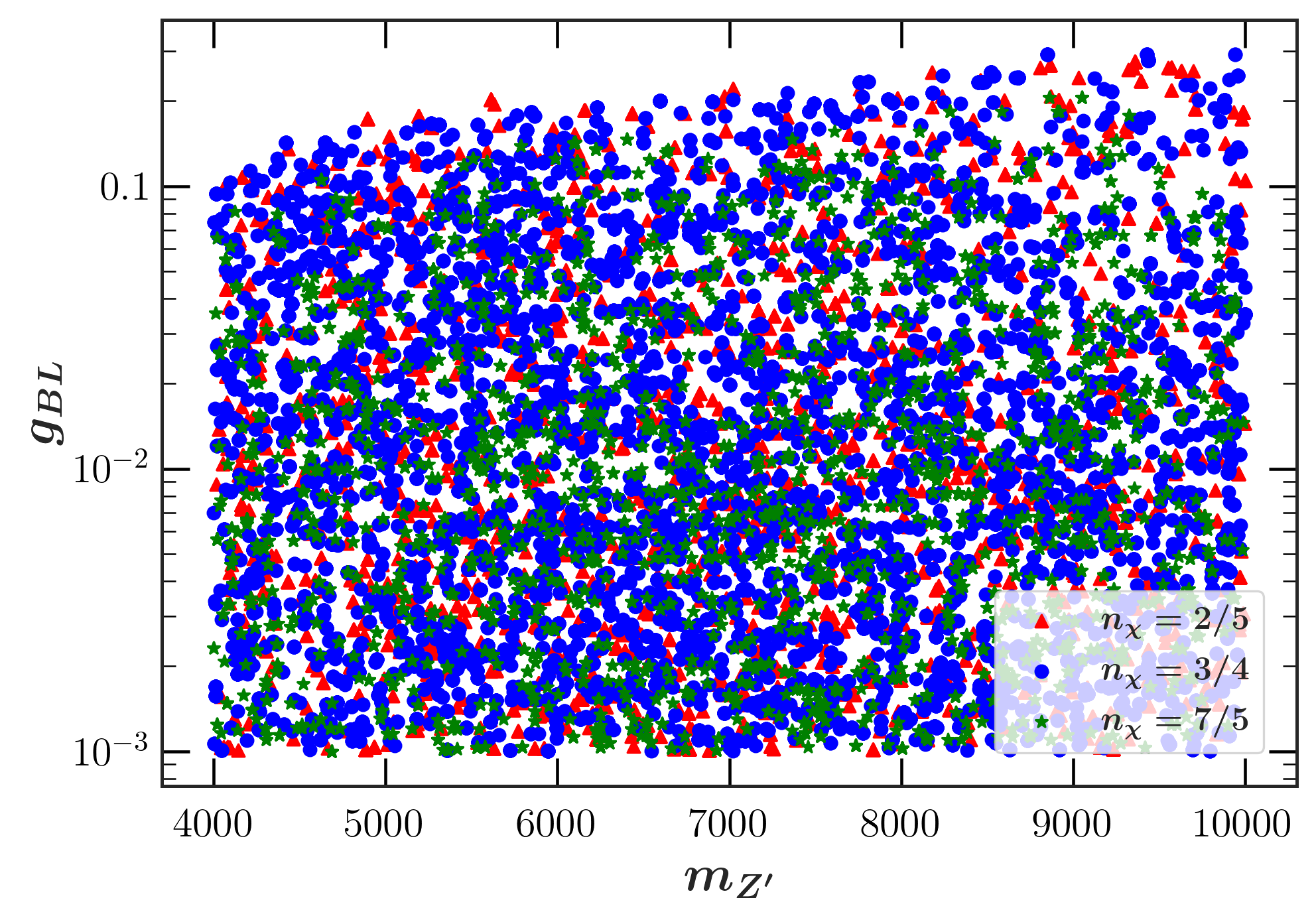}
        \caption{Gauge-sector parameter space.}
        \label{fig:nchi-scan-gauge}
    \end{subfigure}

    \caption{Comparison of the viable parameter space for $n_\chi=2/5$, $3/4$, and $7/5$. The left panel shows the ratio of the scalar and fermionic relic abundances as a function of the dark matter mass ratio. The right panel shows the corresponding viable points in the $(m_{Z'},g_{BL})$ plane. All samples satisfy the same theoretical, collider, relic-density, and direct-detection requirements used in the main numerical analysis.}
    \label{fig:nchi-scans}
\end{figure}

The fixed-parameter comparison above isolates the microscopic dependence of the reaction rates on $n_\chi$. We now ask whether these changes propagate to the full phenomenology once the remaining model parameters are allowed to vary. To this end, we repeat the complete numerical analysis for $n_\chi=2/5$ and $7/5$ and compare the resulting viable samples with the benchmark $n_\chi=3/4$. The resulting comparison is shown in Fig.~\ref{fig:nchi-scans}.

The three samples largely overlap in both projections. The relic composition retains the same qualitative structure found for the benchmark: both scalar- and fermion-dominated solutions occur for $m_S<m_\chi$, whereas the $m_S>m_\chi$ branch is predominantly fermion dominated. Likewise, no clear systematic displacement appears in the $(m_{Z'},g_{BL})$ plane as the charge is varied.

A similarly weak dependence is found in direct detection. Although the $Z'$-mediated amplitudes depend explicitly on $n_\chi$, the relic-fraction-rescaled spin-independent cross sections show no clear systematic shift among the three scans. This is not in conflict with the charge dependence of the gauge contribution, since the observable rates also depend on $g_{BL}$, $m_{Z'}$, the scalar interactions, and the relic fractions $\xi_\chi$ and $\xi_S$.

Taken together, these results show that the explicit charge dependence of the gauge interaction does not translate into an equally strong dependence of the full phenomenology. The clearest effect appears in the gauge-dominated 1100 annihilation rate, where the expected $n_\chi^2$ scaling is reproduced quantitatively. By contrast, the scalar annihilation and conversion rates are much less sensitive to the charge, and the resulting relic composition and viable parameter space remain essentially unchanged. Within the range explored here, the benchmark choice $n_\chi=3/4$ is therefore representative of the broader phenomenology of the model.

\section{Conclusions}
\label{sec:conclusions}

We have proposed an interacting scalar--fermion two-component dark matter model within a gauged $U(1)_{B-L}$ extension of the SM. The dark sector consists of a Dirac fermion $\chi$ and a complex scalar $S$, whose charges allow a renormalizable Yukawa interaction directly connecting the two particles. For suitable charge assignments and $m_S < 2 m_\chi$, both particles are stable without the need for an additional discrete symmetry. The same interaction generates conversion, semi-conversion, and semi-annihilation processes, leading to a genuinely coupled two-component freeze-out.

Our numerical analysis shows that viable solutions exist for both $m_S > m_\chi$ and $m_S < m_\chi$. In particular, semi-conversion can play a central role in depleting the fermionic component while simultaneously feeding the scalar population. The model remains viable over broad non-resonant regions of parameter space, showing that the observed DM abundance can be reproduced without relying on $Z'$ resonance enhancement.

Current constraints still leave sizable viable regions of parameter space. Direct detection provides one of the strongest restrictions on the model, but part of the surviving parameter space remains within the projected reach of next-generation experiments. Finally, varying the dark $B-L$ charge over representative non-exceptional values leaves the overall phenomenology largely unchanged, indicating that the benchmark choice $n_\chi = 3/4$ captures the main features of the model over the range of charges explored.

\section*{Acknowledgments}
The authors thank Giorgio Arcadi, Juri Smirnov, and Josef Pradler for insightful discussions. JPN thanks Patricio Escalona for his immense help with the multidimensional parameter space analysis and for providing valuable code advice, and Guillermo Gambini for helpful discussions about the model-building aspects of this work. 
JPN thanks the support from the Austrian Federal Ministry of Women, Science and Research (BMFWF) via the \textit{Ernst Mach Grant, weltweit} (Award number: MPC-2025-01008), 
the Fundação de Amparo e Promoção da Ciência, Tecnologia e Inovação do Rio Grande do Norte (FAPERN) via the call number 06/2026 - FAPERN/IIP (Process number: 10910022.000493/2024-04.),
the National Council for Scientific and Technological Development (CNPq) (Grant number: 307130/2021-5), and
the Brazilian Federal Agency for Support and Evaluation of Graduate Education (CAPES) (Grant number: 88887.712383/2022-00).
JPN thanks the Marietta Blau Institute for Particle Physics (MBI/ÖAW) for the hospitality during the partial development of this project. This work was supported by the Simons Foundation (Award Number: 1023171-RC).
FSQ is supported by the Humboldt Foundation (Friedrich Wilhelm Bessel Research Award), ANID - Millennium Science Initiative Program - ICN2019\_044, IIF-FINEP grant 213/2024, FAPESP Grants 2021/14335-0 and 2021/00449-4, and CNPq grants 307130/2021-5, 403521/2024-6, 309465/2026-5, and 446711/2025-0.

\bibliographystyle{JHEPfixed.bst}
\bibliography{references}

@article{Qi:2025jpm,
    author = "Qi, XinXin and Sun, Hao",
    title = "{A two-component dark matter model with Z $_{2}$ {\texttimes} Z $_{4}$ symmetry}",
    eprint = "2504.12876",
    archivePrefix = "arXiv",
    primaryClass = "hep-ph",
    doi = "10.1088/1475-7516/2025/08/002",
    journal = "JCAP",
    volume = "08",
    pages = "002",
    year = "2025"
}

@article{Arcadi:2017kky,
    author = "Arcadi, Giorgio and Dutra, Ma{\'\i}ra and Ghosh, Pradipta and Lindner, Manfred and Mambrini, Yann and Pierre, Mathias and Profumo, Stefano and Queiroz, Farinaldo S.",
    title = "{The waning of the WIMP? A review of models, searches, and constraints}",
    eprint = "1703.07364",
    archivePrefix = "arXiv",
    primaryClass = "hep-ph",
    doi = "10.1140/epjc/s10052-018-5662-y",
    journal = "Eur. Phys. J. C",
    volume = "78",
    number = "3",
    pages = "203",
    year = "2018"
}

@article{Abercrombie:2015wmb,
    author = "Abercrombie, Daniel and others",
    editor = "Boveia, Antonio and Doglioni, Caterina and Lowette, Steven and Malik, Sarah and Mrenna, Stephen",
    title = "{Dark Matter benchmark models for early LHC Run-2 Searches: Report of the ATLAS/CMS Dark Matter Forum}",
    eprint = "1507.00966",
    archivePrefix = "arXiv",
    primaryClass = "hep-ex",
    reportNumber = "FERMILAB-PUB-15-282-CD",
    doi = "10.1016/j.dark.2019.100371",
    journal = "Phys. Dark Univ.",
    volume = "27",
    pages = "100371",
    year = "2020"
}

@article{Esch:2014jpa,
    author = "Esch, Sonja and Klasen, Michael and Yaguna, Carlos E.",
    title = "{A minimal model for two-component dark matter}",
    eprint = "1406.0617",
    archivePrefix = "arXiv",
    primaryClass = "hep-ph",
    reportNumber = "MS-TP-14-22",
    doi = "10.1007/JHEP09(2014)108",
    journal = "JHEP",
    volume = "09",
    pages = "108",
    year = "2014"
}

@article{Klasen:2016qux,
    author = "Klasen, Michael and Lyonnet, Florian and Queiroz, Farinaldo S.",
    title = "{NLO+NLL collider bounds, Dirac fermion and scalar dark matter in the B{\textendash}L model}",
    eprint = "1607.06468",
    archivePrefix = "arXiv",
    primaryClass = "hep-ph",
    reportNumber = "MITP-16-065",
    doi = "10.1140/epjc/s10052-017-4904-8",
    journal = "Eur. Phys. J. C",
    volume = "77",
    number = "5",
    pages = "348",
    year = "2017"
}

@article{Abdallah:2015ter,
    author = "Abdallah, Jalal and others",
    title = "{Simplified Models for Dark Matter Searches at the LHC}",
    eprint = "1506.03116",
    archivePrefix = "arXiv",
    primaryClass = "hep-ph",
    reportNumber = "FERMILAB-PUB-15-283-CD, CERN-PH-TH-2015-139",
    doi = "10.1016/j.dark.2015.08.001",
    journal = "Phys. Dark Univ.",
    volume = "9-10",
    pages = "8--23",
    year = "2015"
}

@article{Arcadi:2024ukq,
    author = "Arcadi, Giorgio and Cabo-Almeida, David and Dutra, Ma{\'\i}ra and Ghosh, Pradipta and Lindner, Manfred and Mambrini, Yann and Neto, Jacinto P. and Pierre, Mathias and Profumo, Stefano and Queiroz, Farinaldo S.",
    title = "{The Waning of the WIMP: Endgame?}",
    eprint = "2403.15860",
    archivePrefix = "arXiv",
    primaryClass = "hep-ph",
    doi = "10.1140/epjc/s10052-024-13672-y",
    journal = "Eur. Phys. J. C",
    volume = "85",
    number = "2",
    pages = "152",
    year = "2025"
}

@article{Das:2022oyx,
    author = "Das, Arindam and Gola, Shivam and Mandal, Sanjoy and Sinha, Nita",
    title = "{Two-component scalar and fermionic dark matter candidates in a generic U(1)X model}",
    eprint = "2202.01443",
    archivePrefix = "arXiv",
    primaryClass = "hep-ph",
    reportNumber = "EPHOU-21-017",
    doi = "10.1016/j.physletb.2022.137117",
    journal = "Phys. Lett. B",
    volume = "829",
    pages = "137117",
    year = "2022"
}

@article{DiazSaez:2021pmg,
    author = "D{\'\i}az S{\'a}ez, Basti{\'a}n and Escalona, Patricio and Norero, Sebasti{\'a}n and Zerwekh, Alfonso R.",
    title = "{Fermion singlet dark matter in a pseudoscalar dark matter portal}",
    eprint = "2105.04255",
    archivePrefix = "arXiv",
    primaryClass = "hep-ph",
    doi = "10.1007/JHEP10(2021)233",
    journal = "JHEP",
    volume = "10",
    pages = "233",
    year = "2021"
}

@article{Bhattacharya:2025eym,
    author = "Bhattacharya, Subhaditya and Pradhan, Dipankar",
    title = "{Direct Search signal of two-component Dark Matter}",
    eprint = "2501.17862",
    archivePrefix = "arXiv",
    primaryClass = "hep-ph",
    month = "1",
    year = "2025"
}

@article{Yaguna:2019cvp,
    author = "Yaguna, Carlos E. and Zapata, \'Oscar",
    title = "{Multi-component scalar dark matter from a $Z_N$ symmetry: a systematic analysis}",
    eprint = "1911.05515",
    archivePrefix = "arXiv",
    primaryClass = "hep-ph",
    doi = "10.1007/JHEP03(2020)109",
    journal = "JHEP",
    volume = "03",
    pages = "109",
    year = "2020"
}

@article{Zurek:2008qg,
      author         = "Zurek, Kathryn M.",
      title          = "{Multi-Component Dark Matter}",
      journal        = "Phys. Rev.",
      volume         = "D79",
      year           = "2009",
      pages          = "115002",
      doi            = "10.1103/PhysRevD.79.115002",
      eprint         = "0811.4429",
      archivePrefix  = "arXiv",
      primaryClass   = "hep-ph",
      reportNumber   = "FERMILAB-PUB-08-542-A",
      SLACcitation   = "%%CITATION = ARXIV:0811.4429;%%"
}

@article{Costa:2022oaa,
    author = "Costa, Francesco and Khan, Sarif and Kim, Jinsu",
    title = "{A two-component dark matter model and its associated gravitational waves}",
    eprint = "2202.13126",
    archivePrefix = "arXiv",
    primaryClass = "hep-ph",
    reportNumber = "CERN-TH-2022-022",
    doi = "10.1007/JHEP06(2022)026",
    journal = "JHEP",
    volume = "06",
    pages = "026",
    year = "2022"
}

@article{Queiroz:2024ipo,
    author = "Queiroz, Farinaldo S. and Zamora-Saa, Jilberto and Silva, Ricardo C. and Oviedo-Torres, Y. M.",
    title = "{LHC and HL-LHC bounds on visible and invisible Z' decays in the B-L model}",
    eprint = "2501.00610",
    archivePrefix = "arXiv",
    primaryClass = "hep-ph",
    doi = "10.1103/PhysRevD.111.095021",
    journal = "Phys. Rev. D",
    volume = "111",
    number = "9",
    pages = "095021",
    year = "2025"
}

@article{Ferber:2023iso,
    author = "Ferber, Torben and Grohsjean, Alexander and Kahlhoefer, Felix",
    title = "{Dark Higgs bosons at colliders}",
    eprint = "2305.16169",
    archivePrefix = "arXiv",
    primaryClass = "hep-ph",
    reportNumber = "P3H-23-034, TTP23-018",
    doi = "10.1016/j.ppnp.2024.104105",
    journal = "Prog. Part. Nucl. Phys.",
    volume = "136",
    pages = "104105",
    year = "2024"
}

@article{Alguero:2023zol,
    author = "Alguero, G. and Belanger, G. and Boudjema, F. and Chakraborti, S. and Goudelis, A. and Kraml, S. and Mjallal, A. and Pukhov, A.",
    title = "{micrOMEGAs 6.0: N-component dark matter}",
    eprint = "2312.14894",
    archivePrefix = "arXiv",
    primaryClass = "hep-ph",
    doi = "10.1016/j.cpc.2024.109133",
    journal = "Comput. Phys. Commun.",
    volume = "299",
    pages = "109133",
    year = "2024"
}

@article{Kannike:2012pe,
    author = "Kannike, Kristjan",
    title = "{Vacuum Stability Conditions From Copositivity Criteria}",
    eprint = "1205.3781",
    archivePrefix = "arXiv",
    primaryClass = "hep-ph",
    doi = "10.1140/epjc/s10052-012-2093-z",
    journal = "Eur. Phys. J. C",
    volume = "72",
    pages = "2093",
    year = "2012"
}

@article{Alloul:2013bka,
    author = "Alloul, Adam and Christensen, Neil D. and Degrande, C{\'e}line and Duhr, Claude and Fuks, Benjamin",
    title = "{FeynRules  2.0 - A complete toolbox for tree-level phenomenology}",
    eprint = "1310.1921",
    archivePrefix = "arXiv",
    primaryClass = "hep-ph",
    reportNumber = "CERN-PH-TH-2013-239, MCNET-13-14, IPPP-13-71, DCPT-13-142, PITT-PACC-1308",
    doi = "10.1016/j.cpc.2014.04.012",
    journal = "Comput. Phys. Commun.",
    volume = "185",
    pages = "2250--2300",
    year = "2014"
}

@article{Semenov:2014rea,
    author = "Semenov, A.",
    title = "{LanHEP {\textemdash} A package for automatic generation of Feynman rules from the Lagrangian. Version 3.2}",
    eprint = "1412.5016",
    archivePrefix = "arXiv",
    primaryClass = "physics.comp-ph",
    reportNumber = "LAPTH-234-14",
    doi = "10.1016/j.cpc.2016.01.003",
    journal = "Comput. Phys. Commun.",
    volume = "201",
    pages = "167--170",
    year = "2016"
}

@misc{alguero_2024_10844949,
  author       = {Alguero, Gaël and
                  Belanger, Genevieve and
                  Boudjema, Fawzi and
                  Chakraborti, Sreemanti and
                  Goudelis, Andreas and
                  Kraml, Sabine and
                  Mjallal, Ali and
                  Pukhov, Alexander},
  title        = {micrOMEGAs 6.0},
  month        = mar,
  year         = 2024,
  publisher    = {Zenodo},
  doi          = {10.5281/zenodo.10844949},
  url          = {https://doi.org/10.5281/zenodo.10844949},
}

@article{Gondolo:1990dk,
    author = "Gondolo, Paolo and Gelmini, Graciela",
    title = "{Cosmic abundances of stable particles: Improved analysis}",
    reportNumber = "UCLA-90-TEP-68",
    doi = "10.1016/0550-3213(91)90438-4",
    journal = "Nucl. Phys. B",
    volume = "360",
    pages = "145--179",
    year = "1991"
}

@article{Feng:2017drg,
    author = "Feng, Jonathan L. and Smolinsky, Jordan",
    title = "{Impact of a resonance on thermal targets for invisible dark photon searches}",
    eprint = "1707.03835",
    archivePrefix = "arXiv",
    primaryClass = "hep-ph",
    reportNumber = "UCI-TR-2017-06",
    doi = "10.1103/PhysRevD.96.095022",
    journal = "Phys. Rev. D",
    volume = "96",
    number = "9",
    pages = "095022",
    year = "2017"
}

@article{Yaguna:2021rds,
    author = "Yaguna, Carlos E. and Zapata, {\'O}scar",
    title = "{Fermion and scalar two-component dark matter from a Z4 symmetry}",
    eprint = "2112.07020",
    archivePrefix = "arXiv",
    primaryClass = "hep-ph",
    doi = "10.1103/PhysRevD.105.095026",
    journal = "Phys. Rev. D",
    volume = "105",
    number = "9",
    pages = "095026",
    year = "2022"
}

@article{Queiroz:2019acr,
    author = "Queiroz, Farinaldo S. and Siqueira, Clarissa",
    title = "{Search for Semi-Annihilating Dark Matter with Fermi-LAT, H.E.S.S., Planck, and the Cherenkov Telescope Array}",
    eprint = "1901.10494",
    archivePrefix = "arXiv",
    primaryClass = "hep-ph",
    reportNumber = "IIPDM-2019",
    doi = "10.1088/1475-7516/2019/04/048",
    journal = "JCAP",
    volume = "04",
    pages = "048",
    year = "2019"
}

@article{CMS:2015hra,
    author = "Khachatryan, Vardan and others",
    collaboration = "CMS",
    title = "{Search for a Higgs boson in the mass range from 145 to 1000 GeV decaying to a pair of W or Z bosons}",
    eprint = "1504.00936",
    archivePrefix = "arXiv",
    primaryClass = "hep-ex",
    reportNumber = "CMS-HIG-13-031, CERN-PH-EP-2015-074",
    doi = "10.1007/JHEP10(2015)144",
    journal = "JHEP",
    volume = "10",
    pages = "144",
    year = "2015"
}

@article{CMS:2018amk,
    author = "Sirunyan, Albert M and others",
    collaboration = "CMS",
    title = "{Search for a new scalar resonance decaying to a pair of Z bosons in proton-proton collisions at $\sqrt{s}=13 $ TeV}",
    eprint = "1804.01939",
    archivePrefix = "arXiv",
    primaryClass = "hep-ex",
    reportNumber = "CMS-HIG-17-012, CERN-EP-2018-009",
    doi = "10.1007/JHEP06(2018)127",
    journal = "JHEP",
    volume = "06",
    pages = "127",
    year = "2018",
    note = "[Erratum: JHEP 03, 128 (2019)]"
}

@article{CMS:2021yci,
    author = "Tumasyan, Armen and others",
    collaboration = "CMS",
    title = "{Search for a heavy Higgs boson decaying into two lighter Higgs bosons in the $\tau\tau$bb final state at 13 TeV}",
    eprint = "2106.10361",
    archivePrefix = "arXiv",
    primaryClass = "hep-ex",
    reportNumber = "CMS-HIG-20-014, CERN-EP-2021-094",
    doi = "10.1007/JHEP11(2021)057",
    journal = "JHEP",
    volume = "11",
    pages = "057",
    year = "2021"
}

@article{CMS:2021ctt,
    author = "Sirunyan, Albert M and others",
    collaboration = "CMS",
    title = "{Search for resonant and nonresonant new phenomena in high-mass dilepton final states at $ \sqrt{s} $ = 13 TeV}",
    eprint = "2103.02708",
    archivePrefix = "arXiv",
    primaryClass = "hep-ex",
    reportNumber = "CMS-EXO-19-019, CERN-EP-2021-026",
    doi = "10.1007/JHEP07(2021)208",
    journal = "JHEP",
    volume = "07",
    pages = "208",
    year = "2021"
}

@article{ATLAS:2018sbw,
    author = "Aaboud, Morad and others",
    collaboration = "ATLAS",
    title = "{Combination of searches for heavy resonances decaying into bosonic and leptonic final states using 36  fb$^{-1}$ of proton-proton collision data at $\sqrt{s} = 13$ TeV with the ATLAS detector}",
    eprint = "1808.02380",
    archivePrefix = "arXiv",
    primaryClass = "hep-ex",
    reportNumber = "CERN-EP-2018-179",
    doi = "10.1103/PhysRevD.98.052008",
    journal = "Phys. Rev. D",
    volume = "98",
    number = "5",
    pages = "052008",
    year = "2018"
}

@article{ATLAS:2020fry,
    author = "Aad, Georges and others",
    collaboration = "ATLAS",
    title = "{Search for heavy diboson resonances in semileptonic final states in pp collisions at $\sqrt{s}=13$ TeV with the ATLAS detector}",
    eprint = "2004.14636",
    archivePrefix = "arXiv",
    primaryClass = "hep-ex",
    reportNumber = "CERN-EP-2020-049",
    doi = "10.1140/epjc/s10052-020-08554-y",
    journal = "Eur. Phys. J. C",
    volume = "80",
    number = "12",
    pages = "1165",
    year = "2020"
}

@article{ATLAS:2020tlo,
    author = "Aad, Georges and others",
    collaboration = "ATLAS",
    title = "{Search for heavy resonances decaying into a pair of Z bosons in the $\ell ^+\ell ^-\ell '^+\ell '^-$ and $\ell ^+\ell ^-\nu {{\bar{\nu }}}$ final states using 139 $\mathrm {fb}^{-1}$ of proton{\textendash}proton collisions at $\sqrt{s} = 13\,$TeV with the ATLAS detector}",
    eprint = "2009.14791",
    archivePrefix = "arXiv",
    primaryClass = "hep-ex",
    reportNumber = "CERN-EP-2020-153",
    doi = "10.1140/epjc/s10052-021-09013-y",
    journal = "Eur. Phys. J. C",
    volume = "81",
    number = "4",
    pages = "332",
    year = "2021"
}

@article{ATLAS:2022xzm,
    author = "Aad, Georges and others",
    collaboration = "ATLAS",
    title = "{Search for resonant and non-resonant Higgs boson pair production in the $ b\overline{b}{\tau}^{+}{\tau}^{-} $ decay channel using 13 TeV pp collision data from the ATLAS detector}",
    eprint = "2209.10910",
    archivePrefix = "arXiv",
    primaryClass = "hep-ex",
    reportNumber = "CERN-EP-2022-109",
    doi = "10.1007/JHEP07(2023)040",
    journal = "JHEP",
    volume = "07",
    pages = "040",
    year = "2023"
}

@article{Bhattiprolu:2022ycf,
    author = "Bhattiprolu, Prudhvi N. and Wells, James D.",
    title = "{Depleted Higgs boson: Searches for universal coupling suppression, invisible decays, and mixed-in scalars}",
    eprint = "2204.03435",
    archivePrefix = "arXiv",
    primaryClass = "hep-ph",
    doi = "10.1103/PhysRevD.107.055022",
    journal = "Phys. Rev. D",
    volume = "107",
    number = "5",
    pages = "055022",
    year = "2023"
}

@article{ATLAS:2022vkf,
    author = "Aad, Georges and others",
    collaboration = "ATLAS",
    title = "{A detailed map of Higgs boson interactions by the ATLAS experiment ten years after the discovery}",
    eprint = "2207.00092",
    archivePrefix = "arXiv",
    primaryClass = "hep-ex",
    reportNumber = "CERN-EP-2022-057",
    doi = "10.1038/s41586-022-04893-w",
    journal = "Nature",
    volume = "607",
    number = "7917",
    pages = "52--59",
    year = "2022",
    note = "[Erratum: Nature 612, E24 (2022)]"
}

@article{CMS:2022dwd,
    author = "Tumasyan, Armen and others",
    collaboration = "CMS",
    title = "{A portrait of the Higgs boson by the CMS experiment ten years after the discovery.}",
    eprint = "2207.00043",
    archivePrefix = "arXiv",
    primaryClass = "hep-ex",
    reportNumber = "CMS-HIG-22-001, CERN-EP-2022-039",
    doi = "10.1038/s41586-022-04892-x",
    journal = "Nature",
    volume = "607",
    number = "7917",
    pages = "60--68",
    year = "2022",
    note = "[Erratum: Nature 623, (2023)]"
}

@article{ATLAS:2019erb,
    author = "Aad, Georges and others",
    collaboration = "ATLAS",
    title = "{Search for high-mass dilepton resonances using 139 fb$^{-1}$ of $pp$ collision data collected at $\sqrt{s}=$13 TeV with the ATLAS detector}",
    eprint = "1903.06248",
    archivePrefix = "arXiv",
    primaryClass = "hep-ex",
    reportNumber = "CERN-EP-2019-030",
    doi = "10.1016/j.physletb.2019.07.016",
    journal = "Phys. Lett. B",
    volume = "796",
    pages = "68--87",
    year = "2019"
}

@article{ATLAS:2019fgd,
    author = "Aad, Georges and others",
    collaboration = "ATLAS",
    title = "{Search for new resonances in mass distributions of jet pairs using 139 fb$^{-1}$ of $pp$ collisions at $\sqrt{s}=13$ TeV with the ATLAS detector}",
    eprint = "1910.08447",
    archivePrefix = "arXiv",
    primaryClass = "hep-ex",
    reportNumber = "CERN-EP-2019-162",
    doi = "10.1007/JHEP03(2020)145",
    journal = "JHEP",
    volume = "03",
    pages = "145",
    year = "2020"
}

@article{ATLAS:2019npw,
    author = "Aaboud, Morad and others",
    collaboration = "ATLAS",
    title = "{Search for heavy particles decaying into a top-quark pair in the fully hadronic final state in $pp$ collisions at $\sqrt{s} =$ 13  TeV with the ATLAS detector}",
    eprint = "1902.10077",
    archivePrefix = "arXiv",
    primaryClass = "hep-ex",
    reportNumber = "CERN-EP-2018-350",
    doi = "10.1103/PhysRevD.99.092004",
    journal = "Phys. Rev. D",
    volume = "99",
    number = "9",
    pages = "092004",
    year = "2019"
}

@article{Arcadi:2023lwc,
    author = "Arcadi, Giorgio and Dorsch, Glauber C. and Neto, Jacinto P. and Queiroz, Farinaldo S. and Oviedo-Torres, Y. M.",
    title = "{Probing a dark sector with collider physics, direct detection, and gravitational waves}",
    eprint = "2307.06376",
    archivePrefix = "arXiv",
    primaryClass = "hep-ph",
    doi = "10.1016/j.physletb.2023.138382",
    journal = "Phys. Lett. B",
    volume = "848",
    pages = "138382",
    year = "2024"
}

@article{Planck:2018vyg,
    author = "Aghanim, N. and others",
    collaboration = "Planck",
    title = "{Planck 2018 results. VI. Cosmological parameters}",
    eprint = "1807.06209",
    archivePrefix = "arXiv",
    primaryClass = "astro-ph.CO",
    doi = "10.1051/0004-6361/201833910",
    journal = "Astron. Astrophys.",
    volume = "641",
    pages = "A6",
    year = "2020",
    note = "[Erratum: Astron.Astrophys. 652, C4 (2021)]"
}

@article{Belanger:2014vza,
    author = "B{\'e}langer, G. and Boudjema, F. and Pukhov, A. and Semenov, A.",
    title = "{micrOMEGAs4.1: two dark matter candidates}",
    eprint = "1407.6129",
    archivePrefix = "arXiv",
    primaryClass = "hep-ph",
    doi = "10.1016/j.cpc.2015.03.003",
    journal = "Comput. Phys. Commun.",
    volume = "192",
    pages = "322--329",
    year = "2015"
}

@article{HESS:2022ygk,
    author = "Abdalla, H. and others",
    collaboration = "H.E.S.S.",
    title = "{Search for Dark Matter Annihilation Signals in the H.E.S.S. Inner Galaxy Survey}",
    eprint = "2207.10471",
    archivePrefix = "arXiv",
    primaryClass = "astro-ph.HE",
    doi = "10.1103/PhysRevLett.129.111101",
    journal = "Phys. Rev. Lett.",
    volume = "129",
    number = "11",
    pages = "111101",
    year = "2022"
}

@article{Fermi-LAT:2015att,
    author = "Ackermann, M. and others",
    collaboration = "Fermi-LAT",
    title = "{Searching for Dark Matter Annihilation from Milky Way Dwarf Spheroidal Galaxies with Six Years of Fermi Large Area Telescope Data}",
    eprint = "1503.02641",
    archivePrefix = "arXiv",
    primaryClass = "astro-ph.HE",
    reportNumber = "FERMILAB-PUB-15-081-AE",
    doi = "10.1103/PhysRevLett.115.231301",
    journal = "Phys. Rev. Lett.",
    volume = "115",
    number = "23",
    pages = "231301",
    year = "2015"
}

@article{Silverwood:2014yza,
    author = "Silverwood, Hamish and Weniger, Christoph and Scott, Pat and Bertone, Gianfranco",
    title = "{A realistic assessment of the CTA sensitivity to dark matter annihilation}",
    eprint = "1408.4131",
    archivePrefix = "arXiv",
    primaryClass = "astro-ph.HE",
    doi = "10.1088/1475-7516/2015/03/055",
    journal = "JCAP",
    volume = "03",
    pages = "055",
    year = "2015"
}

@article{CTA:2020qlo,
    author = "Acharyya, A. and others",
    collaboration = "CTA",
    title = "{Sensitivity of the Cherenkov Telescope Array to a dark matter signal from the Galactic centre}",
    eprint = "2007.16129",
    archivePrefix = "arXiv",
    primaryClass = "astro-ph.HE",
    doi = "10.1088/1475-7516/2021/01/057",
    journal = "JCAP",
    volume = "01",
    pages = "057",
    year = "2021"
}

@article{Amiri:2025ras,
    author = {Amiri, Amir and Diaz Saez, Bastian and M{\"o}hling, Kilian},
    title = "{Freeze-in at Low Reheating and Direct Detection of Fermion Dark Matter}",
    eprint = "2511.21520",
    archivePrefix = "arXiv",
    primaryClass = "hep-ph",
    month = "11",
    year = "2025"
}

@article{Haisch:2013uaa,
    author = "Haisch, Ulrich and Kahlhoefer, Felix",
    title = "{On the importance of loop-induced spin-independent interactions for dark matter direct detection}",
    eprint = "1302.4454",
    archivePrefix = "arXiv",
    primaryClass = "hep-ph",
    reportNumber = "OUTP-13-06P",
    doi = "10.1088/1475-7516/2013/04/050",
    journal = "JCAP",
    volume = "04",
    pages = "050",
    year = "2013"
}

@article{Chang:2017gla,
    author = "Chang, Chia-Feng and He, Xiao-Gang and Tandean, Jusak",
    title = "{Two-Higgs-Doublet-Portal Dark-Matter Models in Light of Direct Search and LHC Data}",
    eprint = "1702.02924",
    archivePrefix = "arXiv",
    primaryClass = "hep-ph",
    reportNumber = "NCTS-PH-1707",
    doi = "10.1007/JHEP04(2017)107",
    journal = "JHEP",
    volume = "04",
    pages = "107",
    year = "2017"
}

@article{XENON:2023cxc,
    author = "Aprile, E. and others",
    collaboration = "XENON",
    title = "{First Dark Matter Search with Nuclear Recoils from the XENONnT Experiment}",
    eprint = "2303.14729",
    archivePrefix = "arXiv",
    primaryClass = "hep-ex",
    doi = "10.1103/PhysRevLett.131.041003",
    journal = "Phys. Rev. Lett.",
    volume = "131",
    number = "4",
    pages = "041003",
    year = "2023"
}

@article{LZ:2024zvo,
    author = "Aalbers, J. and others",
    collaboration = "LZ",
    title = "{Dark Matter Search Results from 4.2{\,}{\,}Tonne-Years of Exposure of the LUX-ZEPLIN (LZ) Experiment}",
    eprint = "2410.17036",
    archivePrefix = "arXiv",
    primaryClass = "hep-ex",
    reportNumber = "FERMILAB-PUB-24-0796-V",
    doi = "10.1103/4dyc-z8zf",
    journal = "Phys. Rev. Lett.",
    volume = "135",
    number = "1",
    pages = "011802",
    year = "2025"
}

@article{Baudis:2024jnk,
    author = "Baudis, Laura",
    title = "{DARWIN/XLZD: A future xenon observatory for dark matter and other rare interactions}",
    eprint = "2404.19524",
    archivePrefix = "arXiv",
    primaryClass = "astro-ph.IM",
    doi = "10.1016/j.nuclphysb.2024.116473",
    journal = "Nucl. Phys. B",
    volume = "1003",
    pages = "116473",
    year = "2024"
}

@article{Frank:2025jjt,
    author = "Frank, Mariana and Ghosh, Purusottam and Majumdar, Chayan and Senapati, Supriya",
    title = "{Exploring two component doublet dark matter}",
    eprint = "2505.02816",
    archivePrefix = "arXiv",
    primaryClass = "hep-ph",
    doi = "10.1103/3sxl-lf71",
    journal = "Phys. Rev. D",
    volume = "112",
    number = "5",
    pages = "055014",
    year = "2025"
}

@article{Belanger:2020hyh,
    author = "B{\'e}langer, Genevi{\`e}ve and Pukhov, Alexander and Yaguna, Carlos E. and Zapata, {\'O}scar",
    title = "{The Z$_{5}$ model of two-component dark matter}",
    eprint = "2006.14922",
    archivePrefix = "arXiv",
    primaryClass = "hep-ph",
    doi = "10.1007/JHEP09(2020)030",
    journal = "JHEP",
    volume = "09",
    pages = "030",
    year = "2020"
}

@article{Belanger:2022esk,
    author = "B{\'e}langer, Genevi{\`e}ve and Pukhov, Alexander and Yaguna, Carlos E. and Zapata, {\'O}scar",
    title = "{The Z$_{7}$ model of three-component scalar dark matter}",
    eprint = "2212.07488",
    archivePrefix = "arXiv",
    primaryClass = "hep-ph",
    doi = "10.1007/JHEP03(2023)100",
    journal = "JHEP",
    volume = "03",
    pages = "100",
    year = "2023"
}

@article{Tucker-Smith:2001myb,
    author = "Tucker-Smith, David and Weiner, Neal",
    title = "{Inelastic dark matter}",
    eprint = "hep-ph/0101138",
    archivePrefix = "arXiv",
    reportNumber = "UCB-PTH-00-43, LBNL-47234, UW-PT-00-17",
    doi = "10.1103/PhysRevD.64.043502",
    journal = "Phys. Rev. D",
    volume = "64",
    pages = "043502",
    year = "2001"
}

@article{Queiroz:2026tah,
    author = "Queiroz, Farinaldo S. and Siqueira, Clarissa and Yaguna, Carlos E.",
    title = "{Multi-Component Dark Matter as a Solution to the Galactic Center GeV Excess}",
    eprint = "2603.17095",
    archivePrefix = "arXiv",
    primaryClass = "hep-ph",
    month = "3",
    year = "2026"
}

@article{Belanger:2026asz,
    author = "Belanger, G. and Belyaev, A. and Bernal, N. and Boudjema, F. and Chakraborti, S. and Goudelis, A. and Pukhov, A.",
    title = "{micrOMEGAs 7: Beyond standard cosmology}",
    eprint = "2606.06645",
    archivePrefix = "arXiv",
    primaryClass = "hep-ph",
    reportNumber = "IPPP/26/46",
    month = "6",
    year = "2026"
}

@article{Escalona:2026trm,
    author = "Escalona, Patricio and Neto, Jacinto P. and Neves, M. J. and Ramos, Camila and Suarez, David",
    title = "{Two-Component Dark Matter in the Type-I 2HDM}",
    eprint = "2603.18158",
    archivePrefix = "arXiv",
    primaryClass = "hep-ph",
    month = "3",
    year = "2026"
}

@article{Rodejohann:2015lca,
    author = "Rodejohann, Werner and Yaguna, Carlos E.",
    title = "{Scalar dark matter in the B{\ensuremath{-}}L model}",
    eprint = "1509.04036",
    archivePrefix = "arXiv",
    primaryClass = "hep-ph",
    doi = "10.1088/1475-7516/2015/12/032",
    journal = "JCAP",
    volume = "12",
    pages = "032",
    year = "2015"
}

@article{Bernal:2018aon,
    author = "Bernal, Nicol{\'a}s and Restrepo, Diego and Yaguna, Carlos and Zapata, {\'O}scar",
    title = "{Two-component dark matter and a massless neutrino in a new $B-L$ model}",
    eprint = "1808.03352",
    archivePrefix = "arXiv",
    primaryClass = "hep-ph",
    doi = "10.1103/PhysRevD.99.015038",
    journal = "Phys. Rev. D",
    volume = "99",
    number = "1",
    pages = "015038",
    year = "2019"
}

@article{VanDong:2020bkg,
    author = "Van Dong, Phung and Nam, Cao H. and Van Loi, Duong",
    title = "{Canonical seesaw implication for two-component dark matter}",
    eprint = "2007.08957",
    archivePrefix = "arXiv",
    primaryClass = "hep-ph",
    doi = "10.1103/PhysRevD.103.095016",
    journal = "Phys. Rev. D",
    volume = "103",
    pages = "095016",
    year = "2021"
}

@article{Belanger:2011ww,
    author = "Belanger, Genevieve and Park, Jong-Chul",
    title = "{Assisted freeze-out}",
    eprint = "1112.4491",
    archivePrefix = "arXiv",
    primaryClass = "hep-ph",
    doi = "10.1088/1475-7516/2012/03/038",
    journal = "JCAP",
    volume = "03",
    pages = "038",
    year = "2012"
}

@article{Hoferichter:2017olk,
    author = "Hoferichter, Martin and Klos, Philipp and Men{\'e}ndez, Javier and Schwenk, Achim",
    title = "{Improved limits for Higgs-portal dark matter from LHC searches}",
    eprint = "1708.02245",
    archivePrefix = "arXiv",
    primaryClass = "hep-ph",
    reportNumber = "INT-PUB-17-031",
    doi = "10.1103/PhysRevLett.119.181803",
    journal = "Phys. Rev. Lett.",
    volume = "119",
    number = "18",
    pages = "181803",
    year = "2017"
}

@article{Basso:2010jm,
    author = "Basso, Lorenzo and Moretti, Stefano and Pruna, Giovanni Marco",
    title = "{A Renormalisation Group Equation Study of the Scalar Sector of the Minimal B-L Extension of the Standard Model}",
    eprint = "1004.3039",
    archivePrefix = "arXiv",
    primaryClass = "hep-ph",
    reportNumber = "SHEP-10-16",
    doi = "10.1103/PhysRevD.82.055018",
    journal = "Phys. Rev. D",
    volume = "82",
    pages = "055018",
    year = "2010"
}

@article{Mohapatra:1979ia,
    author = "Mohapatra, Rabindra N. and Senjanovic, Goran",
    title = "{Neutrino Mass and Spontaneous Parity Nonconservation}",
    reportNumber = "MDDP-TR-80-060, MDDP-PP-80-105, CCNY-HEP-79-10",
    doi = "10.1103/PhysRevLett.44.912",
    journal = "Phys. Rev. Lett.",
    volume = "44",
    pages = "912",
    year = "1980"
}

@article{Elor:2015bho,
    author = "Elor, Gilly and Rodd, Nicholas L. and Slatyer, Tracy R. and Xue, Wei",
    title = "{Model-Independent Indirect Detection Constraints on Hidden Sector Dark Matter}",
    eprint = "1511.08787",
    archivePrefix = "arXiv",
    primaryClass = "hep-ph",
    reportNumber = "MIT-CTP-4742",
    doi = "10.1088/1475-7516/2016/06/024",
    journal = "JCAP",
    volume = "06",
    pages = "024",
    year = "2016"
}

@article{Okada:2018ktp,
    author = "Okada, Satomi",
    title = "{$Z'$ Portal Dark Matter in the Minimal $B-L$ Model}",
    eprint = "1803.06793",
    archivePrefix = "arXiv",
    primaryClass = "hep-ph",
    doi = "10.1155/2018/5340935",
    journal = "Adv. High Energy Phys.",
    volume = "2018",
    pages = "5340935",
    year = "2018"
}
\end{document}